\documentstyle[12pt]{article}
\newcommand{\beq}{\begin{equation}}
\newcommand{\eeq}{\end{equation}}
\newcommand{\bea}{\begin{eqnarray}}
\newcommand{\eea}{\end{eqnarray}}

\newcommand{\ra}{\rightarrow}

\newcommand{\gsim}{\lower.7ex\hbox{$
\;\stackrel{\textstyle>}{\sim}\;$}}
\newcommand{\lsim}{\lower.7ex\hbox{$
\;\stackrel{\textstyle<}{\sim}\;$}}

\def\lsim{\mathrel{\rlap{\lower3pt\hbox{\hskip0pt$\sim$}}
    \raise1pt\hbox{$<$}}}         
\def\gsim{\mathrel{\rlap{\lower4pt\hbox{\hskip1pt$\sim$}}
    \raise1pt\hbox{$>$}}}         

\renewcommand{\Im}{{\rm Im}\,}

\newcommand{\aver}[1]{\langle #1\rangle}

\newcommand{\La}{\overline{\Lambda}}
\newcommand{\Lam}{\Lambda_{\rm QCD}}
\newcommand{\mhad}{\mu_{\rm hadr}}

\newcommand{\al}{\alpha}
\newcommand{\be}{\beta}
\newcommand{\as}{\alpha_s}
\newcommand{\GeV}{\,\mbox{GeV}}
\newcommand{\MeV}{\,\mbox{MeV}}
\newcommand{\matel}[3]{\langle #1|#2|#3\rangle}

\newcommand{\eq}[1]{Eq.\hspace*{.15em}(\ref{#1})\hspace*{-.3em}
}

\newcommand{\re}[1]{Ref.~\cite{#1}}
\newcommand{\res}[1]{Refs.~\cite{#1}}

\begin{document}

\begin{titlepage}
\renewcommand{\thefootnote}{\fnsymbol{footnote}}

\begin{flushright}
UND-HEP-98-BIG\hspace*{.03em}1\\
February 1998\\
hep-ph/9804275\\
\end{flushright}

\begin{center} \Large
{\bf Heavy Quark Expansion in Beauty and its Decays}
\end{center}
\vspace*{1.5cm}
\begin{center} {\Large
Nikolai Uraltsev} \\
\vspace{1.2cm}
{\normalsize
{\it Dept. of Physics, Univ. of Notre Dame du Lac, Notre Dame, IN 46556,
U.S.A.}\\
{\small \rm and} \\
{\it St.\,Petersburg Nuclear Physics Institute,
Gatchina, St.\,Petersburg 188350, Russia\footnote{Permanent address}
}\\
}

\vspace*{1.7cm}
{\small \it  Lectures given at the International School of
Physics {\rm Enrico Fermi}\\ ``Heavy Flavour Physics: A Probe of
Nature's Grand Design''\\
Varenna, July 7-18 1997} \vspace*{1.cm}

{\large{\bf Abstract}}\\
\end{center}
I give an introduction to the QCD-based theory of the heavy flavor
hadrons and their weak decays. Trying to remain at the
next-to-elementary level and skip technicalities, I concentrate on the
qualitative description of the most important applications and physical
meaning of the theoretical statements. The numerical results of the
dedicated theoretical analyses of extracting $|V_{cb}|$ are given and
the possibilities to determine $|V_{ub}|$ in future are discussed.  At
the same time I describe in simple language subtle peculiarities
distinguishing actual QCD of heavy quarks from naive quantum mechanical
treatment often applied to heavy flavor hadrons. These subtleties are
often mistreated. Particular attention is paid to the concept of the
heavy quark mass and its evaluation, to the kinetic operator and the
question of the $1/m_Q$ corrections to inclusive widths of heavy flavor
hadrons. I argue that the properly defined $b$ quark mass is known with
a good accuracy from the $\bar b b$ threshold cross section.

\vspace*{.2cm}
\vfill
\noindent
\end{titlepage}
\addtocounter{footnote}{-1}

\newpage

\tableofcontents

\newpage

\section{Introduction}

The Standard Model (SM) has six quarks whose existence  by now
is reliably established in experiment. They are naturally allocated
to three generations
$$
\left(
\begin{array}{cc} u \\ d \end{array}
\right) \qquad
\left(
\begin{array}{cc} c \\ s \end{array}
\right) \qquad
\left(
\begin{array}{cc} t \\ b \end{array}
\right)
$$
Masses of $u$ and $d$ quarks are
only a few $\MeV$, much smaller than $\Lam$, and in most applications
of hadron physics $u$ and $d$
can be considered as massless.  The strange
quark mass is about $150\MeV$, that is, literally only a little smaller
than $\Lam$. Nevertheless, there is ample evidence that treating
$s$-quark as light is justified, and corrections to the (light) $SU(3)$
symmetry are reasonably small. Thus, $m_s$ lies essentially below the
actual typical hadronic QCD scale $\mhad \sim 500\mbox{ to } 700 \MeV$.

Those quarks $Q$ for which $m_Q \gg \mhad \sim (2\,$--$\,3) \Lam$ are
heavy quarks. The sixth $t$ quark is the heaviest, $m_t\approx
170\GeV$. However, it is too heavy. Its width due to the `semiweak'
decay $t\ra b+W^+$ is $\Gamma_t \approx 1\GeV$. It decays too fast for
the $t$-hadrons, the bound states with light quarks to be formed.

The best candidate for application of the Heavy Quark Expansion
(HQE)~\footnote{There exists a commonly accepted abbreviation HQET
which sounds very similar. It denotes the ``Heavy Quark Effective
Theory'', a rather specific approach extensively used for a number of
exclusive transitions.  In the past every theoretical consideration
related to heavy quarks was often not quite consistently called HQET.
The difference will be illustrated later. The term HQE was introduced,
apparently, in \cite{Ds} to distinguish a more general direct
QCD-based approach.} are beauty hadrons. It is heavy enough to
confidently use the expansion in $1/m_b$, yet it appears not too heavy
for the leading corrections to the heavy quark limit to be negligible.

The charmed quark $c$ can be called heavy only with some
reservations.
While in some cases it gives a reasonable approximation, $m_c$ often
appears manifestly too low for a quantitative treatment of charmed
physics in the $1/m_Q$ expansion. There is no universal answer here,
and considerable care must be exercised in such a treatment in every
particular case.

The hadrons we consider are
composed from
one heavy quark $Q$, and a light cloud: a light antiquark $\bar q$, or
diquark $qq$,
together with the gluon medium and light
quark-antiquark pairs. The role of the gluon medium is to
keep all `valence' constituents together, in a
colorless bound state which will be generically denoted by $H_Q$.
Therefore, we have the following simplified picture of a heavy-flavor
hadron.  The heavy quark has a small
size $\sim 1/m_Q$, and is surrounded by a static Coulomb-like color
field $A_0$ at small distances. Non-Abelian self-interaction slightly
modifies the potential, but the non-linearity is driven by the coupling
$\as\left(r^{-1}\right)/\pi$ and is not significant. At larger
distances the self-interaction strengthens, at $R \gsim \Lam^{-1}$ it
is completely nonperturbative: the soft modes of the light fields are
strongly coupled and strongly fluctuate.

What happens when such a hadronic state is disturbed? In weak decays
the standard situation is that the external (to QCD) forces like
$W$-bosons interact with the heavy quark, say, instantaneously replace
the $b$ quark by $c$ quark generally changing its velocity, make it to
move.  Such an event excites first the typical modes of the heavy
quark, hard gluons with $\vec{k}_{\rm typ} \sim m_Q$. Since
$$
\as \left(k_{\rm typ} \right)\sim  \as(m_Q) \ll 1
$$
one can apply {\it perturbation theory} to describe what happens there.

An actual decay process, nevertheless, eventually runs into the
strong-interaction nonperturbative domain of $\vec{k} \sim \omega \sim
\Lam$, where $\omega$ denotes the characteristic frequencies. The final
hadronization dynamics shaping the hadrons we observe in experiment, is
a result of soft nonperturbative physics which is responsible for
confinement. It is important that the complicated final state dynamics
involve $\omega \ll m_Q$. This means that to deal with nonperturbative
effects one can safely use a {\it nonrelativistic expansion} for
the heavy quark.

The main subject of the HQE is nonperturbative physics. The reason is
twofold: first, the perturbative corrections are straightforward. The actual
computations are often cumbersome and, beyond the first-order effects,
typically require sophisticated state-of-the-art technique.
Nevertheless, they are conceptually simple.

The second reason is that, anyway the perturbative corrections are
calculated in the full QCD rather than in the effective low-energy
theory, since the part (often dominant) comes just from the gluon
momenta $k\sim m_Q$ where the nonrelativistic approximation is not
applicable. Still, it is worth noting that the interplay of the
perturbative and nonperturbative effects is quite nontrivial and
involves theoretical subtleties which are not always fully
respected.

The treatment of the nonperturbative effects is a nontrivial problem, and
different methods of QCD are used here. The basic tool for all of them
in heavy quarks is the Wilson operator product expansion (OPE)
\cite{wilson}. Stating briefly,
$$
\mbox{HQE } = \mbox{ OPE } +
\mbox{ nonrelativistic expansion}\;\;.
$$
Unless an analytic solution
of QCD is found, these two ingredients are indispensable for the heavy
quark theory. What do they mean and help us with?

\subsection{Nonrelativistic Expansion}

The main simplification
of a nonrelativistic treatment is that the number of heavy
quarks $n_Q$ and antiquarks $\bar n_Q$ are separately conserved.
Propagation of the nonrelativistic quark
with usual relativistic
Green function for the heavy quark contains also the process
of the $Q\bar Q$ pair creation if the time ordering of the
vertices along the line is reversed somewhere.  In the nonrelativistic
kinematics, however, such
configurations yield a power-suppressed contribution, since the
virtuality of the intermediate state (the energy denominator, in
the language of noncovariant perturbation theory of quantum
mechanics) is of the order of $2m_Q$. In order to observe such
processes as real ones, it would be necessary to supply energy at least
as large as $\omega \approx 2m_Q$.

Heavy quarks can appear also in closed loops, for example, in the
processes of the virtual gluon conversion. For heavy quarks
such effects are also suppressed, $\sim
\vec{k}^2/m_Q^2$ if the gluon momentum $k$ is much smaller than $m_Q$.

As a result, the field-theoretic, or second-quantized
description of the heavy  quark  becomes redundant, and
it is sufficient to resort to its usual quantum-mechanical (QM)
treatment.
For example, the wavefunction of a heavy flavor hadron takes the form
\beq
\Psi_\al\left[\vec{x}_Q,\, \left\{x_{\rm light} \right\} \right]\;\;,
\label{4}
\eeq
where $\vec{x}_Q$ is the heavy quark coordinate, $x_{\rm light}$
generically represents an infinite number of light degrees of freedom;
index $\al$ describes the heavy quark spin.

A relativistic $S=\frac{1}{2}$ particle has four components, i.e.
$\al=1...4\,$. The nonrelativistic spinor $\Psi(x)$ has only two of
them, $\al=1,2$ describing two spin states. The remaining $\al=\{3,4\}$
components describe antiparticles
which decouple in the nonrelativistic theory:
\beq
Q(x) \;=\;
\left(
\begin{array}{cc} \Psi(x) \\ \chi(x) \end{array}
\right) \qquad \qquad
\begin{array}{ll}
\Psi(x) \sim {\cal O}(1) \\
\chi(x) \sim \frac{\vec{p}}{m_Q} \ra 0
\end{array}
\label{6}
\eeq
The nonrelativistic Hamiltonian of the spin-$\frac{1}{2}$
particle is the well known
Pauli Hamiltonian:
\beq
{\cal H}_{\rm Pauli}\;=\;
-g_s A_0 \;+ \;
\frac{\left(i\vec\partial-g_s \vec{A}\right)^2}{2m} \;+\;
\frac{g_s \vec \sigma \vec B}{2m}\;.
\label{8}
\eeq
The last operator was written with a coefficient appropriate
for an `elementary' pointlike
particle. Presence of the interaction generally renormalizes its
strength, the chromomagnetic moment of the heavy quark. In the heavy
quark limit $m_Q\ra \infty$ only the first term survives while the last
two terms describing space propagation and interaction of spin with the
chromomagnetic field $\vec{B}$, disappear. An infinitely heavy quark is
static and interacts only with the color Coulomb
potential.\footnote{Speaking of the Coulomb interaction in QCD we mean
the interaction with the time-like component $A_0$ of the color gauge
potential. It differs from the simple $1/r$ electrodynamic potential.
It is also customary to absorb the coupling $g_s$ into the definition
of the gauge potential, and I often omit it from the expressions.} In
turn, the heavy quark is a source of a static color field independent
of the heavy quark spin. The actual dynamics of the heavy quark spin
reveal itself in full only when the quark interacts with quanta having
large momentum $|\vec{q}\,| \gsim m_Q$.

It is easy to illustrate the $b$ quark spin-independence of the strong
forces in the following way. One can imagine the QCD world where,
instead of the actual $b$ quark there exists a scalar spinless $\tilde
b$ quark with the same mass. The usual spin-$0$ $B$ meson, $b\bar{q}$
would become a spin-$\frac{1}{2}$ particle $\tilde B \sim \tilde b
\bar{q}$. The $\Lambda_b$ baryon $bud$ having spin-$\frac{1}{2}$ would,
in turn become a scalar spinless $\tilde bud$ state. Nevertheless, at
$m_b \ra \infty$ the properties of $B$ and $\tilde B$ or $\Lambda_b$
and $\tilde \Lambda_b$ would become identical. The two degenerate spin
states of $\tilde B$  are counterparts of $B$ and $B^*$ of the actual
QCD.

An immediate question comes to one's mind at this point: what about
the relation between spin and statistics? In such a gedanken operation
we replace spin--$\frac{1}{2}$ hadrons  by $S=0$ ones, and {\it vice
versa}, {\it i.e.} interchange fermions and bosons. It is clear,
however, that for the states or processes with a single heavy quark the
statistics symmetry properties do not play a role.

Such an independence of the strong dynamics of the heavy quark spin is
called the heavy quark {\it Spin Symmetry}.

The rest-mass $m_Q$ of the heavy quark also enters in a trivial way:
the Hamiltonian merely contains $(n_Q+\bar n_Q) m_Q$. Since both are
fixed, it is just an overall additive constant. For a moving quark this
constant is $E=\sqrt{m_Q^2+\vec{p}^{\,2}} = m_Q\sqrt{1+\vec{v}^{\,2}}$.
Therefore, the actual dynamics is not affected by the concrete value of
the mass $m_Q$. This is a {\it Heavy Flavor Symmetry} which states, for
example, the equal properties of charmed and beauty hadrons to the
extent they both can be considered heavy enough.

The heavy flavor symmetry leads also to certain scaling behavior of the
transition amplitudes with the heavy flavor hadrons: the amplitudes
depend on
their velocities rather than on the absolute values of momenta:
\beq
A\left(
P_Q^{in}, p_i^{in};\: P_{Q'}^{out}, p_l^{out}
\right) \;=\;
{\cal A} \left(
\frac{P_Q^{in}}{m_Q}, p_i^{in};\: \frac{P_{Q'}^{out}}{m_{Q'}}, p_l^{out}
\right) \;.
\label{10}
\eeq
Here $P$ denote the momenta of the heavy flavor hadrons and $p$ refer
to other participating particles. It is important to keep in mind that
such a scaling behavior is valid only with respect to the soft part of
the interaction, and no hard gluons with $\vec{k} \sim m_Q$
are involved.

The one-particle (or QM) description of the heavy quark degrees of
freedom is the key simplification of the nonrelativistic expansion. The light
cloud, however, still requires a full-fledged field-theoretic
treatment. Even considering the static limit $m_Q \ra \infty$ where
only interaction with the Coulomb field $A_0$ remains, one should
realize that it is `very quantum' and strongly fluctuate, in contrast
to simple potential QM models where the corresponding potential
$V(x)$ is merely a $c$-number function of coordinates.

For this reason even the quantum mechanics of heavy quarks is highly
nontrivial. Exploiting the symmetry properties of the heavy quark
interactions does not require understanding of these complicated
strong-interaction dynamics. This was the main field of applications at
an early stage of theoretical development of heavy quark physics.
The recent progress is mainly related to a better treatment of basic
properties of this complicated strongly-interacting system~\footnote{In
early papers on heavy quarks the system
of soft components of quark and gluon fields constituting the light
cloud was often negligently referred to as
``brown muck",  which seems unfair. Already in the pre--heavy-quark era
most of the impressive theoretical results leading, for example, to
formulation of QCD as a theory of strong interactions, were related to
ingenious treatment of this soft cloud in the world of light hadrons.}
via application of dynamic QCD methods based on the short-distance
expansion.

\subsection{Operator Product Expansion}

The basic theoretical tool of the heavy quark theory is the Wilson
operator product expansion  \cite{wilson}. It is not so easy to give
a precise definition of what is OPE in a few understandable words. In
particular, I found that many authors often associate with it somewhat
different ideas, or even limit it to concrete technicalities.

The idea of the OPE was formulated by K.~Wilson in the late 60's,
originally in the context of the statistical problems which are closely
related to the field theories in Euclidean space. This idea, in
general, is a separation of effects originating at large and small
distances. The application to real physical processes in
Minkowski space is often less transparent and technically more
complicated, however it is always based on the same underlying
concept.

The original QCD Lagrangian
\beq
{\cal L} =-\frac{1}{4} G_{\mu\nu}^a G_{\mu\nu}^a
+\sum_{q}\bar q (i\not\!\!{D}-m_q) q
 +\sum_{Q}\bar Q (i\not\!\!{D} - m_Q) Q =
{\cal L}_{\rm light}
+ \sum_{Q}\bar Q (i\not\!\!{D} - m_Q) Q
\label{12}
\eeq
is formulated at very short
distances, or, which is the same, at a very high normalization point
$\mu =
M_0$, where $M_0$ is the mass of an ultraviolet
regulator. In other words, the
normalization point is assumed to be much
higher
than all mass scales in the theory, in particular, $\mu\gg m_Q$.
An
effective theory for  describing the
low-energy
properties
is obtained by evolving the
Lagrangian from the high scale $M_0$ down to a lower
normalization point
$\mu$.
This means that we  integrate out, step by step,
all high-frequency modes in the theory thus calculating the
Lagrangian ${\cal L}(\mu )$. The latter is a full-fledged
Lagrangian  with respect to  the soft modes
with characteristic frequencies less than $\mu$.  The hard
(high-frequency) modes determine the coefficient  functions
in ${\cal L}(\mu )$, while the contribution of the soft modes
is hidden in the matrix elements of (an infinite set of)
operators appearing in ${\cal L}(\mu )$.
The value of this approach,  outlined by Wilson long ago
\cite{wilson}, has become
widely recognized and exploited in countless applications.
The peculiarity
of the heavy quark theory lies in the fact that the {\em in} and
{\em out} states contain heavy quarks. Although we
integrate out the  field fluctuations with the frequencies down to
$\mu \ll m_Q$, the heavy quark fields themselves are not integrated
out since we consider the sector with nonzero heavy-flavor
charge. The effective Lagrangian ${\cal L}(\mu )$ acts in this
sector. Since the heavy quarks are neither produced nor annihilated,
any sector with the given $Q$-number is treated separately
from all others.

The simplest and most familiar example of integrating out high-frequency
modes is the low-energy four-fermion weak decay Lagrangian.
The weak decays are mediated by virtual exchanges of $W$
bosons and, in principle, depend on the external momenta via the
momentum flowing through the $W$ boson line:
\beq
A= \frac{g_2^2}{8} V_{cb}V_{ud}^*\,
\bar{u}_c(p_c)\gamma_\mu(1\!-\!\gamma_5) u_b(p_b)\,
\bar{u}_d(p_d)\gamma_\nu(1\!-\!\gamma_5) u_u(p_u)\:
\frac{\delta_{\mu\nu} - \frac{(p_b-p_c)_\mu (p_b-p_c)_\nu}{M_W^2}}
{(p_b-p_c)^2 - M_W^2} \,.
\label{16}
\eeq
The nontrivial denominator signifies propagation of a particle, the $W$
boson. Eliminating explicit $W$ boson from the theory we are left with
amplitudes generated by the local effective Lagrangian:
\beq
{\cal L}_{\rm w} \;=\; -\frac{G_F}{\sqrt{2}}V_{cb}V_{ud}^*
\,\int\; d^4 x\:
\bar{c}(x)\gamma_\mu(1-\gamma_5) b(x)\,
\bar{d}(x)\gamma_\mu(1-\gamma_5) u(x)\; +\; {\cal
O}\left(\frac{p^2}{M_W^2}\right)\;,
\label{18}
\eeq
provided all momenta are much smaller than $M_W$.

This example is particularly simple in a few aspects: the $W$ boson
does not couple to gluons, and it has a large energy $M_W$ even if
it carries a small momentum. The OPE works perfectly in a more general
situation as well: for example, one can integrate out hard
gluons with large momenta (virtualities) even though the gluon field is
massless, and gluons couple to each other. In spite of these
complications, the effect of hard gluons is given by series of local
operators generating simple effective vertices, which are more and more
suppressed when their dimension increases. To state it differently, soft
particles (fluctuations) can ``look inside'' a hard process only for
the price of power suppression.

This basic idea applies to any field theory. It brings in additional
advantages for QCD related to the fact that QCD is a renormalizable
theory. Another convenience is that the emerging effective operators
are gauge-invariant. This reduces their number and leads to many useful
relations among generated amplitudes. Combining the OPE with
nonrelativistic expansion is particularly constraining. For example, if
the heavy quark spin is switched off, there is no independent axial
current, {\it etc}.

Another feature is that in QCD the short-distance physics is governed
by the small running coupling $\as(q^2)$ and, therefore, the
coefficients in front of these operators, affected by the QCD dynamics,
can be calculated in perturbation theory.

Heavy quark physics has an intrinsic large scale $m_Q$. Nevertheless,
the majority of phenomenologically interesting processes are `soft',
that is, essentially dependent on the low-scale dynamics as well. There
are a few exceptions which are the subject of special attention. An
important progress in the heavy quark theory was reached identifying
those processes that are genuinely `hard' in this respect.

\section{Basics of the Heavy Quark Theory}

\subsection{Effective Hamiltonian}

What are virtual particles which must be integrated out to obtain an
effective theory suitable for heavy quarks? In the case of ordinary
weak decays the answer was quite obvious, at least in the tree
approximation: one had heavy virtual $W$ which could virtually appear
only for a short interval of time $\sim 1/M_W$ allowed by the
uncertainty principle. For a heavy quark $Q$ its mass $m_Q$ is a large
parameter, but the quark itself should not be eliminated completely
since it is present in the initial state.

The strategy is described in the
text-books. The nonrelativistic fermion field
$Q(x)$ has four degrees of freedom. Two of them, $\Psi(x)$ in
Eq.~(\ref{6}) are nearly on-shell and two $\chi(x)$ are highly virtual
describing excitation of antiquarks. One needs to integrate out first
the antiparticle fields $\chi(x)$. For simplicity, we consider it in
the rest frame.

Integrating out the antiparticle fields can be done straightforwardly
since the QCD Lagrangian is bilinear in the quark fields. In this way
one would obtain the (tree-level version of the) so-called Lagrangian
of HQET ${\cal L}_{\rm HQET}$.
It is a correct nonrelativistic Lagrangian up to the first order in
$1/m_Q$.

This does not complete the program, however: one yet has a full
`particle' field $\Psi(x)$ which includes all frequencies from $0$ to
$\infty$. Without taking care of this problem we would be in a position
of a theorist in the end of the 20's when the nonrelativistic QM of a
charged particle interacting with the electromagnetic field had been
written, but its usefulness was severely limited by the presence of
ultraviolet (UV) divergences when the corrections due to quantum
fluctuations were considered.  Such a problem does not show up in
the potential models where modes with $k \gg \Lam$ are not excited.
However, in actual QCD all radiative corrections would diverge due to
large momentum gluons.

Therefore, besides the antiparticle fields one needs to integrate out
hard gluons and the high frequency components of the heavy quark field
$\Psi(x)$ itself, those for which $|\vec{k}\,|, \omega > \mu$.
The scale $\mu$ is the normalization point of the effective theory.
Since the configurations we integrate out depend on $\mu$, the
remaining effective low-energy theory cannot but have to be
$\mu$-dependent.

In practice, we want to have $\mu\ll m_b$ and, actually, as low as
possible. On the other hand, $\mu$ must still belong to the perturbative
domain. In practice this means that the best choice routinely adopted
for applications is $\mu\sim \mbox{ a few } \times \Lam$, from $0.7
\mbox{ to } 1\GeV$. All coefficients in the effective Lagrangian
obtained in this way are functions of $\mu$.

The nonrelativistic heavy quark expansion can be performed in somewhat
different ways. The best known examples are NRQCD (Nonrelativistic QCD)
\cite{nrqcd} and HQET \cite{hill,georg} which are mainly used for
describing exclusive heavy flavor transitions or static
characteristics. At first glance, HQET differs from NRQCD only
in the choice of
the zero-order approximation for $m_Q \ra \infty$,
by discarding even the standard QM kinetic term $\vec{p}^{\,2}/2m_Q$
together with the Pauli term $\vec\sigma \vec H /2m_Q$. However, in higher
orders in $1/m_Q$ the expansion has not
been applied quite consistently until recently. The correct way was
advocated by K\"{o}rner {\it et al.} \cite{korner} and followed the
text-book methods elaborated in early days of QM. It thus coincided
with the approach of NRQCD (the practical applications of the latter
usually did not address higher orders in $1/m_Q$ where the
difference emerges). Unfortunately for HQET, for a few years these
justified suggestions were ignored and, with a delay, admitted only
recently \cite{manohfol,grinrot}.

Another peculiarity of HQET as an effective theory is that, at the
technical level, it attempts to ``integrate out'' perturbative
corrections down to $\mu=0$ (except for some special cases). While such
a procedure is not justified and, rather, looks illegitimate, it is
often possible to perform such an integration in the concrete orders in
perturbation theory. These inconsistencies seemed to be curable
technicalities. However, in a number of publications
\cite{neubtasi,lms,ns,manohfol,neubdr} they were canonized and concrete
computational techniques having limited applicability  were promoted
to the status of indispensable elements of HQET itself. As a
result, there is no yet a clear understanding of what exactly is HQET
as an actual quantum field theory (QFT), and this ambiguity is
reflected in the existing definitions of some fundamental
HQET parameters. Inaccurate treatment of such theoretical subtleties
sometimes led to paradoxical claims. For example, it was stated
\cite{lms} that HQET contains a source of the nonperturbative
$1/m_Q$ corrections to inclusive decay widths, contrary to
the theorem established in QCD itself.

To summarize, in treating heavy quarks we separate all strong
interaction effects into `hard' and `soft' introducing a
normalization scale $\mu$. To calculate the effect of the
short-distance (perturbative) physics we use the original QCD
Lagrangian Eq.~(\ref{12}). The soft physics is treated by the
nonrelativistic Lagrangian where the heavy quarks are represented by
the corresponding nonrelativistic fields $\varphi_Q$:
$$
{\cal L}_{\rm eff}=
-\frac{1}{4} G_{\mu\nu}^2
+\sum_{q}\bar q (i\not\!\!{D}-m_q) q
+\sum_{Q} \left\{
-m_Q \varphi_Q^+ \varphi_Q + \varphi_Q^+ iD_0 \varphi_Q  \;-
\right.
$$
\beq
\left.
-\;\frac{1}{2m_Q} \varphi_Q^+ \left(\vec\sigma i\vec{D}\right)^2
\varphi_Q -
\frac{1}{8 m_Q^2}\,\varphi_Q^+ \, \left[
-(\vec D\vec E)+\vec\sigma\cdot \{\vec E\times\vec\pi-\vec\pi\times\vec E\}
\right]\, \varphi_Q \right\}
\label{21}
\eeq
where
\beq
\vec{\pi} \equiv i\vec{D} = \vec{p} - \vec{A}\;,
\qquad
\left(\vec\sigma i\vec{D}\right)^2 =
\left(\vec\sigma \vec{\pi}\right)^2=
\vec\pi^{\,2}+\vec\sigma \vec{B}\;;
\label{22}
\eeq
for simplicity I
wrote only the tree-level $\mu$-independent coefficients.  By the
standard rules one constructs from the Lagrangian (\ref{21}) the
corresponding Hamiltonian.
The heavy quark part takes the
form~\footnote{The leading term $A_0$ is often omitted here. It is then
implied that the time evolution operator is
$\pi_0=i\frac{\partial}{\partial t} + A_0$ rather than
$i\frac{\partial}{\partial t}$ in the usual Schr\"{o}dinger equation.
This can be consistently carried out through the analysis.}
\beq
{\cal H}_Q\,=\,-A_0+\frac{1}{2m_Q}\,({\vec\pi}^2 + \vec\sigma \vec
B)\,+\, \frac{1}{8m_Q^2}\,\left[-(\vec D\vec E)+ \vec\sigma \cdot
\{\vec E\times\vec\pi-\vec\pi\times\vec E\} \right] +{\cal
O}(1/m_Q^3) \;.
\label{ham}
\eeq

Since the external interactions (electromagnetic, weak {\it etc.}) are
given in terms of the full QCD fields $Q(x)$, one needs also the relation
between $Q(x)$ and $\varphi_Q(x)$:
\beq
\varphi=\left(1+\frac{(\vec\sigma\vec\pi)^2}{8m_Q^2}+ ...
\right)\frac{1+\gamma_0}{2}\,Q\;, \qquad \;
\frac{1-\gamma_0}{2} Q = \frac{\not\!\!{\,\pi} }{2m_Q} Q\, .
\label{23}
\eeq

Although obtaining the nonrelativistic Lagrangian is a standard
text-book procedure, let me briefly recall it. One starts with
\beq
{\cal L}_{\rm heavy}^0 = \bar Q(x) (i\not\!\!D -m_Q)Q(x)
\label{24}
\eeq
and factors out of the $Q(x)$ field the ``mechanical'' time-dependent
factor associated with the rest energy $m_Q$:
\beq
Q(x) = {\rm e}\,^{-im_Qt}{\tilde Q} (x) \; .
\label{25}
\eeq
In an arbitrary frame moving with four-velocity $v_\mu$ it
takes the following form:
\beq
Q(x) = {\rm e}\,^{-im_Qv_\mu x_\mu}{\tilde Q} (x) \;.
\label{13a}
\eeq
Then
$$
iD_\mu Q(x) = {\rm e}\,^{-im_Q(v x)}\, \left( m_Q v_\mu +
\pi_\mu\right) \tilde Q (x)\;, \qquad \qquad \pi_\mu
\equiv \hat P_\mu-m_Qv_\mu\;.
$$
The Dirac equation  $\left(i\not\!\!D -m_Q\right)Q = 0$ takes the form
(now the tilde on $Q$ is omitted)
\beq
\frac{1-\gamma_0}{2}\, Q = \frac{\not\!\!{\,\pi} }{2m_Q}\, Q\: ,
\qquad\qquad \pi_0 Q = -\frac{\pi^2 +\frac{i}{2}\sigma G}{2m_Q}\, Q\, .
\label{27}
\eeq
$$
\frac{i}{2} \sigma G =\frac{i}{2} g_s \sigma_{\mu\nu} G_{\mu\nu}\:,
\qquad\qquad
i G^{\mu\nu}= \left[\pi_\mu, \pi_\nu\right] = \left[\hat P_\mu,
\hat P_\nu\right]
$$
which allows one to exclude the small low components
$\frac{1-\gamma_0}{2} Q(x)$ expressing them {\it via} the `large' upper
components $\frac{1+\gamma_0}{2} Q(x)$.

A subtlety emerges on this route  that must be treated properly: at
order $1/m_Q^2$ the time derivative $\partial_0$ appears with a
nontrivial coefficient depending, for example, on the gluon field. This
can be eliminated, and the time derivative returned to its canonical
form performing the Foldy-Wouthuysen transformation
\beq
\varphi(x)=\left(1+\frac{(\vec\sigma\vec\pi)^2}{8m_Q^2} + ...
\right)\frac{1+\gamma_0}{2}\,Q(x)\;.
\label{28}
\eeq

Why does one need this Foldy-Wouthuysen transformation? It is worth
discussing it since HQET uses the different fields
\beq
h(x)= \frac{1+\gamma_0}{2}\,\tilde Q(x)
\label{29}
\eeq
to all orders in $1/m_Q$ to represent the effective low-energy heavy
quark field. As a result, for example, the higher-order terms in the
HQET effective Lagrangian are different from Eq.~(\ref{21}) starting
already $1/m_Q^2$ (in this order they consist of the so-called
Darwin and convection current, or $LS$ interaction). Moreover, $h(x)$
is not mass-independent but contains explicit $1/m_Q^2$, $1/m_Q^3$ ...
terms depending on the external fields.

In principle, it is not a mistake {\em per se}. If one were able to
calculate all HQET Green functions computing the functional integral
with such a HQET Lagrangian exactly, it wouldn't have mattered: any
choice of dynamic variables is legitimate as soon as the proper
reduction technique is applied to obtain the $S$-matrix elements from
the Green functions. In reality, such a possibility is as Utopian for
HQET as for full QCD. Therefore, in practice one can only rely on
various expansion techniques, and here using the correct
nonrelativistic fields incorporating the Foldy-Wouthuysen
transformation and the Lagrangian becomes mandatory, as was realized 70
years ago. Let me briefly illustrate it.

The standard approach in the heavy quark expansion uses the
well-elaborated methods of QM operating with the QM states, in
particular, in constructing the perturbation theory in $1/m_Q$. In QM
the terms in the expansion are expressed {\it via} various
$T$-products, and this holds in the standard form only if the canonical
coordinates are used. Otherwise, somewhat different objects replace
usual $T$-products.  Moreover, the effective Hamiltonian describing the
evolution of the $h$ fields becomes, in a sense, non-Hermitian in
higher orders in $1/m_Q$.

On the other hand, one wants to employ the QFT methods which, in
particular, are used in the OPE. They rely on the equations of motion
for quantum field operators. If a system is described by the Lagrangian
${\cal L}$, one has the classical equations of motion
\beq
\partial_\mu \frac{\delta {\cal L}}{\delta \partial_\mu \varphi(x)}
-\frac{\delta {\cal L}}{\delta \varphi(x)} \;=\;0\;.
\label{32a}
\eeq
However, we always use the very same equations of motion for the
quantum Heisenberg {\it operators}~\footnote{In fact, the classical
equations of motion are often modified due to anomalies. This is a
different aspect related to the UV divergences in the theory. What is
discussed here has nothing to do with divergences and applies even
to the systems with the finite number of degrees of freedom.}
$\varphi(x)$.  The validity of these operator equations is a nontrivial
fact (recall that for any operator $A$ the time derivative by
definition is $i\left[{\cal H}, A\right]$), and is a consequence of an
analogue (generalization) of the Erenfest theorem in QM. This is true,
however, {\it only} if the term with the time derivative has a
canonical (coordinate-independent in QM, of field-independent in the
field theory) form. For the HQET quantum fields the naive equations of
motion do not hold. This general fact was ``empirically'' observed in
\cite{flsgamma} (see also \cite{mw}) in attempts to reproduce in
HQET the $1/m_b^2$ corrections to the inclusive decay widths calculated
earlier directly in QCD. They required modifications of the existed
strategy of obtaining higher-order $1/m_Q$ corrections. This was
interpreted as a `subtlety in applying equations of motion'
\cite{flsgamma,mw}. Surprisingly, the well known origin of this
subtlety has not been realized. Not all related flaws have been
eliminated, however. The expansion of the exclusive heavy flavor
transition at order $1/m_Q^2$ \cite{fn} suffering from the same
inconsistencies has never been revised even though it is used nowadays
by Neubert to claim the most accurate evaluation of the zero recoil
$B\ra D^*$ formfactor. This will be discussed in more detail in
Sect.~3.5.

Finally, when no rigorous QCD evaluation of the nonperturbative
effects can be
made, one often resorts to simplified QM constituent quark models
({\it e.g.}, ISGW \cite{isgw}) where all wavefunction overlaps,
expectation values and other characteristics can be computed. In
applying these computations to the QCD system one certainly must use
the proper relation between the QCD fields and the variables
employed in the model. As was first noted, apparently, by
A.~Le~Yaouanc \cite{leya}, this was not always done properly in HQET
\cite{fn}.

Therefore, it is certain that the Foldy-Wouthuysen transformation is
not an obsolete, mole-beaten technique used only before the advantages
of the modern formulations of the quantum fields theories were
known or appreciated in full. Using the standard nonrelativistic
expansion also often gives a simple derivation, or at least transparent
interpretation of many results. Missing it may easily lead to direct
mistakes.

\subsection{Applications to Spectroscopy of Heavy Flavor Hadrons}

To illustrate the consequences of the heavy quark Hamiltonian, let us
consider the masses of hadrons containing a single heavy quark. This
is not a dynamic question and does not require more than just symmetry
properties of ${\cal H}_Q$. It is described in much detail in a number
of old reviews \cite{reviews}, and I can be brief.

The mass of a hadron $H_Q$ is given by the expectation value of the
Hamiltonian:
\beq
M_{H_Q}\;=\; \frac{1}{2M_{H_Q}} \matel{H_Q}{{\cal H}_{\rm tot}}{H_Q}\;,
\label{27a}
\eeq
and we have
\beq
{\cal H}_{\rm tot}\; =\;{\cal H}_{\rm light}  +{\cal H}_Q + m_Q\;,
\label{28a}
\eeq
where we can expand
\beq
{\cal H}_Q\; =\;{\cal H}_{0} + \frac{1}{m_Q} {\cal H}_1 +
\frac{1}{m_Q^2} {\cal H}_2 +\, ...
\label{29a}
\eeq
with
$$
{\cal H}_{0}\;=\; -\varphi_Q^+ A_0 \varphi_Q \;
\stackrel{\rm QM}{\longrightarrow}\; -A_0(0)\left[\left\{x_{\rm
light}\right\}\right]\;,
$$
$$
\frac{1}{m_Q} {\cal H}_1 \;=\;
\,\frac{1}{2m_Q}\,({\vec\pi}^2 +
\vec\sigma \vec B)\;,
$$
\beq
\frac{1}{m_Q^2} {\cal H}_2 \;=\;
\frac{1}{8m_Q^2}\,\left[-(\vec D\vec E)+
\vec\sigma \cdot
\{\vec E\times\vec\pi-\vec\pi\times\vec E\} \right] \;,
\label{30}
\eeq
{\it etc.} (the first terms in ${\cal H}_2$ is called Darwin and the
second is  the convection current, or $LS$ term).
Therefore,
\beq
M_{H_Q} = m_Q +\bar\Lambda + \frac{1}{2m_Q}
\frac{\matel{H_Q}{{\vec\pi}^2 +\vec\sigma\vec B }{H_Q} }
{2M_{H_Q}}+ ... =
m_Q +\bar\Lambda + \frac{(\mu_\pi^2 - \mu_G^2)_{H_Q}}{2m_Q}
+ ...
\label{32}
\eeq
Here we introduced the notations $\mu_\pi^2$, $\mu_G^2$ for the
expectation values of two $D=5$ heavy quark operators which will often
appear in our discussion:
$$
\mu_\pi^2 =
\matel{H_Q}{{\vec\pi}^2}{H_Q}_{\rm QM} \equiv
\frac{1}{2M_{H_Q}} \matel{H_Q}{\bar{Q}{\vec\pi}^2Q(0)}{H_Q}_{\rm QFT}
$$
\beq
\mu_G^2 =
\matel{H_Q}{\vec\sigma\vec B }{H_Q}_{\rm QM} \equiv
\frac{1}{2M_{H_Q}} \matel{H_Q}{-\bar{Q}
\frac{i}{2}\sigma_{\mu\nu}G^{\mu\nu} Q(0)}{H_Q}_{\rm QFT}\;.
\label{34}
\eeq

The physical meaning of $\mu_\pi^2$ is quite evident:
the heavy quark inside $H_Q$ experiences a {\em zitterbewegung}
due to its coupling to light cloud.
Its average spatial momentum squared is $\mu_\pi^2$. The second
expectation value measures the amount of the chromomagnetic field
produced by the light cloud at the position of the heavy quark.
In principle, the actual heavy hadron states $H_Q$ depend on $m_Q$
{\it via} the $1/m_Q$-suppressed terms of the Hamiltonian. Therefore,
the above expectation values also have such terms. Often it is
convenient to consider the asymptotic values at $m_Q \ra \infty$, and
to use, correspondingly, the eigenstates of the $m_Q \ra \infty$
Hamiltonian ${\cal H}_0+{\cal H}_{\rm light}$.

The value of $\La = \lim_{m_Q \ra \infty} \left(M_{H_Q}-m_Q\right)$ has
the scale of $\Lam$ and depends on the state of light degrees of
freedom. These states generally carry spin $j$. In the limit $m_Q\ra
\infty$ the heavy quark spin decouples since the spin-dependent parts
are present only starting ${\cal H}_1$. Thus, the heavy flavor hadrons
can be classified not only by their total spin $J$ but by the spin of
light degrees of freedom $j$. It would be just the overall spin of a
hadron in the hypothetical world with the spinless heavy quarks
discussed in Sect.~1.1. Unless $j=0$, there are two values of the total
spin $J=j\pm \frac{1}{2}$. The corresponding states form a `hyperfine'
multiplets and are degenerate up to $1/m_Q$ corrections. They are, for
example
$$
j=1/2 \;\; \left\{
\begin{array}{lll} D,\; & B\;\, & J=0\\
D^*, & B^*  & J=1
\end{array}
\right.
\qquad \qquad
j=0\,,\;J=1/2 \qquad \Lambda_c\,,\;\,\Lambda_b
$$
For $\Lambda_Q$-baryons all spin is carried by the heavy quark (up to
$1/m_Q$ corrections). The observed spectroscopy of these states clearly
supports this picture:
$$
M_{\Lambda_b}-M_B \simeq 350\MeV\qquad\qquad M_{B_s}-M_{B^-} \simeq
90\MeV
$$
\beq
M_{\Lambda_c}-M_D \simeq 420\MeV\qquad\qquad M_{D_s}-M_{D^0} \simeq
104\MeV
\label{36}
\eeq

These relations are easily improved including $1/m_Q$ terms,
Eq.~(\ref{32}). The operator $\bar{Q} \vec{\pi}^{\,2} Q$ is
spin-independent and its expectation value is the same for all members
of a hyperfine multiplet. It does not split masses inside the multiplet.

The chromomagnetic operator $\bar{Q} \frac{i}{2} \sigma G Q = - 2
\vec{S}_Q \vec{B}(0)$ depends on the heavy quark spin $\vec{S}_Q$ and
thus lifts degeneracy leading to the hyperfine splitting among the
members of a multiplet. Restricted to a particular
hyperfine multiplet, the chromomagnetic field $\vec{B}(0)$ as an
operator is proportional to the operator of spin of light degrees of
freedom:
$$
\vec{B}\; =\; c \cdot \vec{j}\;.
$$
Therefore,
\beq
\aver{\vec\sigma \vec{B}} \;=\; 2c \aver{\vec{S}_Q \vec{j}}\;=\;
c\left(J(J+1)-j(j+1)-\frac{3}{4} \right)\;.
\label{38}
\eeq
As a matter of fact, it is easy to see that
\beq
\sum_{H_Q} \:\matel{H_Q}{\bar{Q} \vec{\sigma}\vec{B} Q}{H_Q}
\;\equiv \;
{\rm Tr}\,\bar{Q} \vec{\sigma}\vec{B} Q \;=\;0
\label{39}
\eeq
always holds if the summation is performed over a hyperfine
multiplet. Therefore, for example,
\beq
\mu_G^2(B) + 3 \mu_G^2(B^*)\;=\;0\;.
\label{40}
\eeq
In the $\Lambda_b$ family the expectation value of $\bar{b} \frac{i}{2}
\sigma G b$ vanishes.

So far most of the practical applications refer to $B$ mesons; as a
result, usually $\mu_\pi^2$ proper denotes the expectation value of the
kinetic operator just in $B$ or $B^*$. Likewise,  $\mu_G^2\equiv
\mu_G^2(B) = -3\mu_G^2(B^*)\,$.
Experimentally,
\beq
M_{B^*}-M_B\;\simeq \; \left(\frac{1}{3}+1\right)
\frac{\mu_G^2}{2m_b}\;=\; \frac{4}{3}\frac{\mu_G^2}{2m_b} \simeq
46\MeV\;.
\label{41}
\eeq
Neglecting the difference between
$M_B+M_{B^*}$ and $2m_b$ (which is formally an effect of higher order
in $1/m_b$), one can write
\beq
\mu_G^2\;\simeq \; \frac{4}{3}
\left(M_{B^*}^2-M_B^2\right) \;\simeq\; 0.36\GeV^2\;.
\label{42}
\eeq
In charmed mesons $M_{D^*}-M_D\simeq 140\MeV$ which agrees with the
fact that this hyperfine splitting is proportional to $1/m_c$.
The hadron mass averaged over a hyperfine multiplet, {\it e.g.}
$\overline{M} = \frac{3M_{B^*}+M_B}{4}$ is affected at order $1/m_Q$ by
only the kinetic energy term $\mu_\pi^2/2m_b$.

In many applications one needs to know the difference between
$m_b$ and $m_c$.
This
difference is well constrained in the heavy quark expansion. For
example,
\beq
m_b-m_c=\overline{M}_B-\overline{M}_D +
\mu_\pi^2\left(\frac{1}{2m_c}-\frac{1}{2m_b}\right) +
\frac{\rho_D^3-\bar\rho^3}{4}
\left(\frac{1}{m_c^2}-\frac{1}{m_b^2}\right) \;+\; {\cal
O}\left(\frac{1}{m^3}\right).
\label{3.16}
\eeq
Here $\mu_\pi^2$ is the asymptotic expectation value of the kinetic
operator, $\rho_D^3$ is the expectation value of the Darwin term
given by the local four-fermion operator and
$\bar\rho^3 \equiv \rho_{\pi\pi}^3+\rho_S^3$ is the sum of two
positive non-local correlators \cite{optical}. As will be discussed
in the subsequent lectures, all quantities in \eq{3.16} but the
meson masses depend on the normalization point which can be
arbitrary, except that it must be much lower than $m_{c,b}$. In this
way we arrive at
\beq
m_b-m_c \simeq 3.50\GeV \;+\; 40\MeV\,
\frac{\mu_\pi^2-0.5\, {\rm GeV}^2}{0.1\,{\rm GeV^2}} \;+ \; \Delta
M_2\;\;,\;\;\;\; |\Delta M_2|\lsim 0.015\GeV \; .
\label{3.17}
\eeq
We will discuss the values of the hadronic parameters later. Now, let
me mention that the $m_b-m_c$ estimate at $\mu_\pi^2=0.5\GeV^2$ appears
to be in a good agreement with the separate determinations of $m_b$ and
$m_c$ from the sum  rules in charmonia and $\Upsilon$.

The main uncertainty in $m_b-m_c$ is due to that in the value of
$\mu_\pi^2$.
The Darwin term can be reasonably estimated relying on  factorization
\cite{motion}; it is of order $0.1\GeV^3$. The non-local
positive matrix elements  $\rho^3_{\pi\pi}$ and $\rho^3_{S}$ are
expected,
generally speaking, to be of the same order. Altogether, assuming
$|\rho_D^3-\bar\rho^3| \lsim 0.1\GeV^3$, one arrives at the
uncertainty in $m_b-m_c$ due to the higher-order terms $\Delta M_2$
quoted above.

\subsection{Heavy Quark Symmetry for Formfactors}

The amplitudes of the semileptonic weak $b\ra c$ decays are described
by the corresponding transition formfactors. Typical semileptonic $b\ra
c$ decays as they look like in Feynman diagrams are shown in Figs.~1.
The hadronic part of the weak decay Hamiltonian mediating such decays
is
\beq
{\cal H}_{\rm weak} = \int \; {\rm d}^3 x\:
{\rm e}\, ^{i\vec{q}\vec{x}}\:
\bar{c} \gamma_\mu(1-\gamma_5) b(x) \;.
\label{47}
\eeq
It says that $b$ with a momentum $\vec{p}$ is instantaneously replaced
by the $c$ quark with the momentum $\vec{p}-\vec{q}$. The resulting
state hadronizes into eigenstates of the Hamiltonian corresponding to
$m_Q=m_c$, that is, must be projected onto such states.

\begin{figure}
\vspace{3.1cm}
\includegraphics{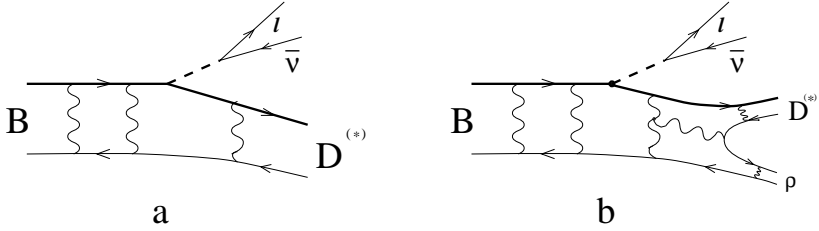}
\caption{\footnotesize
Quark diagrams for the exclusive $B\ra D^{(*)}\:$
({\bf a}) and generic ({\bf b}) semileptonic decays.}
\end{figure}

The space-time picture of the decay is simple for heavy quarks. At $t<0$
the initial $b$ hadron is at rest and constitutes a coherent state of
light degrees of freedom in the static field of the heavy quark. At
$t=0$ the $b$ quark emits the lepton pair with the momentum $\vec{q}$
and transforms into a $c$ quark. The $c$ quark gets the recoil momentum
$-\vec{q}$ and starts moving with the velocity $\vec{v}=-\vec{q}/m_c$.
Such a state is not anymore an eigenstate of the Hamiltonian, and
afterwards undergoes nontrivial evolution. The light cloud can get a
coherent boost along the direction of $-\vec q$ and form again the same
ground-state, or excited hadron. Alternatively, it can crack
apart and produce a few-body final hadronic state.

The heavy quark symmetry {\it per se} cannot help calculating the
amplitudes to create such final states. However, it tells one that the
hadronization process does not depend on the heavy quark spin, or on the
concrete value of the mass $m_Q$ but rather on the velocity of the final
state heavy hadron. This velocity cannot change in the process of
hadronization if only soft gluons are exchanged between the heavy quark
and the light cloud.  This independence holds, of course, only when
$m_Q$ is very large (and the final state quark does not move too fast);
there are various $1/m_Q$ corrections at finite masses.

Let us consider, for example, the ground-state transition $B\ra D$. Its
amplitude depends on the velocity $\vec{v}$ of $D$, $f(\vec{v}^{\,2})$.
The very same function would describe also decays $B\ra D^*$, or
the elastic amplitude of scattering of a photon on the $b$ quark  $B\ra
B$, $B\ra B^*$. Moreover, in the proper normalization $f(0)=1$ holds.
This fact follows from the conservation of the $b$-quark vector
current, for the amplitude at zero momentum transfer measures the total
`$b$-quark charge' of the hadron $n_b -n_{\bar{b}}$. Its origin is
simply understood: if $\vec{v}=0$, the final state is not really
disturbed by movement of the static source. A nontrivial rearrangement
of the light cloud for $B\ra D$ is associated only with
the mass-dependent terms in ${\cal H}_Q$ vanishing when $m_Q \ra
\infty$.

As far as I know, this physical picture and the fact of normalization
of the $B\ra D$ and $B\ra D^*$ transition amplitudes near zero recoil
$\vec{q} =0$ were first discussed in QCD by M.~Shifman and M.~Voloshin
in 1986 \cite{vshqs}.

Let us consider the vector $\bar{b} \gamma_\mu b$ current in $B$ meson.
It is described by the single formfactor $f_+(q^2)$:
\beq
\matel{B(p')}{\bar{b}\gamma_\mu b(0)}{B(p)}\;=\;f_+(q^2)\,
(p+p')_\mu\;, \qquad \qquad q_\mu = p'_\mu - p_\mu
\label{49}
\eeq
(the second structure $(p+p')_\mu$ is forbidden by $T$ invariance or
current conservation). The value $f_+(0)$ measures the total `beauty
charge' of the hadron and is not renormalized by the strong
interaction, $f_+(0)=1$.\footnote{For simplicity, I use the convention
that $B$ consists of $b$ and $\bar{q}$. That is, $B^-$ is a $B$ meson
while $B^+$ is $\bar{B}$.}
Passing to the velocities, we use instead of
$q^2$ the scalar product $v_\mu v'_\mu$:
\beq
(vv')\;=\; \frac{(pp')}{M_B^2} \;=\; 1- \frac{(q^2)}{2M_B^2} \,\ge\,
1\;,
\label{50}
\eeq
and
\beq
f_+(q^2)\;=\; \xi(vv')\;, \;\; \xi(1)=1\;.
\label{51}
\eeq
$\xi(vv')$ is the Isgur-Wise function. The heavy quark symmetry then
states that
\beq
\matel{D(v')}{\bar{c}\gamma_\mu b(0)}{B(v)}\;=
\left(
\frac{M_B+M_D}{2\sqrt{M_BM_D}}
(p+p')_\mu\; -  \frac{M_B-M_D}{2\sqrt{M_B M_D}}
(p-p')_\mu \right) \,\xi(vv')\;.
\label{52}
\eeq
$D^*$ differs from $D$ only by the alignment of the $c$ quark spin.
Taking this into account yields
$$
\matel{D^*(v',\epsilon)}{\bar{c}\gamma_\mu b(0)}{B(v)}\;=\;
i\epsilon_{\mu\nu\al\be}\epsilon^*_\nu v'_\al v_\be \,\xi(vv') \:
\sqrt{M_B M_D^*}
$$
\beq
\matel{D^*(v',\epsilon)}{\bar{c}\gamma_\mu b(0)}{B(v)}\;= \;
\left\{\epsilon^*_\mu (vv'+1)-v'_\mu(\epsilon v)\right\}
\,\xi(vv')\:\sqrt{M_B M_D^*}\;.
\label{54}
\eeq
It is important that these relations are valid in the limit $m_{b,
c}\ra \infty$ and if no short-distance radiative corrections were
present.  The corrections to the symmetry limit are minimal at
$\vec{q}=0$ ($v=v'$, the so-called zero recoil point); numerically the
value of the axial formfactor $F_{B\ra D^*}(0)\simeq 0.9$ \cite{vcb}.
The correction at arbitrary $\vec{v} \sim 1$ are generally quite
significant.

\subsection{Feynman Rules at $m_Q\ra \infty$}

The Feynman rules for heavy quarks are usual propagators and vertices
for nonrelativistic particles where $1/m \ra 0$. If the heavy quark
momentum is $p_\mu=(m_Q+\omega, \; \vec{p})$ then
$$
G(p)\;=\;\lim_{m\ra\infty} \, \delta_{\al\be}\,
\frac{1}{\omega-i\epsilon-\frac{\vec{p}^{\,2}}{2m} }\, \delta_{ij}\;=\;
\frac{\delta_{\al\be}}{\omega-i\epsilon }\, \delta_{ij}
$$
\beq
\Gamma_\mu\;=\; g_s \frac{\lambda^a_{\al\be}}{2} \delta_{\mu 0}\;.
\label{57}
\eeq
Here $a$, $\al$, $\be$ are color indeces and spinor indeces
$i,j$ take values $1$ or $2$. (It is often advantageous
to keep the nonrelativistic
term $\vec{p}^{\,2}/2m_Q$ in the propagator.) These rules follow
immediately from the static Lagrangian
$$
{\cal L}_Q\;=\; \varphi_Q^+ iD_0 \varphi_Q\;, \qquad
D_0=\partial_0-ig_s A_0^a \frac{\lambda^a}{2}\;.
$$
The same nonrelativistic system can be considered in the arbitrary
moving frame, where one can write
\beq
{\cal L}_v\;=\; \varphi_v^+ i(vD) \varphi_v\;.
\label{39a}
\eeq
Instead of the nonrelativistic spinor $\varphi_Q$ one then considers the
``bispinor'' $\varphi_v(x)= \frac{1+\not\!{\:v} }{2}\tilde Q(x)$; the
propagator is written as
\beq
\frac{1+\not\!\!{\,v} }{2}
\frac{m_Q+\not\!\!{\,p}-\not\!\!{\,k}}{m_Q^2-(p-k)^2-i\epsilon}
\frac{1+\not\!\!{\,v} }{2}
\;\longrightarrow \;
\frac{1+\not\!\!{\,v}}{2}  \frac{1}{vk-i\epsilon}
\label{60}
\eeq
$$
\Gamma_\mu\; =\; g_s \frac{\lambda^a}{2} v_{\mu}\;,
$$
where $k=p-m_Qv$. Of course, such a generalization can be
useful only when the initial and final state hadrons have
different velocities, $\vec v \ne \vec v\,'$. In that case the external
(`weak') source carries a large momentum $\vec{q} \sim m_Q \,\delta
\vec{v}\,$.

It must be noted, however, that HQET is not quite consistently
formulated as a quantum field theory for processes where $\vec{v}\ne
\vec v\,'$ when the quantum radiative corrections are really
incorporated. Any change in the velocity of a static source leads to
actual radiation of real hadrons with momenta $\vec{k}$ all the way up
to $m_Q$:
\beq
\frac{{\rm d}w}{{\rm d}\omega} \;\sim \frac{\alpha_s(\omega)}{\omega}
\: (\vec v\,'- \vec v)^2\;;
\label{62}
\eeq
here $\omega$ denotes the radiated energy. The non-Abelian QCD dipole
radiation was considered in \cite{dipole}. On the other hand, effective
theories are called upon to eliminate (`integrate out') all
high-momentum subprocesses. In order to do this one would have to
`integrate out' not only highly virtual, but also real processes, which
clearly is not a completely legitimate procedure.

This is a real difficulty and not a pure mathematical subtlety.
HQET as an effective static theory was originally introduced
\cite{hill} to study the nonperturbative effects in the heavy
flavor hadrons on the
lattice. Suppose we would try indeed to extract physical results for
the amplitudes with a nonzero velocity transfer on a fine enough
lattice using the HQET Lagrangian (\ref{39a}). Since this Lagrangian
describes infinitely heavy quarks, we would have to get zero for any
exclusive amplitude: all they contain a more or less universal
suppression factor
\beq
S \;\sim\;  {\rm e}\,^{-\frac{4\as}{9\pi}(\Delta \vec v)^2
\ln{\frac{m_Q}{\mu}} }
\label{64}
\eeq
and thus must vanish at $m_Q\ra \infty$ ($\mu$ is a hadronic mass scale
serving as an infrared cutoff).
This factor $S$ is the square of the well known Sudakov formfactor
(more exactly, of its nonrelativistic analogue).

Any actual simulation would yield, of course, a final result since
lattices contain an intrinsic UV cutoff, the inverse lattice spacing
$1/a$. The result, therefore, would essentially depend on this cutoff
for any observable process we would try to compute. It is not too
surprising from a general point of view. HQET considers not only
the sectors with different numbers of heavy quarks as completely
separate. Even the sectors with a single heavy quark but with different
velocities are different quantum systems for HQET, with formally
different dynamic degrees of freedom, Hilbert space of states, {\it
etc}. On the other hand, the transition amplitudes at $\vec v\ne \vec
v\,'$ require, for example, calculating the `inner' product of states in
these theories. Such a procedure is not quite legitimate or
well-defined. Hence, it does not come as a surprise that the results
sometimes look unphysical or prove to be fallacious. A closer look at
some formal derivations \cite{neublam,ryzak} reveals that the
subtleties at $\vec v\ne \vec v\,'$ were indeed neglected there.

This means that HQET {\it per se} is not a well-formulated theory
for different velocities beyond the first order in $\vec{v}$ when the
gluon bremsstrahlung appears. The introduction of a ``hard''
factorization scale $\mu$ necessary for any effective theory, becomes
here a nontrivial and not always harmless procedure.
A thoughtful reader notices at this point that the problem roots to the
essentially Minkowskian nature of the processes with actual heavy
quarks.
On the other hand,
in QCD the Sudakov-type bremsstrahlung is to a large extent universal,
the fact analogous to the fundamental factorization of hard and soft
processes in usual OPE formulated in Euclidean domain. Therefore,
an alternative way of consistently formulating the OPE expansion for
these problems seems in principle possible, even when nonperturbative
effects are addressed.

\section{Heavy Quark Expansion in Dynamics}

The heavy quark expansion resulted in many important insights into
physics of heavy flavor hadrons. Already the simple qualitative
considerations or symmetry relations noted in the 80's influenced both
the theoretical understanding and the experimental strategy in studying
$B$ particles. Nevertheless, a true measure of our control over QCD
dynamics is, eventually, how accurately we can extract fundamental
parameters like $|V_{cb}|$, $|V_{ub}|$, $m_b$, $m_c$ \ldots from
experimental properties of hadrons, or predict necessary amplitudes. At
this point we need to address more dynamic aspects of QCD in heavy
flavors.

A major progress over the last years was made in this direction. It is
impossible to review, even briefly, all relevant topics here.
Therefore, I am forced to limit myself to a brief discussion of the
existing situation with, probably, the most intriguing
subjects -- extracting $|V_{cb}|$, $|V_{ub}|$ and
$m_b$. An approximate -- though already nontrivial -- determination of
some of these parameters was made already a decade ago. The accuracy
and a confidence in our estimates now are at a new level. This required
a better understanding of various subtleties of the heavy quark theory
as a quantum field theory derived from QCD, which are absent in
simplified hadronic models. Moreover, explaining itself the origin of
controversy in the literature often requires going into such
`technicalities'. As a result, a consistent review of the situation
requires at this point a more rigorous presentation of the OPE for
heavy quarks, discussing the precise definitions and the status of
determination of the relevant hadronic parameters, existing
uncertainties, {\it etc}. -- and only then application to heavy
flavor phenomenology.

In these lectures I am trying to avoid technicalities in favor of
outlining a more transparent physical picture. To get a feeling of what
we need to know and how accurately, let me start directly with the
question of the precise determination of $|V_{cb}|$ from the total
semileptonic $B$ width, the most reliable up to date method both
theoretically and experimentally. This will naturally lead me to the
problem of the heavy quark masses, kinetic energy {\it etc}. in the
field theory approach. These questions will constitute a significant
part of my lectures. Then we will be able to return to discussing
practical extraction of $V_{cb}$. Such a strategy will lead to a
certain fragmentarity of the presentation, and I apologize for related
inconvenience.

\subsection{Inclusive Widths}

Inclusive widths of the heavy flavor hadrons are examples of the
genuine short-distance processes. The decays proceed at the
space-time intervals $\sim 1/m_b$ (more precisely, inverse energy
release), and the widths are affected by the soft strong dynamics to the
minimal extent. The utility of the OPE is demonstrated by the QCD
theorem \cite{buv,bs}: there are no nonperturbative corrections
to the inclusive widths
of heavy flavor hadrons. It applies to all types of decays:
semileptonic, nonleptonic, radiatively-induced ($b\ra s+\gamma$) {\it
etc}. -- they only must be truly inclusive.

In practical application of this analysis there are some limitations,
the so-called onset of quark-hadron duality must be passed. This
question will be briefly addressed later. The best situation in this
respect exists for semileptonic decays, and we will discuss them now in
detail. It should be noted that the methods of HQET are not
applicable to the inclusive decays. These processes have an intrinsic
large `dynamic' scale of energy release $\sim m_b$, and radically
differ in this respect from the exclusive heavy-flavor formfactors. The
OPE analysis is done directly in QCD utilizing this large kinematic
parameter. Only at the final stage when the
nonperturbative effects are expressed
via the expectation values of the local heavy quark operators, one
employs the nonrelativistic expansion for the $b$ fields \cite{buv}.

It was realized long ago that for heavy enough quarks the inclusive
widths must look similar to the analogous weak decay widths of an
almost non-interacting muon:
\beq
\Gamma(\mu\ra e\nu\bar\nu)\;=\;
\frac{G_F^2 m_\mu^5}{192\pi^3}
\left[z_0\left(\frac{m_e^2}{m_\mu^2}\right) -\frac{\as}{2\pi}
a_1\left(\frac{m_e^2}{m_\mu^2}\right)  +...
\right]\;,
\label{70}
\eeq
where $z_0(x)$ is the phase space factor; the term with $a_1(x)$
accountings for the electromagnetic interaction was calculated already
in 1956 \cite{muon}. Similarly,
\beq
\Gamma_{\rm sl}(B) \;=\; \frac{G_F^2 m_b^5}{192\pi^3}|V_{cb}|^2
\;z_0\left(\frac{m_c^2}{m_b^2}\right) \cdot \kappa
\label{71}
\eeq
(likewise for the $b\ra u$ decays), where all QCD effects, both
perturbative and nonperturbative, are lumped together in the factor
$\kappa$. It was not clear, though, what mass must be used, the quark
mass $m_b$ or the mass of the decaying hadron $M_B$. The numerical
difference is yet quite significant even for $B$ meson:
$M_B^5/m_b^5\approx 1.5 \mbox{ to }2 $, in spite of being
formally a
$1/m_b$ effect:
\beq
\frac{M_B-m_b}{m_b} \sim \frac{\Lam}{m_b}\;,
\qquad \; \frac{M_{\Lambda_b}-M_B}{M_B} \sim \frac{\Lam}{m_b}\;.
\label{72}
\eeq
The OPE analysis in QCD unambiguously states that one must use the
quark masses here, so that the mass factors are the same in all types
of beauty hadrons. In contrast to hadron masses, the inclusive widths
of heavy flavor hadrons do not have relative $1/m_Q$ splitting. This
result was established by Bigi, Shifman, Vainshtein and myself in 1992.

Is this property expected {\it a priori}\,? Yes and no. In the simple
constituent models where the binding energy is neglected,
$M_B=m_b+m_{\rm sp}$. It is obvious that in such systems
the widths are determined by $m_b^5$: the mass of the spectator --
as long as it remains a spectator -- does not enter.

This consideration is, of course, of little relevance for QCD. The
mass of the spectator quarks in $B^-$, $B^0$ or $\Lambda_b$ is
only a few $\MeV$ and we are not interested in such negligible effects.
The actual question is about a few hundred $\MeV$ cloud whose energy,
at least partially, is the interaction energy $E_{\rm pot}$. In
general, this $E_{\rm pot}$ does directly affect the decay width. In
the nonrelativistic models
\beq
M_B \;\simeq\;  m_b + m_{\rm sp} + E_{\rm pot} + E_{\rm kin}\:+
{\cal O}\left(\Lam^2/m_Q\right)
\;,
\label{73}
\eeq
and $E_{\rm pot} \sim E_{\rm kin} \ll m_{\rm sp} \sim \Lam$, so this
effect is less important than mere the rest energy of the spectator
quark, and is routinely discarded. On the contrary, in QCD one can
rather neglect $ m_{\rm sp}$.

At the same time, at least a major part of the mass dependence in
Eq.~(\ref{71}) comes from the phase space available for the final
states (in particular, for leptons), which naively knows nothing about
the quark mass and rather senses the overall hadron mass. It is this
consideration that may seem to be relevant for the QCD picture of
hadrons. It often underlies implicit questioning of the validity
of the QCD analysis resurrecting, in one form or another, the $1/m_Q$
corrections to the widths \cite{alt,flash2,jin,gleb}. And,
nevertheless, the dependence on the binding energy disappears. How it
happens in the language of the summation of the contributions of
separate exclusive final states, was illustrated in \cite{five} using
the so-called heavy quark transition sum rules \cite{optical}. Here,
instead, I'll try to explain the mechanism of the cancellation on a
transparent toy example.

Let us consider the semileptonic decay $b\ra c$ in the kinematics where
leptons carry away a major portion of the energy release, and the final
$c$ quark is moving slowly (this is the so-called small-velocity (SV)
kinematics first suggested as a theoretical tool by Shifman and
Voloshin in \cite{vshqs}).  The initial $b$ quark undergoes a binding
color Coulomb potential $V(0)$ which is of order $\Lam$, and this, in
principle, affects the decay probability. However, the final $c$ quark
is produced inside the very same potential, and thus the energy of
all final states is shifted by the same amount. The equality of
the potential binding force in the initial and final state is
guaranteed in this case by equality of the color `charges' of the
initial $b$ and final $c$ quarks which, in turn, is a consequence of
the conservation of the color flow. Thus, it is easy to see that
whatever strong is the binding energy by the Coulomb interaction, in
such a case it is canceled between the initial and final state
interaction due to the `color charge' conservation.

The situation may seem different when one goes beyond the SV kinematics
and the final quark is moving fast. First, the interaction of the final
relativistic quark is not given by mere the Coulomb potential but
depends on all components of the gauge potential. Second, the quark
promptly leaves the vicinity of the point where it was produced, and the
potential in other space locations is quite different. Moreover, if the
interquark potential grows at large distances, whatever large $m_b$ is
chosen, the final quark can never escape and would remain confined with
the initial light cloud. In contrast, for the interaction vanishing at
large distances the quark momentarily becomes `free'. This may naively
seem to contradict OPE since no assumptions about the
large-distance behavior of the interaction was involved in proving the
cancellation of the binding energy effects.

Nevertheless, the cancellation is complete in this case as well. The
gauge nature of the QCD interaction and the conservation of the color
current ensure that the integrated effect of the interaction with the
potential $A_\mu$ is the same for slow and relativistic particle. A
closer look at the second problem also shows that no actual
contradiction is present. Since the act of the decay happens for a
small time interval $\sim 1/E_{\rm rel}$, its probability is sensitive
only to the potential at the distances $\sim 1/E_{\rm rel}$ from the
origin. Details of the interaction at larger distances merely determine
the spectrum of the final states and how exactly the overall probability
is allocated over the different final states, but not the total decay
probability. For example, if the interaction vanishes beyond a
distance $\sim \Lam^{-1}$, the final states
belong to the continuous spectrum being represented by plane waves with
some distortion near the origin.  If the potential is confining, there
are no real quasifree decays but the transitions to the bound states in
such a potential. The sum of these probabilities reproduces the decay
width into the quasifree quark states.

This picture explains the OPE result for QCD. The latter, of course, is
more general and does not rely on an assumption about QM potential,
{\it etc}. The key point is that the decay is almost instantaneous, its
time is of the order of $1/m_b$. It is for this reason the total rate
is determined by the potential at $|\vec{x}\,| \sim 1/m_b$, and the
width is expanded in local operators $\bar{b} O_i b$ where $O_i$
contain the QCD fields at $\vec x=0$ and their derivatives.
In essence, the perturbative corrections represent the expansion in the
singularity of the interaction at $x=0$ (recall, $V(x)\propto
\as/|\vec{x}\,|$), and the power nonperturbative corrections are
related to the terms of the Taylor expansion of the interaction near
$x=0$.\footnote{It was suggested in \cite{inst} that the effects of
duality violation are related to the singularities of the interaction
at finite distances $x$ from the origin, in the complex plane. I do
not discuss this aspect here.} This is a quite general property and
does not depend on the underlying nature of the interaction.

The conservation of the color flow and the gauge nature of the QCD
interaction of quarks leads to an additional property: the leading term
in the Taylor expansion, the `potential' $V(0)$ does not appear as a
result of the exact cancellation between the effects of the initial and
final state interactions. This is a meaning of the theorem about the
absence of $1/m_Q$ corrections in the inclusive widths.

This is an example of a general QM statement known as the KLN theorem.
In the simple words, it states that all ``sufficiently inclusive''
transition probabilities must be {\it completely} infrared (IR)
insensitive.\footnote{I deliberately do not explain here the precise
meaning of both the assumption and of the statement, what is IR
insensitive. It can be found in the dedicated papers \cite{kln,zahkln}.}
However, to be ``sufficiently inclusive'', the process, in general,
requires a summation over the degenerate initial states as well, which
is not realized in any experiment. The degree to which the KLN theorem
is valid for actual ``insufficiently inclusive'' transitions is {\it
a priori} unknown. The OPE -- when it applies -- allows to
quantify this sensitivity.

It is worth emphasizing that the absence of $1/m_Q$ corrections in the
inclusive widths is not an automatic consequence of the KLN theorem or
any other general QM fact. In dynamic realizations of
strong interactions other than QCD -- for example, due to exchange of
scalar particles -- such corrections would be present and depend
non-universally on various kinematic details. That is why, for example,
the summation of the exclusive semileptonic transition probabilities in
the ISGW model \cite{isgwidth} did not reproduce the partonic width,
especially in the $b\ra u$ decays.

After the general illustrations given above it is easier to outline the
main technical steps in calculating the widths: one only needs to
believe that a particular inclusive width is indeed given by the local
operators.\footnote{One possibility to show it formally is described
in detail, {\it e.g.} in \cite{inst}. Historically, an important role
here was played by paper \cite{mirage}.} Then we can write
\beq
\Gamma_{H_b}=\frac{1}{2M_{H_b}}
\matel{H_b}{c_1 \bar{b} b + c_G \bar{b}\frac{i}{2}\sigma G b +
\sum c_i\bar{b} O_i b +...}{H_b}
\label{80}
\eeq
where $c$ are short-distance coefficient functions and $\bar{b} O_i b$
denote local operators of $D=6$ and higher. Note that there is no
independent operator of $D=4$: the width must be a Lorentz scalar, and
the only possible operator $\bar{b} \gamma_\al iD_\al b$ reduces to
$m_b \bar{b} b$ by the QCD equations of motion. This is the counterpart
of the cancellation of the Coulomb potential $A_0$ is the previous
discussion: gauge invariance states that $A_0$ must always enter
together with $\partial_0$, and this combination applied to the $b$
quark field is free of any dynamics to order $\Lam$ by virtue of
equations of motion (that is, including the interaction of the initial
$b$ quark).

The expectation values of the operators in Eq.~(\ref{80}) scale like
$\Lam^{d-3}$ (more exactly, it can be a power of the
normalization scale $\mu$, {\it i.e.} $\mu^{d-3}$). Therefore, the
relative contribution to the width of higher-dimension operators scale
like $(\Lam, \mu)^{d-3}/m_Q^{d-3}$.

Eq.~(\ref{80}) is not the whole story, however, since the leading
operator $\bar{b}b$ describing the free quark decay is also
affected by the nonperturbative dynamics. Its expectation
value, however, can be
expanded in $1/m_b$:
$$
\matel{H_b}{\bar bb}{H_b} = \matel{H_b}{\bar b\gamma_0 b}{H_b} +
\frac{\matel{H_b}{\bar b\left(\pi^2+\frac{i}{2}\sigma G \right)
b}{H_b}}{2m_b^2}+
{\cal O}(1/m_b^4)\;=
$$
\beq
=\;
2M_{H_b} \left(1-\frac{\mu_\pi^2-\mu_G^2}{2m_b^2} +...\right)
\;.
\label{81}
\eeq
The expectation value of $\bar b \gamma_0 b$ is exactly unity, which
completes the proof of absence of $1/m_Q$ corrections. Eq.~(\ref{81})
was also derived by Bigi, Shifman, Vainshtein and myself and first
given in \cite{buv}.\footnote{Recently this relation was attributed by
Neubert \cite{neubincl} to the later papers, in particular, to
\cite{fn}.  However, there was no such a relation there.} It is worth
noting that the term $-\frac{\mu_\pi^2}{2m_b^2}$ also has a transparent
physical meaning \cite{prl}: it is nothing but a Lorentz dilation of
the decay due to nonrelativistic `shivering' (so-called Fermi motion)
inside the hadron through the binding effects.

Including the leading corrections, the semileptonic width has the
following form \cite{buv,bs,dpf,prl}:
\beq
\Gamma_{\rm sl} = \frac{G_F^2 m_b^5}{192\pi^3} |V_{cb}|^2 \left\{z_0
\left(1-\frac{\mu_\pi^2-\mu_G^2}{2m_b^2} \right)
-2\left(1-\frac{m_c^2}{m_b^2} \right)^4\frac{\mu_G^2}{m_b^2}
-\frac{2}{3} \frac{\as}{\pi} a_1 + ...
\right\}
\label{crad6}
\eeq
where ellipses stand for higher order perturbative and/or power
corrections,  $z_0$ and $a_1$ depend on $m_c^2/m_b^2$.
The value of $\mu_\pi^2$ is not yet measured in experiment. There
exists an inequality $\mu_\pi^2 > \mu_G^2$ which essentially
constraints the interval of its possible values.
The status of $\mu_\pi^2$ will be addressed later.
Irrespectively of the exact value of $\mu_\pi^2$, the
$\, 1/m_b^2$ corrections to $\Gamma_{\rm sl}$ are rather small, about
$-5\%$ and lead to the increase in the extracted value of
$|V_{cb}|$ by $2.5\%$. The impact of the higher order power corrections
is negligible.

A rather detailed most accurate numerical analysis of the semileptonic
widths is given in \cite{rev}, Sect.~8.1. Let me briefly summarize it.
The leading-order ${\cal O}(\as)$ perturbative corrections are known
from the QED calculations in muon decay \cite{muon}. The $1/m_b^2$
nonperturbative
corrections were calculated in the above mentioned papers and decrease
the width by about $5\%$. The $1/m_b^3$ corrections were evaluated in
\cite{bds,grekap} and are at the level of $1\%$.  They are negligible
for extracting $|V_{cb}|$.
The part of the higher-order perturbative corrections
associated only with running of $\as$ in the first-order loop diagrams,
the so-called BLM corrections \cite{blm}, were calculated to order
$\as^2$ \cite{wiseblm} and to all orders \cite{bbbsl}.
Their impact appeared to be small.
The remaining,
nontrivial non-BLM (`genuine') two-loop $\as^2$ corrections were
expected to be moderate. They were recently evaluated in a series of
papers by Czarnecki and Melnikov \cite{czarm} and were confirmed
to be small. These were highly nontrivial sophisticated technical
analyses often considered unfeasible even a few years ago. A
resummation of certain enhanced non-BLM higher-order corrections was
performed in \cite{five}. At present, no significant theoretical
uncertainty remains in the perturbative corrections to the semileptonic
width.

Evaluating the theoretical prediction we get
$$
|V_{cb}|=0.0419\left(\frac{{\rm BR}(B\rightarrow
X_c\ell\nu)}{0.105}
\right)^{\frac{1}{2}}\left(\frac{1.55\,\rm
ps}{\tau_B}\right)^{\frac{1}{2}}
\cdot \left(1-0.012\frac{(\mu_\pi^2-0.5\GeV^2)}{0.1\GeV^2}\right)
\times
$$
\beq
\left(1-0.01\frac{\delta m_b(\mu)}{50\MeV}\right)
\left(1+0.006 \frac{\as^{\overline{\rm MS}} (1\GeV)-
0.336}{0.02}\right)
\left(1+0.007\frac{\bar\rho^3}{0.1\GeV^3}\right).
\label{w12}
\eeq
Here $m_b(\mu)$ is a certain short-distance mass of the $b$ quark which
is the subject of a special discussion in the next section, $\bar\rho$
is a positive non-local correlator mentioned before (its expected scale
is $0.1\GeV^3$). The main theoretical uncertainty at the moment resides
in the value of $\mu_\pi^2$. We arrive thus at the model-independent
evaluation
$$
|V_{cb}|=0.0419
\left( \frac{{\rm BR}(B\rightarrow X_c\ell\nu)}{0.105}
\right) ^{\frac{1}{2}}
\left( \frac{1.55\,\rm ps}{\tau_B}\right) ^{\frac{1}{2}}
\times
$$
\beq
\left( 1-0.012\frac{(\mu_\pi^2-0.5\,\rm GeV^2)}{0.1\,\rm
GeV^2}\right)
\cdot \left( 1 \pm 0.015_{\rm pert} \pm 0.01_{m_b} \pm
0.012
\right) \, ,
\label{w20}
\end{equation}
where the last error reflects $m_Q^{-3}$ and higher
power corrections as well as possible deviations from duality.
The quoted uncertainty associated with $m_b$ can be viewed as rather
conservative; this assertion will be explained in the sections to
follow.

Similar to the treatment of $\Gamma (B\ra X_c \ell \nu )$,  it is
straightforward to relate the value of $|V_{ub}|$ to the total
semileptonic width $\Gamma(B\ra X_u\, \ell\nu)$
\cite{upset}:
\beq
|V_{ub}|=0.00465\left(\frac{{\rm BR}(B^0\rightarrow
X_u\ell\nu)}{0.002}
\right)^{\frac{1}{2}}\left(\frac{1.55\,\rm
ps}{\tau_B}\right)^{\frac{1}{2}}
\cdot \left(1 \pm 0.025_{\rm pert} \pm 0.03_{m_b}
\right)
\;.
\label{22r}
\eeq
The dependence on $\mu_\pi^2$ is practically absent here. The
theoretical uncertainty in $\Gamma(B\ra X_u\, \ell\nu)$ is relatively
insignificant. The major problem is experimental, how to measure ${\rm
BR_{sl}}(b\ra u)$. For a long time it seemed unfeasible. Nevertheless,
a possibility of extracting it from experiment with a reasonable
confidence is being discussed since recently, and may eventually yield
an accurate determination of $|V_{ub}|$ on this route.

I'll return to the question of $\Gamma_{\rm sl}$ and $|V_{cb}|$ later
in the summary. Before proceeding to a more theoretical issues, it seems
necessary to note that rather controversial opinions existed in the
literature regarding the reliability of calculation of $\Gamma_{\rm
sl}$  and, in particular, the impact of higher-order perturbative
corrections. Often they were claimed to be too large or even
uncontrollable \cite{update,wiseblm,neubmori,neubchi}. In reality, the
apparent instability was associated with not quite physical formulation
of the problem and the confusion rooted to the problem of the
definition of the heavy quark mass.  As a matter of fact, after all
refinements and thorough analysis of the corrections, we are basically
back to square one.  The expression for $|V_{cb}|$ given in
Ref.~\cite{vcb} has changed by less than one percent (for the same
input values of the experimental parameters).  What became clearer in
the last two years is the acceptable range of the underlying key QCD
parameters entering the heavy quark expansion, and the size of
theoretical corrections. I cannot help mentioning in this respect that,
meanwhile, the value of $\tau_B$ (at least used by theoreticians as an
input) drifted from $1.32 \pm 0.09\,{\rm ps}$ \cite{ls} to $1.57
\,{\rm ps}$.

\subsection{Problem of $m_Q$ in QCD}

The heavy quark mass is certainly one of the most important starting
parameters of the heavy quark theory. The practical necessity to know
it well is illustrated by the fact that the decay widths of the beauty
hadrons are proportional to a high power of the mass:
\beq
\Gamma(B, \Lambda_b) \; \sim \; m_b^n\;,\qquad n=5 \gg 1\;.
\label{90}
\eeq
All uncertainties in $m_b$ are magnified by the factor $n$. On the
other hand, one can turn vices into virtues developing $1/n$ expansion
and performing resummation of the large-$5$ terms. This was
accomplished in \cite{five}. Let me mention only one, the perturbative
aspect.

In general, the perturbative series for, say, $\Gamma_{\rm sl}$ have
$n$-enhanced terms:
\beq
\Gamma_{\rm sl}^{\rm pert} \;\propto
\; 1\,+\,n\,d_1\frac{\as}{\pi}  \,+\,n^2
d_2\left(\frac{\as}{\pi}\right)^2 \,+\,n^3
d_3\left(\frac{\as}{\pi}\right)^3 \,+\:...  \label{91} \eeq where
$d_1,\: d_2\: ...\, \sim 1$. Even at $n=5$ and a moderate value of
$\frac{\as}{\pi}=0.1$ the higher-order terms are rather significant to
be neglected already if a $5\%$ accuracy in $V_{cb}$ is targeted.
Fortunately, this series can be readily summed up \cite{five}, and the
remaining corrections are small. However, once again it requires an
accurate understanding of what are $m_b$ and $m_c$.

Meanwhile, quite controversial estimates could be found in the
literature for the numerical values of the charm and, in
particular, bottom quark masses. As a result, the majority of the HQET
applications, for example, were focused on the quantities which were
mainly independent on $m_b$. HQET, claiming to have finally
clarified the issue of what is the heavy quark mass, actually admitted
that its determination with a sensible accuracy was impossible in
practice.

In appeared that these `practical' difficulties had a deep
theoretical reason. HQET was based on the concept of the ``pole''
mass of the heavy quark $m_Q^{\rm pole}$. However, in QCD-like theories
the ``pole'' mass is not a physical notion. Moreover, remaining a
purely perturbative construction, it cannot {\it theoretically} be
defined with the necessary accuracy \cite{gurman}, the irreducible
uncertainty in defining its value is of the order of typical hadronic
scale:
\beq
\delta m_Q^{\rm pole}\;=\; \mbox{few units}\,\times \Lam\;.
\label{93}
\eeq
No wonder all attempts to extract this ill-defined value yielded
controversial results.

\subsubsection{What is $m_Q$?}

In quantum field theory the object we begin our work with is the
Lagrangian formulated at some high scale $M_{ 0}$.
The mass $m_0$  is a parameter in this
Lagrangian; it enters on the same footing as, say,
the bare coupling constant  $\alpha_s^{(0)}$ with the only difference being
that it
carries dimension. As with any other
coupling, $m_0$ enters in all  observable quantities in a certain
combination with  the
ultraviolet cutoff $M_{0}$, which is universal for a renormalizable
theory.

The mass parameter $m_0$ by itself is not observable, like $\as^{(0)}$.
For calculating observable quantities   at the scale $\mu \ll  M_{0}$
it is usually convenient to relate $m_0$ to some  mass  parameter
relevant to the scale $\mu$. Integrating out momentum scales above
$\mu$ converts $\as^{(0)}$ into $\as(\mu)$ -- and likewise $m_0$ into
$m(\mu)$.  Such $m(\mu)$ is not something absolute since depends on
$\mu$. It is either used on the same footing as $\as(\mu)$ or, in the
final expressions, is eliminated in favor of some suitable observable
mass.  For example, in quantum  electrodynamics (QED) at low energies
({\it i.e.}  $E\ll m_e$) there is  an obvious ``best" candidate: the
actual mass of an  isolated  electron, $m_e$. In  the perturbative
calculations it is determined  as the position of  the pole in the
electron Green function (more exactly, the beginning of the cut). The
advantages are evident:  $m_e$ is  gauge-invariant and experimentally
measurable.

The analogous  parameter for heavy quarks in QCD is
referred to as the pole quark mass, the position  of the pole of the
quark Green function. Like $m_e$ it is gauge  invariant.  Unlike QED,
however, the quarks do not exist as isolated objects (there are no
states with the quark quantum numbers in the  observable spectrum, and
the quark Green function beyond a given order has neither a pole nor a
cut).  Hence, $m^{\rm pole}$ cannot be directly measured;  $m^{\rm
pole}$ exists only as a theoretical construction.

In principle, there is nothing wrong with using $m^{\rm pole}$
{\em in perturbation theory} where it naturally appears
in  the Feynman graphs
for the quark Green functions, scattering amplitudes and so on. It
may or may not be convenient, depending on concrete goals.

The pole mass in QCD is perturbatively infrared stable, order by
order, like in QED.
It is well-defined to every given
order in perturbation theory. One cannot define it to all orders,
however; the sum of the series does not converge to a definite number.
In a sense, the pole mass is {\em not}
infrared-stable nonperturbatively.
Intuitively this is clear: since the quarks are
confined in the full theory, the best one can do  is to define the
would-be pole position with  an intrinsic uncertainty of order $\sim
\Lam$ \cite{gurman}.

Based on experience in QED or ordinary QM, non-existence of
$m_Q^{\rm pole}$ may seem counter-intuitive. Even rigorous inequalities
like $\La > 232\MeV$ were claimed \cite{manoh} for the `fundamental
HQET parameter', the difference $\La=M_B-m_b^{\rm pole}$ (the flaw was
pointed out in \cite{gurman}). The confusion did not dissipated
completely and still surfaces in the literature. In the
review article on the quark masses in PDG-96 by A.~Manohar
it was stated that there existed a third, the ``HQET'' mass
$m_Q^{\rm HQET}$ which
differed from $m_Q^{\rm pole}$ starting $\as^2$ in QCD with massless
(light) flavors \cite{pdg}. That unsubstantiated claim is erroneous;
the mass used by HQET coincided with $m_Q^{\rm pole}$ to any concrete
order of perturbation theory. Moreover, it is not clear what
normalization point could have had the coupling $\as$ in that
difference: the renormalizability would leave as an option only the
mass of the light quark\ldots I hope that such statements will be
corrected in future issues of PDG. The ``HQET'' mass is as ill-defined
as the ``pole'' mass.

Employing the perturbation theory, we start with the free quark
propagator
$$
G(p)= \frac{1}{m_Q(\mu)-\not\!\!{\,p} }
$$
($m_Q(\mu)$ is the parameter entering the Lagrangian)
which has a pole at $p^2= m_Q^2(\mu)$ and, therefore, describes a
particle with mass $m_Q(\mu)$. Accounting for the gluon exchanges to
the first order in $\as$ adds the diagram Fig.~2 and the pole moves
to $p^2\simeq \left(m_Q(\mu)+\frac{4\as}{3\pi}\mu\right)^2 $ describing
now a particle with the mass $m_Q^{(1)}\simeq
 m_Q(\mu)+\frac{4\as}{3\pi}\mu$, {\it etc}. In any order of the
perturbation theory we see a quark pole and the corresponding particle
with certain mass, differing from the mass in the Lagrangian. This mass
is just the ``pole'' mass. Clearly, it depends on the order of
perturbation theory one considers, and on a concrete version of
the employed expansion:
\beq
m_Q^{(k)} \;=\; m_Q(\mu )\;\sum_{n=0}^{k}\,
C_n\left(\frac{\mu}{m}\right) \,\left(\frac{\as (\mu )}{\pi}\right)^n
\; , \;\;\; C_0=1\;.
\label{100}
\eeq
It is tempting to define the `actual' pole mass as a sum of the series
(\ref{100}), and this used to be a standard assumption in HQET. It
appears, however, that the sum does not converge to a reasonable number
and cannot be defined with the necessary accuracy in a motivated way,
but suffers from an irreducible uncertainty of order $\Lam$.

\begin{figure}
\vspace{2.8cm}
\includegraphics{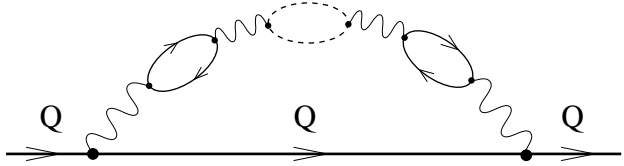}
\caption{\footnotesize
Perturbative diagrams leading to renormalization of the heavy quark
mass. The contribution of the gluon momenta below $m_Q$ expresses the
classical Coulomb self-energy of the colored particle. The number of
bubble insertions into the gluon propagator can be arbitrary
generating corrections in all orders in $\as$. The factorial growth
of the coefficients
produces the IR renormalon uncertainty in $m_Q^{\rm pole}$ of order
$\Lam$.}
\end{figure}

Before explaining the origin of this perturbative uncertainty, let me
note that such a definition of the quark mass is nothing but equating
it with the lowest eigenvalue of the (perturbative) Hamiltonian in the
sector with the number of heavy quarks $1$. The eigenvalues of the
Hamiltonian, however, are {\it not} a short-distance quantity, and
their calculation in the perturbative expansion is not adequate. This
is in the nice correspondence with the observed fact that no isolated
heavy quark exists in the spectrum of hadrons which contains instead
$B,\,B^{**},\, \Lambda_b,\, B_s\,$\ldots with masses differing
from each other by amount ${\cal O}(\Lam^1)$.

The physical reason behind the perturbative instability of the long-distance
regime is the growing of the interaction strength $\as$. One can
illustrate this instability in the following
transparent way.  Consider the energy stored in the
chromoelectric field in a sphere of radius $R \gg 1/m_Q$
around a static color source,
\beq
\delta {\cal E}_{\rm Coul} (R) \propto
\int _{1/m_b \leq |x| < R} d^3x {\vec E}^{\,2}_{\rm Coul}
\propto {\rm const} - \frac{\as (R)}{\pi} \frac{1}{R}\, .
\label{102}
\eeq
This energy is what one adds to the bare mass of a heavy particle to
determine what will be the mass including the QCD interactions.
The definition of the pole mass amounts to setting
$R \ra \infty$; {\it i.e.},
in evaluating the pole mass one undertakes to integrate the
energy density associated with the color source over
{\em all} space assuming that it has the Coulomb form.
In real life the color interaction becomes strong at
$R_0 \sim 1/\Lam$; at such distances the
chromoelectric field
has nothing to do with the Coulomb tail. Thus, one cannot include the
region beyond $R_0$ in a meaningful way. Its contribution
which is of order $\Lam$, thus, has to be considered
as an irreducible uncertainty which is power-suppressed relative
to $m_Q$,
\beq
\frac{\delta_{\rm IR}m_Q^{\rm pole}}{m_Q}\; = \;
{\cal O}\left(\frac{\Lam}{m_Q}\right)\; .
\label{3.6}
\eeq

Exactly this behavior is traced formally in the perturbation theory. In
the nonrelativistic
regime when the internal momentum $|\vec{k}\,|\ll m_Q$ the
expression for the diagram Fig.~2 is simple,
\beq
\delta m_Q \sim - \frac{4}{3}
\int \frac{d^4k}{(2\pi )^4 i k_0} \frac{4\pi \alpha_s}{k^2} =
\frac{4}{3}
\int \frac{d^3\vec k}{4\pi ^2} \frac{\alpha_s}{\vec k^2}\, .
\label{3.2}
\eeq
The running of the coupling is generated by dressing the
gluon propagator by virtual pairs and leads to
\beq
\delta m_Q \simeq
\frac{4}{3}
\int \frac{d^3\vec k}{4\pi ^2} \frac{\alpha_s (\vec k^2)}{\vec k^2}\, .
\label{3.2a}
\eeq
Since
\beq
\alpha_s(k^2) =
{\alpha_s(\mu ^2)}\left\{1 + \frac{\alpha_s(\mu ^2)}{4\pi} b
\,\ln{\frac{k^2}{\mu ^2}}\right\}^{-1} , \; \; b = \frac{11}{3} N_c -
\frac{2}{3}n_f \, ,
\label{run}
\eeq
we can expand
$\alpha_s(k^2)$ in a power series in $\alpha_s(\mu ^2)$ and easily
find  the $(n+1)$-th order contribution to $\delta m_Q$,
\beq
\frac{\delta m_Q^{(n+1)}}{m_Q} \sim \frac{4}{3}
\frac{\alpha_s(\mu)}{\pi} \, n!  \left(
\frac{b \alpha_s(\mu)}{2\pi}\right) ^n \; .
\label{3.4}
\eeq
We observe that
the coefficients grow factorially and contribute with the same sign.
Therefore, one cannot define the sum of these contributions even using
the trick with the Borel transformation.  The best one can do is to
truncate the series judiciously. An optimal truncation leaves us with
an irreducible uncertainty $\sim {\cal O}(\Lam)$
\cite{pole,bbpole}. The above perturbative corrections are example of
the so-called infrared renormalons \cite{renorm}.

This uncertainty can be quantified. A formal Borel resummation of such
non-summable series leads to the result which literally has an imaginary
part which can be taken as a measure of the uncertainty. The
imaginary part for the series Eq.~(\ref{3.4}) is
\beq
\Im m_Q^{\rm pole}\;=\; \frac{8\pi}{3b}{\rm e}\,^{5/6}\,
\Lam^{\overline{\rm MS}}\;.
\label{103}
\eeq
It became conventional to assign to the irreducible uncertainty the
formal imaginary part divided by $\pi$. Even with this minimal choice
\beq
\delta m_Q^{\rm pole}\;=\; \frac{1}{\pi}
\left| \Im m_Q^{\rm pole}\right| \;=\;
\frac{8}{27}{\rm e}\,^{5/6}\,
\Lam^{\overline{\rm MS}}\;\simeq \;0.7 \Lam^{\overline{\rm MS}}\;.
\label{103a}
\eeq

Thus, the perturbative
expansion {\em per se} anticipates the onset of the nonperturbative
regime (the impossibility of locating the
would-be quark pole to accuracy better than $\Lam$).  Certainly, the
concrete numerical value of the uncertainty
in $m_{\rm pole}$
obtained through renormalons is not trustworthy.
The renormalons do not represent the dominant
component of the infrared dynamics.
However, they are a clear indicator of the presence of
the
power-suppressed nonperturbative effects, or infrared instability of
$m_{\rm pole}$; the very fact that there is a correction ${\cal
O}(\Lam /m_Q)$ is beyond any doubt.

It is worth noting that the pole mass was the first example
where a quantity which is perturbatively infrared-stable was shown
not to be stable nonperturbatively at the level $\Lam^1$.
The observation of Refs.~\cite{pole,bbpole} gave impetus
to dedicated analyses of other perturbatively infrared-stable
observables
in numerous hard processes  without OPE, in particular, in  jet
physics. Such nonperturbative infrared contributions, linear in
$\Lam /Q$ were indeed  found shortly after
in thrust and many other jet characteristics (for a review and a
representative list of references see e.g. \cite{VB}).

To demonstrate that the problem of divergence of the
$\alpha_s$ series for $m_Q^{\rm pole}$ is far from being academic,
let us examine how the `perturbative' contribution to the $b$ quark
mass looks numerically:
$$
m_b^{\rm pole} = m_b (1\GeV) +
\delta m_{\rm pert}(\mbox{below}\:1\GeV) \simeq
$$
\begin{equation}
4.55 \, {\rm GeV}
+ 0.25\, {\rm GeV} + 0.22\, {\rm GeV} + 0.38\, {\rm GeV} +
1\, {\rm GeV} + 3.3\, {\rm GeV}  + ...  ,
\label{numbers}
\end{equation}
where $m_b (1\, {\rm GeV}) $ is the running mass at $\mu = 1 \GeV$,
and $\delta m_{\rm pert}$ is the perturbative
series taking account of the loop momenta from $1\GeV$ down to zero.
It is quite obvious that  the corrections start to blow up already in
rather low orders! Their size is consistent with the estimate
Eq.~(\ref{103a}).
Expressing observable infrared-stable quantities (e.g.
the inclusive semileptonic width  $B\ra X_u \ell\nu $) in terms of the
pole mass will necessarily entail large coefficients in the $\alpha_s$
corrections compensating for the explosion of the coefficients in
$m_b^{\rm pole}$. I will return to this point later in Sect.~3.5.

Summarizing, the pole mass {\em per se} does  not appear in  OPE for
infrared-stable quantities.  Such expansions  operate with the
short-distance (running)  mass. Any  attempt to
express the  OPE-based results in terms of the pole mass creates a
problem   making the Wilson coefficients ill-defined theoretically and
poorly convergent numerically.

Since it is impossible to relate $m_Q(\mu)$ and $m_Q^{\rm pole}$ to the
necessary accuracy, it is clear that either $m_Q^{\rm pole}$ or
$m_Q(\mu)$ must be irrelevant for the $1/m_Q$ expansion, for example,
for calculating the widths. Which one? This question was formulated and
answered in early 1994 in Ref.~\cite{pole}. In agreement with the
qualitative discussion given above, the answer is: $m_Q^{\rm pole}$ is
irrelevant and must be replaced everywhere by $m_Q(\mu)$. In the OPE,
the IR part of the pole mass is not related to any local operator
$\bar{Q} O_i Q$, and does not enter any observable calculable in the
short-distance expansion. Since this IR piece does not enter
observables, it cannot, in turn, be determined experimentally. The
numerical instability of various attempts to pinpoint the value of
$m_Q^{\rm pole}$ is a result of using the perturbative expansion for the
effects originating from the nonperturbative domain.

The above facts were later illustrated in detail in Ref.~\cite{bbz} in
a concrete model for the higher-order corrections to the semileptonic
widths obtained via the so-called bubble resummation of the one-loop
perturbative diagrams. The corresponding IR renormalon uncertainty
disappeared from the perturbative corrections when the width was
expressed in terms of the short-distance heavy quark mass.\footnote{A
similar cancellation was claimed to be independently obtained
afterwards also in \cite{ns}. However, there is no a single calculation
of the perturbative correction to the width in that paper.}

It is important to note that the irrelevance of the pole mass goes
beyond the problems with the IR renormalon contributions illustrated
above. Even if there existed some way to define reasonably the sum of
the pure perturbation series for $m_Q^{\rm pole}$, or asymptotic states
with single-quark quantum numbers with finite energy existed in a
strong-interaction theory like QCD, this mass still would have been
inadequate for constructing the effective field theory, to the extent
the difference with $m_Q(\mu)$ cannot be neglected numerically.

A properly constructed perturbative treatment suitable for the
$1/m_Q$ expansion
incorporates only gluons with $|\vec{k}| \gsim \mu$ which are not
`resolved' and are included into the heavy quark field wavefunction
corresponding to the heavy quark field $Q_{(\mu)}(x)$ normalized at the
scale $\mu$. The normalization point $\mu$ can be changed: descending
from $\mu$ down to $\mu_1<\mu$ one has to integrate out newly-resolved
gluons with $\mu_1<|\vec{k}| < \mu$. For example, the Coulomb field
associated with such gluons increases the mass of the quark by the
amount
\beq
\delta m_Q=
\int_{\mu_1<|\vec{k}\,|<\mu} \frac{d^3\vec k}{4\pi ^2} \,\frac{4}{3}\,
\frac{\alpha_s (\vec k^2)}{\vec k^2}\; .
\label{104}
\eeq
The pole mass clearly appears when $\mu_1\ra 0$. From the OPE point of
view, it is an attempt to construct an effective theory with the
normalization scale $\mu=0$ formulated, nevertheless, still in terms
of quarks and gluons.  Speaking theoretically, one can imagine a limit
of small $\mu$ which would correspond to integrating out all modes down
to $\mu=0$ in evaluation of the effective Lagrangian. It would be
nothing but constructing the $S$-matrix of the theory from which one
can directly read off all conceivable amplitudes. Clearly, it could
have been formulated in terms of physical mesons and baryons but not of
quarks and gluons! Such inconsistencies stemmed from postulating in
HQET technicalities like using the dimensional regularization as an
indispensable tool to formulate the theory at the quantum level.

In the Wilson OPE one uses $m_Q(\mu)$ with $\Lam \ll \mu\ll m_b$. I'll
illustrate later that just such a mass can be accurately measured in
experiment. It {\it is} $\mu$-dependent:
\beq
\frac{{\rm d}m_Q(\mu)}{{\rm d}\mu}\; =\; -\;\frac{16}{9}
\frac{\as(\mu)}{\pi}\;- \;
\frac{4}{3} \frac{\as(\mu)}{\pi} \frac{\mu}{m_Q}
\; + \; {\cal O} \left( \as^2, \as\frac{\mu^2}{m_Q^2}\right)\;;
\label{112}
\eeq
the higher order perturbative corrections were computed recently
\cite{dipole}.
There are different schemes for defining $m_Q(\mu)$ (similar to
renormalization schemes for $\as$), and the coefficients above are
generally different there. As long as a concrete scheme is adopted,
there is no ambiguity in the numerical value of $m_Q(\mu)$.
Instead of the HQET parameter $\La$ in QCD one has
$\La(\mu)=\lim_{m_Q\ra \infty} M_{H_Q}-m_Q(\mu)$. The value of
$\La(\mu)$ is of the hadronic mass scale if $\mu$ does not scale with
$m_Q$.

There exists a popular choice of a short-distance mass, the so-called
$\overline{\rm MS}$ mass $\bar{m}(\mu)$. The $\overline{\rm MS}$ mass
is not a parameter in the effective
Lagrangian;
rather it is a certain {\em ad hoc} combination of the  parameters
which is
particularly convenient in the perturbative calculations using
dimensional
regularization. Its relation to the perturbative pole mass is known to
two loops \cite{gray}:
\beq
m_Q^{\rm pole}\;=\;\bar{m}_Q(\bar m_Q)\left\{1+\frac{4}{3}
\frac{\as(\bar m_Q)}{\pi} + (1.56\,b
-3.73)\left(\frac{\as}{\pi}\right)^2
+...\, .
\right\}
\label{m15}
\eeq
At $\mu \gsim m_Q$ the $\overline{\rm MS}$ mass coincides,
roughly speaking,
with the running Lagrangian  mass seen at the scale $\sim
\mu$. However, it becomes rather meaningless at  $\mu  \ll  m_Q$:
\beq
\bar m_Q(\mu) \;\simeq\;\bar{m}_Q(\bar m_Q)\left\{1+\frac{2\as}{\pi}
\ln{\frac{m_Q}{\mu}} \right\}
\; .
\label{114}
\eeq
It logarithmically diverges when $\mu/m_Q \ra 0$.
For this reason $\bar m(\mu)$ is not
appropriate
in  the  heavy quark theory where  the possibility of
evolving down to a low normalization  point,  $\mu \ll  m_Q$,
is crucial. Otherwise, for example, $M_{H_Q}-\bar m_Q \propto m_Q$ and
does not stay constant in the heavy quark limit.

The reason for this IR divergence is that the $\overline{\rm MS}$
scheme technically attempts to determine the (perturbative) running of all
quantities by their divergences calculated when the space-time
dimension approaches $D=4$.  The divergence can emerge only at $k \ra
0$ or $ k\ra \infty$. For the mass, the IR divergences are absent, and
the dimensional regularization is sensitive only to the UV divergence
at $k\gg m_Q$. Studying only $1/(D-4)$ singularities, it is in principle
unable to capture the change of the regime at $\mu \lsim m_Q$ and
assumes the same running in this domain. The
actual running below $m_Q$ is much slower. Such a mistake was made, for
example, in \cite{neubtasi} where the logarithmic running of the mass
was stated for arbitrary scale.

The properly defined short-distance masses always exhibits an
explicit linear $\mu$-dependence similar to Eq.~(\ref{112}) at $\mu
\ll m_Q$.  The perturbative pole mass, order by order, would correspond
to the limit $\mu \ra 0$. However such a limit does not exist.

\subsubsection{Which mass is to be used?}

Since $m_Q^{\rm pole}$ does not exist as a well-defined mass parameter,
a different, short-distance mass must be used. The normalization point
$\mu$ can be arbitrary as long as $\mu \gg \Lam$. It does not
mean, however, that all masses  are equally
practical,  since the  perturbative series are necessarily truncated
 after a few  first  terms. Using an inappropriate scale makes
numerical approximations bad. In particular, relying on
$\bar{m}_Q(m_Q)$
in treating the low-scale observables can be awkward. The
following
toy example illustrates this point.

Suppose, we would like to exploit QCD-like methods in solving the
textbook
problem of the positronium mass, calculating the standard Coulomb
binding energy. To use  the result written in terms  of the
$\overline{\rm MS}$ mass one would, first, need  to evaluate $\bar
m_e(m_e)$. This  immediately leads to technical problems: $\bar
m_e(m_e)$ is known only to the order $\alpha^2  m_e$; therefore,
the
``theoretical uncertainty"  in $\bar m_e(m_e)$ would generate the
error bars $\sim \alpha^3 m_e$ in the binding energy, i.e.  at
least
$0.01\,{\rm  eV}$. Moreover, without cumbersome two-loop
calculations
one would not know the binding  energy  even at a few ${\rm eV}$
level
-- although, obviously, it is known to a much better
accuracy (up to $\al^4 m_e$ without a dedicated field-theory
analysis),
and the result
\beq
M_{P}\;=\; 2m_e(0)\; - \; \frac{\al^2m_e}{4} \;+\;
{\cal O}(\al^3 m_e)
\label{m18}
\eeq
is obtained without actual loop calculations!

Thus, working with the  $\overline{\rm MS}$ mass
we would have to deal here with the ``hand-made" disaster.
The reason is obvious. The relevant momentum scale in this
problem is the inverse Bohr radius,
$\mu_B = r_B^{-1}\sim \alpha m_e$. Whatever contributions emerge
at   much shorter distances, say, at $\mu^{-1}\sim m_e^{-1}$,
their sole role is
to  renormalize  the low-scale parameters $\al$ and $m_e$. If the
binding energy is  expressed in terms of these parameters,  further
corrections are  suppressed by powers  of $\mu_B/\mu$. Exploiting
the
proper  low-energy  parameters $\alpha(0)$, $m_e(0)$ one
automatically
resums `tedious' perturbative corrections.  In essence, this  is  the
basic idea of the effective Lagrangian  approach,  which,
unfortunately,  is often forgotten.

In the perturbative determination of $m_b(\mu)$, for example, the
importance of higher-order corrections becomes governed by
$\left(\frac{\as}{\pi}\right)^k \mu\,$ rather than
$\left(\frac{\as}{\pi}\right)^k m_b$. Since $\mu\ll m_b$, this is a
significant improvement.

Needless to say, it is the
high-scale masses that appear directly in the processes at high
energies. In particular, the inclusive width $Z\ra b \bar b$ is
sensitive to $m_b(M_Z)$;  using $\overline{\rm MS}$  mass normalized at
$\mu\sim M_Z$ is appropriate here.  On the contrary, the inclusive
semileptonic decays $b\ra c\,\ell \nu$ are rather low-energy in this
respect \cite{five}, and, to some extent, that is true even for $b\ra
u$.

The construction of the running low-scale heavy quark mass suitable for
the OPE to any order in perturbation theory can be done in a
straightforward way. However, the Wilsonian approach implies
introduction a cutoff on the momenta of the gluon fields. Since gluon
carries color, its momentum is not a gauge-invariant quantity, and such
a mass typically appears not gauge-invariant. More accurately,
obtaining the same $m_Q(\mu)$ requires somewhat different
cutoff rules in different gauges.

Even though this is not a real
problem for the theory, it is often viewed as a disadvantage. To get
rid of this spurious problem, a manifestly gauge-invariant definition
of the running mass was suggested in \cite{five,blmope} which is
formulated only in terms of observables. Moreover, it is
convenient for the OPE in the $1/m_Q$ expansion since more or less
directly enters many relevant processes, {\it e.g.} heavy flavor
transitions or the threshold heavy flavor production. Its relation to
the $\overline{\rm MS}$ mass is known with enough accuracy
\cite{dipole}. To avoid ambiguities I always use this definition unless
other convention is indicated explicitly.

It is important to note the following fact. In the relativistic theory
for a particle with mass $m$ one always has $p^2=m^2$, that is,
$E=\sqrt{m^2+\vec{p}^{\,2}}$. In the nonrelativistic expansion,
therefore,
\beq
E\;=\; m\,+\,
\frac{\vec{p}^{\,2}}{2m}\,-\,\frac{\vec{p}^{\,4}}{8m^3}\:+\;...\;;
\label{m20}
\eeq
all coefficient are fixed in terms of powers of $m$ by Lorentz
invariance.
As will be mentioned in Sect.~3.3.4, in applications to heavy quarks it
is typically advantageous to use such a Wilsonian cutoff which preserves
usual QM properties for the price of violating Lorentz invariance (the
reasons are discussed in \cite{blmope}). Then the mass parameters in
Eq.~(\ref{m20}) generally become different:
\beq
{\cal H}_Q\;=\; m_0 \varphi_Q^+ \varphi_Q \: - \:
\varphi_Q^+ A_0 \varphi_Q \: + \: \frac{1}{2m_2}\,
\left((i\vec D)^2 +
c_G \vec\sigma \vec B\right)\; - \; ...
\label{m22}
\eeq
even for a ``quasifree'' quark in the effective theory. Of course, the
differences between $m_i$ appear only due to perturbative corrections:
$$
m_i(\mu)\,-\, m_k(\mu) \;=\; {\cal O} (\alpha_s \mu)\;.
$$
Moreover, the differences can be calculated perturbatively and are
completely free of any IR effects existing below the scale $\mu$. For
example, in our scheme $m_0(\mu) \simeq m_2(\mu)+\frac{4}{9}
\frac{\as(\mu)}{\pi}$. We use the mass $m_2(\mu)$, the mass that enters
the kinetic energy operator $\vec{p}^{\,2}/2m$; it is the most
important mass for heavy quark transitions.

This calculable difference between different ``masses'' for the same
quark must be properly accounted for in the OPE analysis. Probably, the
most obvious place where it is important is the $\bar{Q} Q$ threshold
physics discussed in Sect.~3.2.3. While in the short-distance expansion
the free quark threshold starts at $2 m_0(\mu)$, the
propagation of heavy quarks or the bound state dynamics are actually
determined by $m_2(\mu)$. The shift in the position of the threshold
which serves as a reference point for energy in the nonrelativistic
system must be properly taken into account.

Concluding this section, I would like to make a side remark
concerning
the $t$-quark mass \cite{rev}. The peculiarity of the $t$ quark is that
it has a significant width $\Gamma_t \sim 1 \GeV$ due to its weak
decay.  The perturbative position of the pole in the propagator is,
thus, shifted into the complex plane by $-\frac{i}{2} \Gamma_t$.  The
finite decay width of the $t$ quark introduces a {\em  physical}
infrared cutoff for the infrared QCD effects \cite{khoze}. In
particular, the observable decay characteristics do not have ambiguity
associated with the uncertainty in the pole mass discussed above. The
uncertainty cancels in any physical quantity  that can be measured.
That is not the case, however, in the position of the pole of the
$t$-quark propagator in the complex plane (more exactly, its real
part). The quark Green function is not observable there, and one would
encounter the very same infrared problem and the same infrared
renormalon. The latter does not depend on the absolute value of the
quark mass (and whether it is real or have an imaginary part). Thus, in
the case of top, one would observe an intrinsic -- but artificial  --
infrared renormalon uncertainty of several hundred $\MeV$ in attempts
to relate the peak in the physical decay distributions to the position
of the propagator singularity in the complex plane.

While this fact will pose no practical problem any time soon, it will
do so in the future, in particular when analyzing top production at
linear ${\rm e^+e^-}$ colliders.
In my opinion, there is no point in expressing the fundamental
experimental findings in terms of unphysical parameters which,
additionally, cannot be defined even theoretically.
The observables can be conveniently
expressed in terms of a mass $m_t$ defined at the scale $\mu \approx
\Gamma_t$ that is  free from the  renormalon ambiguities. It can be
defined and measured without intrinsic limitations. It is worth trying
to come to an agreement in advance what kind of top quark mass should
be listed in the PDG tables, and thus avoid the problems which
accompanied attempts to pinpoint the masses of $c$ and $b$ quarks.

\subsubsection{The numerical values of $m_c$ and $m_b$}

The mass of the $c$ quark at the scale $\sim m_c \sim 1\GeV$ can
be obtained from
the charmonium sum rules \cite{SVVZ}, $m_c(m_c)\; \simeq\; 1.25 \GeV$.
The result to some extent is affected by the value of the gluon
condensate.  To be safe, we conservatively ascribe a rather large
uncertainty,
$$
m_c(m_c) = 1.25 \pm 0.1 \, \mbox{GeV}\, .
$$
There are reasons to believe that the precision
charmonium sum rules actually determine the charmed quark
mass to a better accuracy.

An accurate measurement of $m_b$ is possible in the $\bar bb$
production in ${\rm e^+e^-}$ annihilation. Since we want to know $m_b$
with comparable or better absolute precision, both the data and
calculations, at first glance, must have increased accuracy. The data
are available, however, only below and near the threshold. Certain
integrals (moments) of the cross section over this domain are
particularly sensitive to the low-scale mass $m_b(\mu)$ with $\mu$ in
the interval $1$ to $2 \GeV$. A brief description of the method can be
found in the review \cite{rev}; the details are given in the original
papers \cite{SVVZ,volmb}.

The value of $m_b$ extracted from the fit, taken literally, has a very
small uncertainty, and the actual error bars lie in the accuracy of
theoretical formulae, in particular, in the perturbative corrections.
Nevertheless, the leading Coulomb corrections have been resummed, and
the expected dominant BLM-type corrections accounted for. We argued in
\cite{rev} that the uncertainty in $m_b(\mu)$ with $\mu\approx 1 \GeV$
must not be worse than $30\MeV$, and suggested that a $\pm 50\MeV$
assessment yields a safe interval:
\beq
m_b(1\GeV)\;=\;4.64\GeV \pm  0.05\GeV\, .
\label{120}
\eeq
The definition  corresponds to the concrete renormalization scheme
\cite{blmope,five,rev} which we consistently use.  In other words,
the ``one-loop" pole mass is \beq m_b^{(1)}\;=\;4.64\GeV
\;+\frac{16}{9}\frac{\as}{\pi}\cdot 1\GeV \;\simeq \; 4.83\GeV
\label{m29}
\eeq
at $\as=0.336$. The result is, of course, sensitive to
the choice of $\mu$.
The corresponding value of $\La(1\GeV) \approx
0.6\,{\rm GeV}$.

The heavy quark masses can be measured, in principle,  by studying
the distributions in
the semileptonic $B$ decays  \cite{prl,volspec}. Such analyses
were undertaken recently \cite{chern,gremm}. Unfortunately, the
data are not good enough yet to yield a competitive determination. On
the theoretical side, there are  potential problems with higher-order
corrections due to not too large energy release $m_b-m_c\simeq 3.5\GeV$
and/or relatively small mass of  the $c$ quark. In particular, the
effect of higher-order power corrections can be noticeable. The result
of analysis reported in Refs.~\cite{chern,gremm} is compatible with the
value (\ref{120}) within the uncertainties.  Unfortunately, the results
are quoted for the pole mass obtained in a certain approximation and,
thus, only the errors of the fit itself are sensible. The
theoretical uncertainties were not even addressed consistently.

I think that direct determinations of $m_b(m_b)$ will
long suffer from the uncertainties at least $100$ to $200 \MeV$. This
is a minimal scale of the second-order corrections in the high-energy
measurements.  In order to extract $m_b$ one has to consider
short-distance observables which essentially depend on $m_b$. Exact
calculation of the ${\cal O}(\as^2)$ corrections with massive particles
is extremely difficult. Simultaneously, the absolute accuracy in the
measurements has to be very good for a competitive determination of
$m_b$.

A much larger uncertainty in the $b$ quark mass is often cited in the
literature, and it would not be justified to avoid discussing this
controversy . For example, Neubert inflates it up to $\pm 300\MeV$
\cite{neubmori,neubchi}. It is not clear where such an uncertainty
comes from; I suspect that it merely refers to the variation of
numerical values one would get using {\em different schemes} to define
theoretically what is the heavy quark mass. It makes no more sense
than, say, calling the difference in the values of $\overline{\rm MS}$
and ${\rm MS}$ strong couplings at the same scale the uncertainty in
$\as$.

Some scepticism to the determination of $m_b$ rose after
Ref.~\cite{pich} obtained -- using the analysis similar to the
Voloshin's \cite{volmb} -- a significantly lower value of $m_b$. The
new analysis was not quite correct, however. Its deficiencies were
discussed in detail recently in \cite{kuhn}. Among other things, it was
illustrated that including the missed resonant contributions in the
used sum rules reproduces the Voloshin's numbers. Nevertheless, the
general sceptical attitude seems to prevail up to now, and I feel
appropriate to present some complimentary arguments why there hardly is
a room for large uncertainties here. The controversial estimates
once again seem to be related to the attempts to determine nonexisting
pole mass of the $b$ quark.

In determination of $m_b$ from ${\rm e^+e^-} \ra b \bar b$ the
experimental input is moments $I_n$ of $R_b(s)$ which is
related to the cross section. The dispersion relations equate them with
the
derivatives of the correlator of the $b$ quark vector currents $\Pi_b$
at $q^2=0$:
$$
\frac{2\pi^2}{n!}\, \Pi_b^{(n)}(0) \; =\; I_n\; = \;
\int\,\frac{ds\,R_b(s)}{s^{n+1}}
\simeq
$$
\beq
M_{\Upsilon(1S)}^{-2(n+1)}\, \int \: ds\:R_b(s)\,
\exp\left\{
-(n+1)\left(\frac{s}{M_{\Upsilon(1S)}^2}-1 \right)\right\} \simeq
\label{3.12}
\eeq
$$
2M_{\Upsilon(1S)}^{-2n-1}\, \int \: dE\:
R_b\left((M_{\Upsilon(1S)}+E)^2\right)\:{\rm e\,}^{-\frac{E}{\Delta}}
\;,\qquad \;
\Delta \,\simeq\, m_b/n\;.
$$
The moments $I_n$ at large $n\gsim 10$ are very sensitive to the value
of $m_b$:
\beq
I_n\; \propto \;
{\rm e}\,^{-2n \frac{m_b-\frac{M_{\Upsilon(1S)}}{2} }{m_b} }\;.
\label{3.13}
\eeq
Moreover, the leading nonperturbative effects are described by the
vacuum gluon condensate $\aver{G_{\mu\nu}^2}$, so these corrections
are very small unless $n$
is taken too large to push the relevant momentum scale $\sim
m_b/\sqrt{n}$ down to a hadronic scale. However, one has to pay the
price of resumming the enhanced perturbative Coulomb corrections for such a
convenience. In the potential interaction the expansion parameter in
the nonrelativistic problem is $\as/|\vec{v}\,|$
rather than $\as$ itself, and typical $|\vec{v}\,|\sim
1/\sqrt{n}$.  The Coulomb corrections also contain numerically large
factors. For example, the resummation of the first-order Coulomb
exchanges yields the factor ranging from $10$ to $50$ for $n=10$ to
$20$ (roughly speaking, it is ${\rm e}\,^{\frac{4}{3}
\sqrt{n}\sqrt{\pi} \as}$ in the relevant domain $\as\sqrt{n} \sim 1$)
\cite{volmb}. Naively, without a precise resummation in the theoretical
expressions for the moments one cannot get anything reasonable for the
$b$ quark mass. Even worse, since the effect of resummation is so
large, one may naturally harbor doubts whether the higher-order effects
left over are really harmless.  In my opinion, all such fears are
exaggerated.

In reality, the majority of the analyses treated the sum rules for the
moments not quite consistently from the basic theoretical perspective,
regarding the role of the infrared effects. The OPE states that the
(leading) effect of the IR QCD dynamics in the moments $I_n$ is given by
the gluon condensate $\aver{G^2}$ {\it only} if the fixed $m_b$ is a
{\it short-distance} mass $m_b(\mu)$. The Coulomb potential
interaction, on the other hand, is resummed in the standard QM way
which is written in terms of the pole masses. In this case the naively
used OPE statement is not applicable, and the impact of the IR domain
on $I_n$ is much larger for large $n$. Keeping $m_b^{\rm pole}$ fixed
in the theoretical expressions rather adds a huge factor $\sim
{\rm e}\,^{2n \delta m_b^{\rm pole}/m_b}$ to $I_n$, where $
\delta m_b^{\rm pole}$ is the shift in the pole mass due to switching
on the $\as$-corrections in the IR domain of momenta below a certain
scale $\mu$. For essentially one-loop calculations, for instance,
$\delta m_b^{\rm pole} \sim \frac{\as}{\pi}\mu$. Any modification of
the used potential at large distances (small exchanged momenta) would
lead actually to a drastic change in $I_n\left(m_b^{\rm pole}\right)$.
This would be just a reflection of the fact that the moments in terms
of the short-distance mass remain the same, and the difference between
the pole mass and $m_b(\mu)$ changes.

If instead one keeps $m_b(\mu)$ fixed, the dependence of $I_n$ on the
IR part of gluon exchanges disappears. This can can be checked
explicitly. It is important to remember, however, that this is a
remarkable property of QCD. Exchange of massless scalar particles, for
example, can perfectly imitate the Coulomb interaction of
nonrelativistic particles, but the relative sign of the Coulomb
interaction and the corrections to the self-energy would be the
opposite for them. This remark suggests that one should be cautious
applying conclusions drawn from {\it ad hoc} nonrelativistic models to
the QCD analysis of the $\Upsilon$ system.
\vspace*{.4cm}

{\it The quantum mechanical interpretation}
\vspace*{.2cm}

\noindent
Let us illustrate in the simple language the OPE statements referred to
above. The largest and most dangerous in practice IR contribution to
the pole mass is linear in the momentum scale, Eq.~(\ref{3.2}). The
expression for the mass shift, actually, is nothing but
self-interaction $\frac{1}{2} V_{\rm IR}(0)$ where $V_{\rm IR}$ is the
heavy quark potential mediated by the gauge interactions with momenta
below certain $\mu \ll m_b$ \cite{gurman,pole}. Yet the mass of the
$\bar{b}b$ system includes also the same Coulomb interaction between
quark and antiquark.  Since for the colorless $\bar{b}b$ state the sum
of color ``charges'' is zero, these effects cancel each other for the
quanta with wavelength less than the interquark spacing $r$. Therefore,
for the Fourier transform of the potential $V(\vec{q}\,)$ in terms of
which
\beq V(0) \;=\; \int\; \frac{d^3 \vec{q}}{(2\pi)^3}\:V(\vec{q}\,)
\;,
\label{3.20}
\eeq only the components with $|\vec{q}\,| \gsim 1/r$
contribute. The softer exchanges are suppressed by powers of the
multipole factor $\vec{q}^{\,2} r^2$.

This, of course, automatically emerges in all calculations. Let us
single out the effect of gluon exchanges with $|\vec{q}\,| < \mu$:
\beq
V_{\rm IR}(r) \;=\; - \int_{|\vec{q}\,| < \mu}
\; \frac{d^3 \vec{q}}{(2\pi)^3}\:V(\vec{q}\,)\:
{\rm e\,}^{-i\vec{q}\vec{r}}\;=\;
-V_0\,+\, \frac{1}{6} r^2 \mu^2 \,V_2 \;-\; ... \;,
\label{3.22}
\eeq
$$
V_0\;=\; \int_{|\vec{q}\,| < \mu}
\; \frac{d^3 \vec{q}}{(2\pi)^3}\:V(\vec{q}\,)\;,\qquad\;
V_2\;=\; \int_{|\vec{q}\,| < \mu}
\; \frac{d^3
\vec{q}}{(2\pi)^3}\:\frac{\vec{q}^{\,2}}{\mu^2}\,
V(\vec{q}\,)\;, \qquad ...
$$
(the minus sign reflects the fact that the second particle is an
antiquark, {\it i.e.} negative $C$ parity of the vector current).
If quarks reside at distances much smaller than $1/\mu$, the soft
potential is just a constant. Its sole role is only to shift the energy
of all $\bar{b}b$ states by a constant amount $-V_0$. It does not
affect wavefunctions and, therefore, does not modify the coupling of
the virtual photon in $\rm e^+e^-$ annihilation to these states.

However, a constant potential $A_0$ cannot change the energy of
the neutral system, whether the field is classical or quantum. It means
that just the opposite shift is present in the sum of masses of $b$ and
$\bar{b}$ renormalized by the same gauge interaction. Indeed, explicit
expression (\ref{3.2}) shows that $\delta m_b$ amounts to a half of
$V_0$, so that the change in the sum of masses cancels the constant
term in the potential for closely spaced $\bar{b}b$. Once again the IR
effect distinguishing the pole mass from $m_b(\mu)$ disappears from the
short-distance quantity!

For the potential mediated by exchanges of a hypothetical scalar
$\phi(x)$, the ``scalar charges'' of $b$ and $\bar{b}$ would have the
same sign, so their sum is nonzero. The quark
self-energy would not cancel
anymore the shift in the potential but rather double it. In the OPE
language this contribution to the moments $I_n$ (and, eventually, to
$m_Q$) would correspond to the expectation value of $\phi^2$.
In QCD the operator $A_0^2$ is forbidden by gauge invariance.

The above reasoning, while illustrating the cancellation of the IR
effects present in the pole mass, is too naive in many aspects. If
these were the only grounds for the QCD analysis, it would have been
hardly relevant at all, at least for $b$ particles. The actual space
separation between quarks even in the lowest $\Upsilon(1S)$ state is
not much smaller than $\Lam^{-1}$ which is still strongly affected by
the nonperturbative dynamics. For higher states the situation is even
less favorable.

Additionally, the relation between the pole mass and the constant term
in the soft potential does not hold in QCD even in perturbation
theory in high orders, where the definition of the potential itself
becomes nontrivial (the subtle large-time effects in the heavy quark
potential not governed by the scale of the momentum transfer were
discussed long ago in \cite{dine}). As a matter of fact, the heavy
quark potential is not a truly short-distance quantity \cite{dipole},
either in coordinate or momentum space at $r\ll \Lam^{-1}$ or
$|\vec{q}\,| \gg \Lam$.

Nevertheless, the OPE ensures that the IR dynamics affect determination
of the short-distance mass $m_b(\mu)$ even less than can be concluded
from the naive potential description. The reason is that the existing
strategy \cite{SVVZ} suggested already for charmonium (where
perturbative potential treatment is out of question) studies genuinely
short-{\it distance} inclusive quantities like moments $I_n$
and not energies or wavefunctions of
individual states. This leads, actually, to a limited applicability of
the simple potential picture for soft power corrections and to
disappearance of unwanted complications mentioned above.
For example, there are no effects $\sim \Lam^3$ (or $\mu^3$, in
general, for $\mu\ll m_b/n$) in the moments $I_n$ which
naively can be anticipated from the curvature $V_2$ of the potential in
Eq.~(\ref{3.22}).

The key point is that the heavy quarks are pairly produced from a point
with the limited energy $\Delta \sim m_b/n$; the pair thus exists for
typical time~\footnote{In this analysis ``energy'' $\Delta$, time $t$,
velocity {\it etc.} are Euclidean or, actually, complex. This does not
change the qualitative reasoning described here.}
$t \sim 1/\Delta$. The quark velocity is small, $v\sim
\sqrt{\Delta/m_b}$. Even though the time separation between $\bar{b}b$
creation and annihilation can be large, $\gsim \Lam$, the quarks still
cannot move apart to a significant distance, and typical $r\sim vt\sim
1/\sqrt{\Delta m_b} \sim \sqrt{n}/m_b$ is small compared to $\Lam$.
Their interaction with gluon fields is then only dipole, {\it i.e.}
weak, and the pair propagates through an almost transparent medium. This
applies to both nonperturbative field configurations and to
perturbative gluons with $|\vec{q}\,| < 1/r$. Therefore, whatever
complicated is the structure of the bound states in an interquark
potential or behavior of the wavefunctions, the heavy quark
propagation between the points of creation and annihilation will be
perturbative.

It is also clear that the applicability of the potential picture for
calculating the corrections to the moments in the fiducial domain is
limited. The literal potential description is adequate when
particles reside long time at a certain distance. This is not the case:
since the quarks are produced pointlike, they remain close to each
other. Already the dipole interaction grows with distance, therefore the
static interaction properly accounts for a relatively small `middle'
part of the lifetime of the pair. The period of expansion from
the point of creation to the typical distances is not
static.  As a result, the naive potential estimates do not always hold
literally \cite{volpsi}.

The actual short-distance OPE expansion is applicable when $\Delta \sim
1/t \gg \Lam$. It predicts a small leading IR effect
$\sim \aver{G_{\al\beta}^2}/(\Delta^3 m_b)$. There is no a
potential-generated contribution $\sim \aver{r^2}\,t\, \mu^2 V_2
\sim \mu^2 V_2 /(\Delta^2 m_b)$ which would dominate at $t\ll
\Lam^{-1}$. This can be interpreted as the fact that the soft IR part
of the potential even at small distances (large spacelike momentum
transfer) depends on the time interval and is suppressed for the
potential ``switched on'' for only a too short period of time.
Similarly, the above mentioned long-distance effects in the
large-$\vec{q}$ potential $V(\vec{q}\,)$ associated with the large time
scale nonlocalities in $V$ do not emerge here.

At $ \Delta \lsim \Lam$ the literal OPE does not apply. Nevertheless
one still can use the usual multipole expansion in space coordinates.
The effect, in general, is not expressed in terms of local in time
expectation values but depends on the time correlators. A simple
estimate for the IR effect
$$ \aver{r^2}\, t \; \aver{\int_0^t d\tau
\vec{E}(0,0)\vec{E}(\tau,0)}\; \sim \;\frac{
\aver{\vec{E}^{\,2}}}{\Delta^3 m_b}
$$
qualitatively agrees with the
OPE at $\Delta \sim \Lam$. This correction is still small.

It is important that, although the literal OPE is justified at $m_b/n
\gg \Lam$, the smallness of the IR contributions and applicability of
the perturbative treatment of moments $I_n$ requires only $p_b \sim
\sqrt{\Delta m_b} \sim m_b/\sqrt{n} \gg \Lam$. The energy resolution
$\Delta = m_b/n$ can be much better than $\Lam$ for large $m_b$, but
the duality between the short-distance expressions for the moments and
their actual saturation by hadronic states will still hold. The
dedicated discussion of the quarkonium production near threshold and
OPE can be found in the old paper \cite{volpsi}.
\vspace*{.3cm}

After this digression in the meaning of the OPE for the moments of the
spectral function let us return to extraction of $m_b$.
Does the above
understanding help in practice? We know that the moments
are sensitive to a short-distance mass $m_b(\mu)$, and it means that
the QCD corrections to it must be moderate. The effective scale $\mu$
depends on $n$ and for the fiducial range $8\,-\,20$ used in
\cite{volmb} is $1$ to $2\GeV$. Then let us, for a moment, leave out
the resummation machinery of Voloshin at all and attempt to determine
$m_b(1\GeV)$ from simply the {\it tree-level} theoretical expressions,
in the same range $8$ to $20$. The result lies in the interval from
$4.58$ to $4.64\GeV$. Even without accounting for any QCD corrections
one gets the correct result with a $50\MeV$ accuracy!

If the ``bare'' result for $m_b$ is $m_b \simeq 4.6\GeV$, one should
more consistently evaluate the effect of the Coulomb resummation in
$I_n$ comparing them with the (bare) moments evaluated with this
short-distance mass. If we do this, the apparent impact of resummation
changes dramatically: the resummation increases moments by a factor
ranging only from 3 to 5. In fact, for a more natural, higher
renormalization point $\mu\simeq 1.5\GeV$ the impact becomes even
smaller and changes a little with variation of $n$. For $n\gg 1$ it is
this $n$-dependence that pinpoints the value of $m_b$.

Of course, before incorporating the perturbative
effects one cannot accurately check the normalization point dependence
of $m_b(\mu)$ and answer the question to which exactly $\mu$ such a
tree-level determination of $m_b$ refers. This requires account for the
perturbative corrections in the
theoretical predictions for the moments. It is
clear that this scale lies in the interval $1$ to $2\GeV$ (depending on
$n$), and is neither $m_b$ nor what would lead to $m_b^{\rm pole}$.

We see that already the tree-level computation yields the correct value
of the short-distance mass $m_b(1\GeV)$ with a $50\MeV$ accuracy. The
precision is further improved by the straightforward Coulomb
resummation. The existing analysis \cite{volmb} is additionally
improved by incorporating the next-to-leading $\as^{n+1}/|\vec{v}\,|^n$
corrections and the presumably dominant BLM $\as^2$ effects. Therefore,
there is no room in sight for larger uncertainty in determination of
$m_b(\mu)$. This explains why the most labor-consuming analysis of the
next-to-leading
effects in \cite{volmb} which was
necessary for getting a sensible determination of $\as$,  led to a
modest change in the numerical value of $m_b$ about only $\sim
15\MeV$, as compared to the ten-year old analysis of Ref.~\cite{old}.

One of the problems inflating the apparent uncertainties is that most
of the analyses of the $\Upsilon$ family attempt to determine
$m_b^{\rm pole}$. The uncertainty here, indeed, cannot be decreased
below $\sim 200\MeV$.  Using the $\overline{\rm MS}$ mass, on the other
hand, makes mandatory accounting for $\as^2$ effects exactly, which
seems unfeasible.

The way which would manifest smallness of the perturbative corrections
to extraction of $m_b(\mu)$ at any stage is to follow the idea of
the Wilson OPE more literally. Namely, one first gets rid of the high
gluon momenta $k\gsim m_b$ passing to the effective low-energy
nonrelativistic theory of $b$ and $\bar{b}$. For example, apart from
evolving $m_b$ to a lower scale, a certain finite renormalization of
the $\gamma^*\bar{b}b$ coupling is generated. This is typically
accomplished in the existing analyses to a certain extent. However we
should also separate out the momenta $|\vec{k}\,| < \mu$ from the
low-energy theory,  in particular from the potential. Their effect on
the moments $I_n$ is small and, if necessary, can be restored by adding
the OPE contribution in the form of $\aver{G^2}_\mu^{\rm pert}$
(for small $n$) or the perturbative part of the correlators in the
dipole expansion. Fitting the moments one would get directly
$m_b(\mu)$.

Since only a relatively small domain of gluon momenta, between $1$ to
$1.5\GeV$ and about $4\GeV$ would remain, the effect of the
exchanges to be computed will be very moderate. For example, the
potential will be peeled off from both hard and long-range parts, the
spectrum of states will be modified mildly, {\it etc.}

As a matter of fact, a similar procedure was already done by Voloshin
in early 1995. Instead of using the literal Coulomb potential with
small non-Abelian corrections, the low-momentum part was removed,
\beq
V_\mu(\vec{r}\,)\;=\; \int_{|\vec{q}\,|>\mu}
\frac{d^3 \vec{q}}{(2\pi)^3}\;
{\rm e\,}^{-i\vec{q}\vec{r}}\, V(\vec{q}\,)
\label{3.30}
\eeq
which yielded a certain $\mu$-dependent potential fast decreasing and
oscillating at large distances. Since the problem with the
Landau pole was absent at $\mu \gg \Lam$, it was possible to use the
all-order BLM-resummed potential instead of only simple BLM scale
fixing. The quark Green functions were calculated in this potential,
and the theoretical expression for the moments explicitly depended on
$\mu$. Fitted $m_b(\mu)$  depended on $\mu$, and the normalization
point dependence of mass agreed well with the perturbative
$\mu$-dependence similar to Eq.~(\ref{112}).

This analysis just led to the result
Eq.~(\ref{120}) (with the uncertainty about $20\MeV$); this
investigation, however, has not been officially published. (It was
referred to already in paper \cite{upset}, Ref.~23.) In the subsequent
papers I always used and quoted this evaluation of $m_b(1\GeV)$; it did
not actually differ from the value given in \cite{volmb} within the
stated error bars. I hope that an accurate analysis of this type will
eventually appear.

Completing the discussion of determination of $m_b$ I would like to
comment briefly on another outcome of the analysis of the
near-threshold ${\rm e^+e^-} \ra b \bar b$  cross section. The overall
normalization of the moments $I_n$ is also sensitive to the
value of $\as$ at the scale $1$ to $2\GeV$, which suggests a way to
accurately determine it. It was done in \cite{volmb} and resulted in a
value on a smaller side corresponding to $\as(M_Z)\simeq 0.110$, with a
very small quoted error. It is possible, however, that the actual
uncertainty in the short-distance $\as$ is larger here. Since
$\Upsilon$'s determine $\as$ at a relatively low scale, the effect of
higher-order terms can be rather significant. In other words, one
determines accurately here a certain effective coupling. Although it
can be, in principle, related to the standard $\overline{\rm MS}$
coupling, this relation is not known with enough precision.
Moreover, since the $\overline{\rm MS}$ coupling is quite unphysical
being selected to satisfy only the requirements
of convenience in technical computations, the higher-order terms in the
relation between the two couplings are expected to be large. The
perturbative relations can be poorly convergent -- if, in particular,
$\Lam$ happens to be on the higher side of the existing estimates. For
this reason, in my opinion it may be premature yet to assign a too
small error in the determination of the $\overline{\rm MS}$ coupling
$\as^{\overline{\rm MS}}$ evolved to higher energies.

At the same time, it is important that the associated uncertainty in
$\as$ does not affect $m_b(\mu)$ with $\mu \approx 1.5\GeV$. Other
masses like $m_b^{\overline{\rm MS}}(m_b)$ or various approximations for
$m_b^{\rm pole}$ are bound to exhibit sensitivity to it since involve
a significant $\as$-dependent evolution from that intermediate value
of $\mu$.

\subsection{ $\mu_\pi^2$ and $\mu_G^2$}

The heavy quark masses $m_c$, $m_b$, being the key parameters in the
HQE, are to a large extent `external' to the properties of the
effective low-energy theory itself. There are two nonrelativistic heavy quark
operators in the Hamiltonian; their expectation values Eq.~(\ref{34}) in
the heavy meson $B$ play a key role in many applications. In contrast
to $m_Q$, they are determined by the QCD dynamics itself. We cannot yet
calculate theoretically their values from first principles of QCD since
it would require more or less exact solution of QCD in the strong
coupling regime. Instead, we can try to measure them extracting from
known properties of hadrons.

The value of $\mu_G^2$ is known: since
$$
\frac{1}{2m_Q} \bar{Q}
\frac{i}{2}
\sigma_{\mu\nu} G_{\mu\nu} Q
$$
describes the interaction of the heavy quark spin with the light cloud
and causes the hyperfine splitting between $B$ and $B^*$,
\beq
\mu_G^2\; \simeq\;\frac{3}{4}\:2m_b(M_{B^*}-M_B) \;\simeq\;
\frac{3}{4}\,(M^2_{B^*}-M^2_B)\;\approx\; 0.36\GeV^2\;.
\label{p7}
\eeq
In actual QCD $\mu_G^2$ logarithmically depends on the normalization
point; usual one-loop diagrams yield
\beq
\mu_G^2(\mu')\; \simeq \;
\left(
\frac{\as(\mu')}{\as(\mu)}
\right)^{\frac{3}{b}}\,  \mu_G^2(\mu)\;\;.
\label{140}
\eeq
In the mass relation given above the operator is normalized at the
scale $\mu \simeq m_b$:
\beq
\frac{1}{m_Q} {\cal H}_1^{\rm spin}\;=\; -C_G(\mu)\frac{\left(
\bar{Q} \frac{i}{2}\sigma G Q\right)_\mu }{2m_Q}\;, \qquad
C_G\simeq
\left(
\frac{\as(m_Q)}{\as(\mu)}
\right)^{\frac{3}{b}}\: .
\label{141}
\eeq
Evolving perturbatively to the
normalization scale $\mu \sim 1\GeV$ slightly enhances the
value of $\mu_G^2(\mu)$, but this
effect is numerically insignificant, and we usually neglect it.

The kinetic expectation value $\mu_\pi^2$ has not been measured yet. It
enters various distributions in semileptonic decays or processes of the
type $ b \ra s+\gamma$. However, it remains rather uncertain at the
moment.
First  attempts to extract it from the semileptonic
distributions were reported, however so far the outcome suffers from
large, partially artificial uncertainties and is inconclusive. The
inconsistency in treating the quark masses propagates to the kinetic
operator; since it is a subleading $1/m_Q$ effect, making up for this
is already more tricky than for $m_b$ itself.

Historically, the first attempt to determine the average of the kinetic
operator from the QCD sum rules was Ref. \cite{neubold} where a
negative (!) value  $\approx-1\GeV^2$ was obtained. The treatment of
the sum rules was not legitimate, however (the result was eventually
retracted by the author.) Shortly after, a more thoughtful application
of the QCD sum rules yielded \cite{pp} $\mu_\pi^2\approx 0.6\GeV^2$.
The prediction was later refined by the authors \cite{ppnew}
\beq
\mu_\pi^2\;=\; (0.5\pm 0.15) \GeV^2 \; .
\label{p10}
\eeq
Meanwhile a model-independent lower bound was established
\cite{motion,volpp,vcb,optical}
\beq
\mu_\pi^2\;>\;\mu_G^2\approx 0.4 \mbox{GeV}^2
\label{p11}
\eeq
which  constrained  possible values of $\mu_\pi^2$.
It is worth emphasizing that this inequality takes place for
$\mu_\pi^2$ normalized at any point $\mu$, provided
$\mu_G^2$ is normalized at the same point.
For large $\mu$ it becomes uninformative; so, it is in our best
interests to use it at $\mu$ = several units $\times \Lambda_{\rm
QCD} \lsim 1 \GeV$.

Recently, a new QCD sum rules evaluation of $\mu_\pi^2$ was undertaken
in \cite{neubp} and was claimed to
yield the result $0.1\pm 0.05\GeV^2$; irrespective of the central
value, the quoted error bars are clearly unrealistic.

The expectation value of the kinetic operator was estimated in
the quark models, with also a controversial spectrum of predictions:
the nonrelativistic ISGW model was stated to predict for $\mu_\pi^2$
the value $0.27\GeV^2$; the relativistic
quark model \cite{rel} gave about $0.6\GeV^2$  or even
slightly larger; the estimates of Ref.~\cite{hwang} yield a close value
$0.5\GeV^2$.

The origin of the difference between the QCD sum rule evaluations
\cite{pp,ppnew} and \cite{neubp} was explained in detail in \cite{rev}.
While Ball {\it et al}. considered the sum rules for $\mu_\pi^2$
directly, the recent approach by Neubert calculated some other
transition amplitude which, however reduces to the expectation value
of the kinetic operator by equations of motion. In this respect, these
two methods {\it a priori} are equally justified and must have yielded
the same result provided the calculations were exact. However, the
techniques of the QCD sum rules is based on certain approximations, and
they are quite different for these quantities. In particular, it refers
to the contribution of the excited states which is a kind
of a theoretical background, and has to be suppressed as much as
possible.

Not much is known about these contributions in QCD. The
simple model considerations suggest \cite{rev} that the bias can be more
significant for the sum rules of \cite{neubp} than in the traditional
approach by Ball {\it et al}. The general QCD sum rule experience also
states that the sum rules operate in better environment when there is a
continuum contribution imitating the excited states than when it is
absent (as in the correlators considered in \cite{neubp}). The actual
scale of associated uncertainty in $\mu_\pi^2$ in each approach remains
questionable. It is obvious that the actual uncertainty in the analysis
of \cite{neubp} is grossly underestimated and in reality must be
increased by a large factor. The error bars quoted in \cite{ppnew} may
also require some revision.

Before mentioning the problem of the quark models, let me
briefly discuss the inequality $\mu_\pi^2 > \mu_G^2$. Its
physical meaning will be explained in Sect.~3.4.1. The most transparent
way to illustrate it is based on the sum rules for $\mu_\pi^2$ and
$\mu_G^2$ \cite{third,rev}:
\beq
\frac{\mu_\pi^2}{3} \;=\;\;\;\;\;
\sum_n \epsilon_n^2|\tau_{1/2}^{(n)}|^2 +
2\sum_m \epsilon_m^2|\tau_{3/2}^{(m)}|^2\, ,
\label{4.3}
\eeq
\beq
\frac{\mu_G^2}{3}\; =
-2 \sum_n \epsilon_n^2|\tau_{1/2}^{(n)}|^2 +
2 \sum_m \epsilon_m^2|\tau_{3/2}^{(m)}|^2 \, ,
\label{4.4}
\eeq
where $\tau_{1/2}^{(n)}$ and $\tau_{3/2}^{(m)}$  are the standard
notations for the transition amplitudes of the
ground state $B$ meson to
the excited `$P$-wave' states with spin of the light cloud
$j=\frac{1}{2}$ and $\frac{3}{2}$, respectively (for explicit
definition see, e.g., \cite{rev}).
The excitation energies
$\epsilon_k$ are defined as $\epsilon_k=M_{H_b^{(k)}}-M_B$.
The products $\epsilon_k \tau^{(k)}$ are merely the matrix elements of
the momentum operators $\vec\pi = \bar{b}(i\vec{D}\,)b(0)$ between the
ground state and the corresponding excited states \cite{optical,rev}.
These sum rules belong to a whole family; I will need later
another, so-called optical sum rule \cite{volopt} for
the difference between the meson and the quark masses:
\beq
\La(\mu) =
2\left( \sum_n \epsilon_n|\tau_{1/2}^{(n)}|^2 + 2\sum_m
\epsilon_m|\tau_{3/2}^{(m)}|^2 \right)\;.
\label{4.2}
\eeq
The two
first sum rules immediately lead to
\beq
\mu_\pi^2\;-\; \mu_G^2 \;=\; 9
\sum_n \:\epsilon_n^2\, |\tau_{1/2}^{(n)}|^2\;>\;0 \; .
\label{4.19}
\eeq
In QCD both $\mu_\pi^2$ and $\mu_G^2 $ depend on normalization point
$\mu$.  The inequality holds for arbitrary $\mu$ as long as
renormalization of both operators is introduced in the same way.

The values of $\tau$'s, in particular, $\tau_{1/2}$ are not well
known at present. The QCD sum rules \cite{bsrho}
as well as the ISGW model \cite{isgw}
predict for the lowest states $\tau_{1/2}^{(1)}\approx
0.25\;\mbox{to}\;0.3$ while the corresponding $\epsilon_1$ is expected
to be about  $400\MeV$.  Assuming $\mu_G^2 \approx 0.4\GeV^2$ we then
have~\footnote{In the nonrelativistic oscillator model the transition
amplitudes vanish for all but the first excitations.} $\mu_\pi^2\gsim
0.5\GeV^2$.

\subsubsection{Kinetic operator in the nonrelativistic models}

Let me briefly describe the situation with the
nonrelativistic quark models. The calculation of the average square of
the spacelike momentum in the ISGW model yield a value close to
$0.27\GeV^2$, notably smaller than the model-independent
bound~\footnote{In the field theory where the perturbative radiative
corrections are included, certain alternative definitions of quantum
operators $O_\pi$, $O_G$ differing in subtraction of the perturbative
effects, may modify the inequality. It cannot be violated, however,
if the radiative effects are absent or neglected, as in the quark
models.} Eq.~(\ref{4.19}).  The reason is that in the nonrelativistic
systems the chromomagnetic field $\vec{B}$ is much smaller than
$\vec{p}^{\,2}$ and the spin-orbital interaction is negligible. As a
result, one additionally has the spectator spin degeneracy of the light
cloud states, and $\epsilon_{1/2}=\epsilon_{3/2}$,
$\tau_{1/2}=\tau_{3/2}$ for all excitations. The sum rule (\ref{4.4})
would yield $\mu_G^2 \simeq 0$ in agreement with vanishing $\vec{B}$ in
the hadronic scale.

The nonrelativistic constituent quark models typically yield
reasonable description of actual mesons despite their admittedly
simplified nature. Why does it fail here? The actual reason lies in an
implicit assumption of what is the counterpart of the kinetic operator
in these models. In QCD $\mu_\pi^2$ is the expectation value of
$\vec{\pi}^{\,2}$, and $\vec{\pi} =i\vec{D}-\vec{A}$, with the vector
gauge potential $\vec{A} \sim \Lam$.  In the nonrelativistic models
$\vec{A} \ll A_0 \sim \Lam$ and is neglected, so that the
identification of the covariant momentum with the ordinary momentum
operator
\beq
\vec{\pi} \;\ra\; i\vec \partial\,=\, \vec p\;, \qquad
\mu_\pi^2\: \ra \: \aver{\vec p^{\,2}}
\label{150}
\eeq
is assumed as a self-evident thing. However, it is by far not as
harmless assumption as it might seem. Since the momentum operators
commute, one has
\beq
\mbox{Quark model: }\;
\left[\pi_j,\pi_k\right]\;=\; 0\;,
\label{151}
\eeq
whereas in QCD
\beq
\left[\pi_j,\pi_k\right]\;=\; iG_{jk}\;=\; - i\epsilon_{jkl} B_l
\;\simeq\; -\frac{2}{3} i \epsilon_{jkl} \; j_l \cdot 0.4\GeV^2
\;.
\label{152}
\eeq
The fact itself of noncommutativity of the momentum operators leads to
the stated lower bound on $\mu_\pi^2$, and this noncommutativity is an
{\it exact} statement of QCD. The nonrelativistic quark models like the
ISGW model, identifying $\pi_i$ with $p_i$, violate this commutation
relation. This is as dangerous as modifying the basic uncertainty
principle of QM $[x_j,p_k]= i\hbar \delta_{jk}$.

This, in fact, is the main problem of quark models: the chromomagnetic
field is experimentally known to be of primary importance for the
properties of the light cloud in $B$ mesons, but it does not naturally
fit the nonrelativistic description of valence quarks.

Can the utility of the models like ISGW proved to be reasonable for
the wealth of other hadron properties, be rescued? In my opinion, the
sum rules provide such a possibility even without actual
rebuilding the model like changing its Hamiltonian or explicitly
introducing the gauge potential. Indeed, we do not expect the model to
yield grossly distorted values of the transition probabilities
$\tau^2$ or the excitation energies $\epsilon_k$. Accounting
for the relativistic spin-orbital interaction is rather expected to
reasonably split the values of $\epsilon^{(k)}_{1/2}$ and
$\tau^{(k)}_{1/2}$ from $\epsilon^{(k)}_{3/2}$ and
$\tau^{(k)}_{3/2}$ which otherwise are equal. The sum rule
Eq.~(\ref{4.4}) says that, most probably, the model systematically
overestimated the value of $\epsilon^{(n)}_{1/2}\tau^{(n)}_{1/2}$
and/or underestimated $\epsilon^{(m)}_{3/2}\tau^{(m)}_{3/2}$. Let us
denote
\beq
\sum_m \epsilon_m^2|\tau_{3/2}^{(m)}|^2\;=\; (1+\al)\,
\sum_n \epsilon_n^2|\tau_{1/2}^{(n)}|^2 \:, \qquad \al>0\:,
\label{154}
\eeq
where $\al$ is a positive constant, presumably of order $1$.
Next, we assume that the ISGW model approximately correctly evaluated
the average sum of the products $|\epsilon^{(k)} \tau^{(k)}|^2$. By
virtue of the sum rule (\ref{4.3}) it amounts to
\beq
(3+2\al)\, \sum_n
\epsilon_n^2|\tau_{1/2}^{(n)}|^2 \;\approx \;\frac{\aver{p^2}_{\rm
ISGW}}{3}\;\approx \; 0.09\GeV^2\;.
\label{155}
\eeq
Eq.~(\ref{4.19}) then yields \beq
\mu_\pi^2\; \simeq\;
\mu_G^2\,+\, \frac{\aver{p^2}_{\rm ISGW}}{1+\frac{2}{3}\al}\;.
\label{156}
\eeq
Even varying the value of $\al$ from $0.5$ to $3$ we get $\mu_\pi^2$
ranging from $0.5\GeV^2$ to $0.6\GeV^2$ (I put $\mu_G^2=0.4\GeV^2$).

Of course, the assumptions made above are not very rigorous.
Nevertheless, I believe that such a way to cure the deficiency of the
nonrelativistic
constituent quark models is reasonable. It is fair to say that
the value of $\mu_\pi^2$ in the QCD-compatible model motivated by ISGW
is $(0.5\pm 0.1)\GeV^2$, unless one is ready to accept that the ISGW
model provides a strongly distorted description of $B$ mesons.

It is important to keep in mind that in actual QCD the expectation
values of the kinetic operator also depend on the normalization point,
the effect which is not present in the simple quark models. The
prediction obtained in such models refer to a relatively low
normalization point $\mu\approx 0.7\,\mbox{to}\,1\GeV$. In HQET an
analogue of $\mu_\pi^2$ is called $-\lambda_1$. If the light cloud in
the hadrons were a simple QM system without actual gluonic
degrees of freedom, then $-\lambda_1$ would be equal to $\mu_\pi^2$.
The short-distance (perturbative) effects distinguish them making
$\mu_\pi^2$ scale-dependent. In contrast to $\mu_\pi^2$,
$-\lambda_1$ is
postulated to be a fundamental scale-independent quantity, from which
``all perturbative contributions'' are subtracted. In this respect, it
is a direct analogue of the pole mass of the heavy quark, and is
ill-defined as well. It makes no sense to state the exact value of
$-\lambda_1$ since it can be chosen arbitrarily within a few
units times $\Lam^2$. On the contrary, $\mu_\pi^2(\mu)$ is well defined
at all $\mu$. At $\mu \gg \Lam$ its $\mu$-dependence is given by
perturbation theory \cite{optical,dipole}
\beq
\frac{{\rm d}\mu_\pi^2(\mu)}{{\rm d}\mu^2}\; = \; \frac{4}{3}\,
\frac{\as\left({\rm e}^{-5/3+\ln{2}} \mu\right)}{\pi}
- \left(\frac{\pi^2}{2}-\frac{13}{4} \right)
\left(\frac{\as}{\pi}\right)^2 \,+\,{\cal O}(\as^3)\;.
\label{160}
\eeq

As a matter of fact, a constructive definition of $-\lambda_1$ has
never been given. What is used in the literature is an implicit
assumption that one must subtract from $\mu_\pi^2(\mu)$ its formal
perturbative part, that is, integrate Eq.~(\ref{160}) down to $\mu=0$
order by order in $\as$. It is well known that such a procedure cannot
be performed correctly, the resulting series is divergent while its
sum is meaningless and the value of $-\lambda_1$ strongly depends on
the way one viciously defines the sum or truncates the series. To
illustrate it numerically, for $\mu=1\GeV$ the series would
look like
\beq
-\lambda_1= \mu_\pi^2(\mu)- \frac{4}{3} \frac{\as}{\pi} -...\,=\,
\mu_\pi^2(1\GeV)-0.12\GeV^2 -
0.36\GeV^2 -\,...  \; ;
\label{crad34}
\eeq
the higher-order terms grow and constitute already in the low orders
shifts larger than the quantity in question. Certainly, the number
assigned to $-\lambda_1$ can be arbitrary compared to $\mu_G^2$: no
rigorous bound can be derived for the quantity which is not possible to
define.

\subsubsection{Hard QCD and normalization point dependence}

The sum rules (\ref{4.3})--(\ref{4.4}) express $\mu_\pi^2$
and $\mu_G^2$ as the sum of observable quantities, products of the
hadron mass differences and transition probabilities. The observable
quantities are scale-independent. How then $\mu_\pi^2$ and $\mu_G^2$
happen to be $\mu$-dependent?

The answer is that in actual quantum field theory like QCD the sums
over excited states are generally UV divergent when $\epsilon_k \gg
\Lam$, and not saturated by a few lowest states with contributions
fading out fast
at large $\epsilon_k$ as in ordinary QM. The contributions of states
with $\epsilon_k \gg \Lam$ are dual to what one calculates in
perturbation theory using its basic objects, quarks and gluons. The
latter, of course, yield the continuous spectrum and the corresponding
transition probabilities do not depend on the initial bound state. They
can be evaluated perturbatively using isolated quasifree heavy
quarks as the initial state. The final states are heavy quarks and a
certain number of gluons and light quarks. The difference between the
actual hadronic and quark-gluon transitions resides at low excitation
energies.

In order to make the sum rules meaningful, we must cut off the sums at
some energy $\mu$ which then makes the expectation value
$\mu$-dependent. The simplest way is merely to extend the sum only up
to $\epsilon_n<\mu$; this is the convention I consistently use. For
analytic computations it is often convenient to use the exponential
factor ${\rm e}^{-\epsilon_n/\mu}$, which is essentially the Borel
transform of the related correlation functions. The concrete values of
$\mu_\pi^2$ and $\mu_G^2$ depend, of course, on the adopted scheme.
Since at large $\mu$ the cutoff factors differ only in the
perturbative domain, the difference between various renormalization
schemes can be calculated perturbatively.

The high-energy tail of the transitions to order $\as$ is given by the
quark diagrams in Figs.~3 with
$$
\sum_k ...\; \ra \; \int \frac{{\rm d}^3\vec k}{2\omega}
$$
where $(\omega,\vec{k})$ is the momentum of the real gluon. The
amplitudes are just a constant proportional to $g_s$, and performing the
simple calculations we arrive at the first-order term in the evolution
of $\mu_\pi^2(\mu)$, Eq.~(\ref{160}). Purely perturbatively, the
continuum analogues of $\tau_{1/2}$ and $\tau_{3/2}$ are equal and a
similar `additive' renormalization of $\mu_G^2$ is absent.

\begin{figure}
\vspace{3.4cm}
\includegraphics{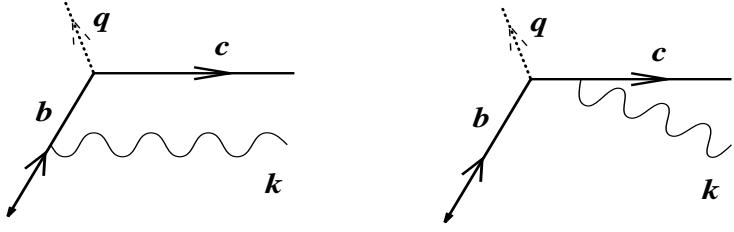}
\caption{\footnotesize
Perturbative diagrams
determining the high-energy asymptotics of the heavy quark transition
amplitudes and renormalization of the local operators.}
\end{figure}

The perturbatively obtained evolution equations (\ref{160}),
(\ref{140}) allow one to determine the asymptotic values of
$\tau_{1/2}$ and $\tau_{3/2}$  at $\epsilon \gg \Lam$:
\beq
\sum_{\mu< \epsilon_n <\mu+\Delta}
\epsilon_n^2|\tau_{1/2}^{(n)}|^2 \;+
2 \!\! \sum_ {\mu< \epsilon_m
<\mu+\Delta}\epsilon_m^2|\tau_{3/2}^{(m)}|^2\, \simeq\,
\frac{8}{9}\frac{\as(\mu)}{\pi}\: \mu \, \Delta\;,
\label{170}
\eeq
\beq
\sum_{\mu< \epsilon_m <\mu+\Delta}
\epsilon_m^2|\tau_{3/2}^{(m)}|^2 -
\sum_{\mu< \epsilon_n <\mu+\Delta}
\epsilon_n^2|\tau_{1/2}^{(n)}|^2 \, \simeq \,
-\frac{3\as(\mu)}{2\pi}\, \frac{\Delta}{\mu}
\left\{
\sum_{\epsilon_m <\mu} \epsilon_m^2|\tau_{3/2}^{(m)}|^2 -
\sum_{\epsilon_n <\mu} \epsilon_n^2|\tau_{1/2}^{(n)}|^2
\right\}.
\label{171}
\eeq
The minimal value of $\Delta$, a duality interval depends on the
considered energy. When $\mu$ is in the resonant zone, only a little
above the first prominent resonances, $\Delta$ is of order $\Lam$.
At large values of $\mu$ where the spectrum is smooth, $\Delta$ can be
taken smaller.

Such asymptotic equations, even incorporating higher-order
corrections, are valid only for large enough $\mu$, above the onset of
duality where the perturbative expressions are applicable. Although formally
$\mu_\pi^2(\mu)$ and $\mu_G^2(\mu)$ are defined at arbitrary $\mu$,
below a certain scale their running is completely different from the
perturbative renormalization. In particular, they both {\it vanish}
when $\mu\ra 0$ while staying finite for any value of $\mu$. It is in
a sharp contrast with the HQET definition of, say, $\mu_G^2(\mu)$ (it is
called $3\lambda_2$) which would exhibit a singularity at some value of
$\mu$ where the $\overline{\rm MS}$ coupling approaches the Landau
singularity. The actual running of $\mu_G^2(\mu)$  has nothing to do
with the formal perturbative expression $-\gamma_G \mu_G^2(\mu)$ at
small $\mu$. The same general remark refers also to the definition of
the Isgur-Wise function.

\subsubsection{Where is the caveat?}

Although the inequality $\mu_\pi^2(\mu) >\mu_G^2(\mu)$ holds for any
$\mu$, it may, or may not be practically informative. The value of
accounting for the so-called ``condensate'' nonperturbative corrections
when the perturbative effects are present is a dynamic question of
QCD, and it cannot be answered {\it a priori} by only mathematical
manipulations. The dynamic challenge emerges for the sum rules as
the question at which scale duality sets in, that is, what
is the minimal value of $\mu$ where the perturbatively obtained
expression for the amplitudes (in particular, $\mu$-dependence) start
to hold for the actual hadrons to a reasonable accuracy. For example,
at large enough $\mu$ the inequality $\mu_\pi^2(\mu) >\mu_G^2(\mu)$ is
trivial:  the kinetic expectation value is then dominated by the large
perturbative piece $\frac{4\as(\mu)}{3\pi}\mu^2$ while
$\mu_G^2(\mu)\sim \Lam^2$. At which minimal $\mu$ the value of
$\mu_G^2(\mu)$ is already a quantity about $0.4\GeV^2$? We can be
reasonably confident that at $\mu$ as large as $m_b\sim 4\GeV$ it
holds. Is it still as significant at the scale $\mu \lsim 1\GeV$, or
its value is mainly saturated at $\epsilon_k\gsim m_c$, we do not know
for sure.  This is the basic question we must learn to determine the
applicability of the heavy quark expansion.

It is tempting to optimistically interprete the well-satisfies
experimentally relation
$$
M_{D^*}^2-M_D^2\;\simeq \; M_{B^*}^2-M_B^2
$$
suggested by the $1/m_{c,b}$ expansion as an evidence for a safe early
onset of duality. It should be noted, however, that the same difference
persists also for $K^*-K$ and even $\rho-\pi$ splitting, which certainly
cannot be attributed to a heavy quark symmetry and must have an
independent dynamic origin.  The question, therefore, remains open.

If, alternatively, the saturation is delayed, one cannot count on
any quantitative application of the heavy quark symmetry to charmed
hadrons, including the symmetry relations for the formfactors. I
consider the question of how the sum rules for $\mu_G^2$ (and the
similar sum rules for $M_B-m_b$, $M_{\Lambda_b}-m_b$) are saturated
one of the most topical issues in the HQE. The first step toward
obtaining the answer from experiment has been recently done in
\cite{tau32exp}. The uncertainties are still large, however. Only the
leading $1/m_Q$ terms in the amplitude were considered; all they
have $1/m_c$ corrections expected to be quite sizeable. The outcome
is yet inconclusive -- stretching the uncertainties in one direction
it is possible to reasonably saturate
$\mu_G^2$ even at as low scale as
$600\MeV$.  The space and time limitations do not allow me to dwell on
this issue here in any detail.  It is very interesting to understand
also on which hadronic states the significantly larger $\La$ for the
$\Lambda_b$ baryon is saturated.  Unfortunately, the necessary
experimental information is practically absent here.

\subsubsection{Theoretical passions around the kinetic operator}

The question of the field-theoretic definition of the kinetic operator
used to be a subject of controversial opinions in the literature; even
now some incorrect statements remain. It makes sense to briefly address
these problems here.

A constructive definition of an effective theory and the composite
operators in it, or defining renormalized operators in full QCD
requires eliminating the high-energy modes of the quark and gluon
fields. It can be done in somewhat different ways. For
instance, even fixing the normalization scale to be $\mu$ one still can
cut off modes higher than, say, $2\mu$ or $\mu/2$. Of course, the
difference can be more essential. This is illustrated in the most
transparent way on the example of the perturbative corrections.

Let us consider the kinetic operator $\bar Q (i\vec D)^2Q(0)$ and keep
in mind that $m_Q\ra \infty$. The operator is perfectly defined at the
semiclassical level when the gauge field is a fixed external
($c$-number) function of coordinates. However, the quantum fluctuations
of $A_\mu$ generate divergent corrections: interacting with the gauge
quanta the heavy quark acquires a recoil which diverges in the ultraviolet
when integrated over the frequencies of the quanta.
For example, two
diagrams in Fig.~4  describe the order-$\as$ corrections to the
operator whose expectation value generally becomes nonvanishing even
without any external field:
$$
\left(\bar Q (i\vec D)^2Q(0)\right)_{\rm pert} \;=\;
\frac{4}{3}\,g_s^2\left(I_a+I_b\right)\;.
$$
The corresponding Feynman integrals are
\beq
I_a\;=\; \int\; \frac{{\rm d}^4k}{(2\pi)^4 i} \:\frac{1}{k^2}\,
\frac{\vec{k}^{\,2}}{k_0^2} \;, \qquad
I_b\;=\; -\int\; \frac{{\rm d}^4k}{(2\pi)^4 i} \:\frac{3}{k^2}
\;.
\label{A1}
\eeq
The integrals are quadratically divergent in the ultraviolet and
somehow must be cut off.

\begin{figure}
\vspace{2.9cm}
\includegraphics{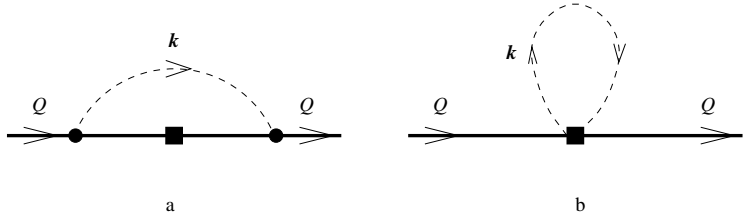}
\caption{\footnotesize
Feynman diagrams contributing to the one-loop
renormalization of the kinetic operator $\bar{Q} (i{\vec D})^2 Q$.
Dashed line denotes gluon, and dark box represents the operator.}
\end{figure}

The UV regularization can be done in the different ways. If one wants
to preserve the QM description of the effective theory, it must be
local in time and no cut over the timelike component of $k_0$ of the
gluon momentum is allowed (a more detailed discussion can be found in
\cite{blmope}). We can thus insert $\vartheta(\mu^2-\vec{k}^{\,2})$ in
the integrand, that is, cut off over the spacelike components of the
gluon momentum. It would preserve such natural properties of the
effective theory  as positivity of transition probabilities, {\it
etc}. -- even for momenta up to $\mu$. This is the regularization
suggested in \cite{optical}, and the explicit expressions I used in
these lectures always refer to it. With such a regularization
$3I_a=-I_b$, and Eq.~(\ref{160}) is reproduced.

In calculating the one-loop perturbative diagrams it is often more convenient to
impose a cutoff as a function of (Euclidean) $k^2$ -- that is, first to
perform the Wick rotation $k_0=ik_4$, and then add the factor
$\vartheta(\mu^2-k^2)$ or any other cutoff function of
$\vec{k}^{\,2}+k_4^2$. For example, this is the way the cutoff is
introduced in the Pauli-Villars scheme which regulates the integrals
subtracting the similar diagrams with fictitious gluons having large
masses. The direct computation yields that if, instead of integrating
over $k_0$ one first averages over the directions of Euclidean
momentum $k$, then $I_a=-I_b$, and the perturbative contribution
vanishes.

This fact was noted in \cite{bbz} where the authors
employed essentially the
Pauli-Villars--type regularization and observed vanishing of the
first-order renormalization of the kinetic operator.\footnote{Actually,
I showed in \cite{blmope} that this fact holds to all orders in the
Abelian theory without light flavors, and up to $\al^3$ in QED with
light flavors.}
It stimulated speculations that this fact could be not accidental but
have a deep reason which would ensure vanishing of the renormalization
to all orders. These speculation were intensively lobbied for some time
(see, {\it e.g.} \cite{invis}). In particular, they served a ground for
the hope that there existed a ``proper'' regularization scheme in which
there were no perturbative corrections to the kinetic
operator at all, and which thus
would provide a possibility to define unambiguously  an ``absolute''
value for $-\lambda_1$, as it is desired in HQET. In reality, there
were no reason for such a miraculous expectation. Eventually, even
Neubert had to admit \cite{67rev} that $-\lambda_1$ is an unphysical
and indefinite parameter.

The proponents of such a possibility blamed the physical
regularization described above for violating Lorentz invariance -- the
cutoff over $\vec{k}$ allegedly is not Lorentz-invariant in contrast
to the cutoff in $k^2$. The nonvanishing renormalization was attributed
to this `defect'. Such a claim is erroneous in this context, for
$\vec{k}^{\,2}$ is as Lorentz invariant as $k^2$ since there is an
intrinsic Lorentz vector $v_\mu$ in the problem, and $\vec{k}^{\,2}=
(vk)^2-k^2$.  The idea to invent a ``Lorentz-invariant'' (that is,
$v$-independent) regularization is misleading since the kinetic
operator itself
$\bar Q \left\{(iv D)^2-(iD)^2\right\}Q(0)$ is explicitly
$v$-dependent. To state it differently, already the HQET heavy quark
field $h_v(x)$ itself is not Lorentz-invariant by construction
Eq.~(\ref{13a}), and it obeys the $v$-dependent equation of motion
$i(vD)\,h_v(x)=0$.

The question of renormalization in higher orders could have been
answered calculating the two-loop radiative corrections to the kinetic
operator. However, it would require first to define a ``proper
Lorentz-invariant'' cutoff procedure. As explained above, there is no
natural way to define it in general, and it could have been done
only in the technical way at the diagrammatic level. The naive
Pauli-Villars regularization is not possible since the gluon mass
violates non-Abelian current conservation; the result becomes
gauge-dependent starting two loops where the difference with the
Abelian case first energies. Moreover, the momentum itself of a color ed
particle (gluon) is a gauge-dependent quantity. Therefore, the
requirement of a ``Lorentz-invariance'' of the cutoff explicitly
conflicts with gauge invariance. Physically, the problem is quite
evident: it is impossible in the non-Abelian theory with
self-interacting gluons to separate the virtual gluons from the static
Coulomb field of the heavy quark, and the latter is
Lorentz-noninvariant.

Thus, there is no alternative definition of the kinetic operator in QCD
beyond the first loop which would be free of ``Lorentz-noninvariance''
of the physical scheme of Ref.~\cite{optical}.

An analysis of the two-loop renormalization of the kinetic operator was
carried out recently by Neubert \cite{67rev} in a scheme which was
claimed to be free of these problems, in the contradiction with the
no-go arguments given above. A closer look reveals, however, that the
analysis is wrong in the basic aspects.

According to the expectations, the two-loop diagrams with the
non-Abelian gluon interaction did not vanish for any symmetry reason
and led to the divergent Feynman integrals which had to be regularized.
To assign them a definite meaning the author used a dispersion integral
with the cutoff over the {\em energy} of intermediate states. It was
failed to realize that it was exactly the same regularization procedure
that is consistently used in the physical definition criticized by the
author, and thus should have been considered not invariant to the same
extent. The claims of its ``Lorentz-invariance'' were wrong.

More to the point, contrary to the assertions of the paper the results
{\rm are} gauge-dependent. It was forgotten that the integrated
discontinuities are gauge-invariant only if the external color
particles are on-shell and only if all diagrams were incorporated.
Both these requirements were not respected in the ``regularization'' of
Ref.~\cite{67rev}. In particular, the ``vanishing'' diagrams discarded
for symmetry reasons -- if regulated in the same way -- would yield
nonzero contributions.

As a result, the procedure used in Ref.~\cite{67rev} cannot even be
considered as a consistent regularization scheme. Leaving aside the
gauge-dependence, one cannot use different schemes for the order-$\as$
and order-$\as^2$ corrections or apply different cutoff rules for
different groups of diagrams in the single order. Moreover, it must be
realized that the regularization of the part of non-Abelian two-loop
diagrams was performed in the same dispersion method as suggested in
\cite{optical}, which would lead to non-trivial renormalization of the
kinetic operator already to order $\as$. The whole analysis then makes
no sense and is irrelevant even if were made consistently.

Summarizing the ``puzzle'' of the perturbative renormalization of the
kinetic operator, the question of whether there exists any motivated
`invariant' non-Abelian regularization beyond the first loop, and if
yes then whether the operator is renormalized in this scheme, remains
open. The claims of answering it have no grounds. Personally, though, I
do not think it has any practical relevance, and am not sure in the
theoretical relevance either.

A completely defined regularization scheme was suggested and discussed
in \cite{optical,blmope,five}; another possibility is using lattice
versions of QCD with heavy quarks. The latter is not
``Lorentz-invariant'' in the discussed sense either. Moreover, it
manifestly violates the rotational invariance. The
cutoff over the gluon momenta in the physical scheme we adopt does not
look in high orders as simple as in the first loop. Moreover, it
necessarily has a nontrivial form for particular graphs to be
gauge-invariant.  The renormalized operator itself is
manifestly gauge-invariant since can be written in terms of the
observable transition probabilities.  Finally, the technical
feasibility of such a renormalization scheme was demonstrated in
\cite{xi,dipole} where the complete two-loop renormalization of the
kinetic operator was computed.

\subsubsection{Regularization and renormalization}

Continuing a more theoretical part of my lectures, I feel necessary to
discuss the difference between regularization and renormalization. The
relevance of this question is not confined by the heavy quark theory
and is more general. The confusion among these two notions is
particularly apparent in heavy quarks, and often is the reason behind
contradicting statements, especially regarding the role of
the dimensional regularization.

In a few words, the regularization is used as a technical tool to
make all intermediate results of the calculations finite when a finite
final answer is split, in the process of calculations, into simpler
pieces which separately are divergent. Therefore,  any consistent
regularization is possible and nothing depends on it. Renormalization
is the actual eliminating of actual divergences constructing an
effective theory with a real ultraviolet cutoff. All effective
couplings and operators do depend on the renormalization procedure.
I'll try to explain it here in more detail.

QED, QCD and the SM in general are renormalizable theories (strictly
speaking, they are perturbatively renormalizable). It means that, fixing
a number of observables corresponding to the bare parameters of the
Lagrangian we can, in principle, determine other observables. For
example, in QED we fix $m_e$, $\al_{\rm em}(0)$ defined {\it via} the
static Coulomb potential at microscopic distances, {\it etc}. All
interesting cross sections and probabilities can be expressed in their
terms. In renormalizable theories all such relations are eventually
finite.

Technically, however, in perturbation theory we usually proceed
indirectly. Namely, the observables are calculated {\it via} the
(bare) parameters of the Lagrangian, and then the quantum corrections
associated with loop effects are typically UV divergent. This reflects
the fact that the complete definition of the theory  requires
introducing an ultraviolet cutoff, and the bare couplings depend on
this cutoff in a divergent way when $\Lambda_{\rm UV} \ra \infty$.

At the same time, the relations between the physical input parameters
characterizing the theory and the bare constants are divergent as well.
Renormalizability means that the divergencies here are exactly the same
and cancel when the two steps are performed together. The presence of
the ultraviolet cutoff in any order of perturbation theory affects the
final result only {\it via} inverse powers of $\Lambda_{\rm UV}$, so
that the dependence on it disappears when  $\Lambda_{\rm UV}\ra
\infty$.

Subtracting or dividing infinities is not a correct mathematical
procedure {\it per se} and, if not done accurately, may lead to an
arbitrary result. The standard way is to {\em regularize} the involved
Feynman integrals to make them finite -- though dependent on the UV
regulator -- at any stage. The regularization involves certain
modifications of the theory -- for example, introducing extra heavy
particles in the Pauli-Villars scheme. As long as it is ensured that
the final physical relations do not depend on them, such modifications
are legitimate. The regularized expressions can be safely manipulated
since they are well-defined numbers. The final relations appear to be
finite and regularization-independent, all infinities must cancel. The
regularization can be removed at the final stage.

The regularization is thus a purely technical element inherent to the
standard way to treat the perturbative integrals. It is related to
splitting the finite, convergent expressions into separately divergent
pieces which are merely easier to calculate. All regularization schemes
are equally acceptable since the result does not depend on them.
It must be stressed, however, that we speak here about the observables.
Similar regularization procedure is often applied also to infrared
problems when an observable is split into parts which separately can be
IR divergent.

One of the most technically convenient regularization
schemes is dimensional regularization (DR) \cite{DR}. Its idea is to
calculate Feynman diagrams in a theory formulated in arbitrary
dimension $D$ where the integrals are well-defined. It has an
additional advantage of preserving gauge-invariance at any value of the
regularization parameter $\epsilon = 4-D$.

Quite different problems must be solved when one needs to construct an
effective theory integrating out the UV degrees of freedom above a
certain scale $\mu$. This procedure really modifies the theory at the
scale $\mu$ and above. For example, the spectrum of a lattice gauge
theory, while describing actual hadrons at small energies, has little in
common with the perturbative QCD at energies $E\gsim 1/a$. All composite
operators in the effective theory are well-defined and do not suffer
from divergences. They are {\em renormalized}. Even the literal
perturbative expansion does not require (UV) regularization in the
effective theory at any step. The renormalization depends, on the
contrary, on the way the cutoff is introduced. The bare couplings, {\it
e.g.} $\as^{(0)}$ are concrete finite numbers in the effective
theory, depending on $\mu$ .

Since all objects of the effective theory are finite, it can be used
also to regularize the perturbative calculations. Although the result
explicitly depends on $\mu$, for the renormalizable bare Lagrangians all
the dependence resides in the terms inversely proportional to powers of
$\mu$ and disappears at $\mu\ra \infty$. Moreover, it is often
convenient to use $\as(\mu)$, $m_q(\mu)$ {\it etc}., in particular in
QCD as an intermediate parameters.

However, can one really use the standard regularization techniques in
the opposite direction, namely, to construct an effective theory and
define renormalized operators? The answer depends on the scheme and, in
general, is ``No''. The details of regularization are crucial here,
contrary to regularization of purely perturbative calculations. In
particular, I'll demonstrate that the most convenient DR is, as a rule,
unsuitable for constructing renormalized operators.

First of all, the main application of the effective theories (say, for
heavy quarks) is nonperturbative effects. While the DR perfectly handles
regularization of Feynman integrals generated by the perturbative
expansion of
the formal QCD functional integrals in arbitrary (complex) dimension, it
is so far impossible to formulate QCD as a complete quantum field
theory including nonperturbative effects in non-integer dimensions. Even
if it were done, the physical point $D=4$ would have been a point of
essential singularity
and had extremely complicated analytic structure
in $\epsilon$ reflecting various phases of the theory. It would not be
possible to define extrapolation to $D=4$ subtracting only a few simple
rational functions of $D$.

Thus, it is clear that the DR as a purely perturbative construction
helps nothing in defining the nonperturbative operators in an effective
theory -- a task which, say, the lattice regularization does
successfully. One can try to limit consideration to only the perturbative
expansion.  There the DR proved to be very useful in many
aspects like applications of renormalization group. However, even
perturbatively the DR typically fails to perform the renormalization of
operators in the way necessary for the Wilson OPE. The only exception
are operators which are not subject to (powerlike) mixing with
lower-dimension operators and have only logarithmic anomalous
dimensions. An example of such operator is the chromomagnetic operator
$O_G$. Their perturbative normalization is given by the (Euclidean)
integrals of the type
\beq
J\;=\; \int\; \frac{{\rm d}^4k}{(2\pi)^4}
\:g_s^2\, \frac{1}{k^4}\; =\; \int\; \frac{k^2{\rm d}k^2}{16\pi^2} \:
g_s^2\, \frac{1}{k^4}\;.
\label{180}
\eeq
The denominators are in fact somewhat
different in the infrared since they include external particle momenta
$p_{\rm ext}$, and the integrals can be IR convergent. However, they
are logarithmically divergent at $k^2\ra \infty$. The DR, roughly
speaking, multiplies the integrand by
$\left(k^2/\mu^2\right)^{\frac{D-4}{2}}$:
\beq
J_D(\mu)\;=\;
\int\; \frac{{\rm d}^D k}{(2\pi)^D}\: \mu^{4-D}\,g_s^2\:
\frac{1}{k^4}\; = \; \int\;
\frac{k^{D-2}{\rm d}k^2}{(4\pi)^{D/2}\Gamma(D/2)} \:\mu^{4-D}
\,g_s^2\: \frac{1}{k^4}\;.
\label{182}
\eeq
At ${\rm Re}\, D<4$ the integral becomes convergent but contains the
pole $1/(D-4)$. The DR in the perturbative calculations suggests merely
to subtract from $J_D(\mu)$ this pole. The minimal subtraction scheme
of the DR, for example, subtracts the combination
$1/(D-4)+\frac{\gamma}{2}-\frac{1}{2}\ln{4\pi}$ containing the
Euler-Mascheroni constant:
\beq J_D^R(\mu)\;=\; J_D(\mu) - \frac{g_s^2}{16\pi^2}
\left( \frac{2}{D-4} +\gamma -\ln{4\pi}\right)\;.
\label{183}
\eeq
In the complete perturbative calculations this UV pole must finally cancel
in observables, and {\it ad hoc} removing it when passing to
$J_D^R(\mu)$ does not change the final answer at $D\ra 4$.  In the
renormalization procedure, on the contrary, we want to define such
integrals themselves. The quantity $\bar J^R(\mu)=\lim_{D\ra 4}
J_D^R(\mu)$  is finite and is the renormalized integral in the
$\overline{\rm MS}$ scheme.

Does $\bar J^R(\mu)$ has a more transparent meaning? Yes, it is easy to
see that it approximately coincides with the original integral
Eq.~(\ref{180}) if the integration is cut off in the UV at $k^2\sim
\mu^2$:
\beq
\bar J^R(\mu)\;=\;
\int^{k^2<\tilde\mu^2}
\; \frac{{\rm d}^4k}{(2\pi)^4} \:g_s^2\,
\frac{1}{k^4}\qquad
\mbox{ with }\; \tilde\mu^2 = {\rm e} \cdot \mu^2\;\gg \;p_{\rm ext}^2
\label{185}
\eeq
(the relation between $\mu$ and $\tilde\mu$ is not universal and
depends also on the order of perturbation theory). Thus, for logarithmic
integrals the DR effectively works as an UV cutoff in the effective
theory, at least in the low orders.

Unfortunately, in all other cases the DR fails as a renormalization
procedure. It can be seen on the example of Feynman integrals
encountered in renormalization of the kinetic operator or mass of the
nonrelativistic particle, the QCD operator $G_{\mu\nu}^2$ {\it etc}.
which are power-divergent.  Already in one loop simple power
integrals of the types
\beq
J^{(2)}\;=\; \int\; \frac{{\rm d}^4k}{(2\pi)^4} \:g_s^2\,
\frac{1}{k^2}\;,
\qquad \int\; \frac{{\rm d}^4k}{16\pi^2} \:  g_s^2\;\,,
\label{187}
\eeq {\it etc.} appear. They are not regularized
by an infinitesimal shift of the dimension. The expressions
can be formally
defined only at complex $D$ far away $D=4$, at ${\rm Re}\,D=2$,
${\rm Re}\, D=0$ and so on (depending on the dimension of the operator),
where they actually vanish. The only conceivable way to
define such integrals as analytic functions of $D$ is to have all them
zero.  This can be anticipated {\it a priori}:  these
integrals must have a positive dimension of mass which can be provided
only by $\mu$ accompanying the coupling $g_s$ at $D\ne 4$.  However,
$D=4$ is not a singular point for such integrals.

A direct way to see this fact is to analyze how these `renormalized'
integrals would enter calculations of observables. They describe the IR
part of the complete Feynman integrals in the domain $k\ll m_Q$. Since
the latter are safely IR convergent, regularization merely does not
change the IR part and, therefore, includes it completely. No room
remains for the additional contribution of such operators in
perturbation theory.

As a result, except for the special case of the logarithmic
renormalization, the DR merely considers the perturbative integrals for
the operators zero irrespective of any details, and this corresponds to
setting the Wilsonian cutoff $\mu=0$. In reality in higher loops, for
example in the kinetic operator, the resulting integrals by themselves
are not sensibly renormalized at any $D$, even at ${\rm Re}\, D=2$.

To summarize this digression into the theoretical subtleties, one
should clearly distinguish regularization and renormalization which
mean quite different things. Renormalization introduces nonvanishing
corrections even in the finite quantities or theories, where
regularization is not necessary.  While the DR is a consistent and one
of the most technically convenient regularization schemes, it is
unsuitable for renormalization of composite operators in the
effective theories.

\subsection{$|V_{cb}|$ from the Zero Recoil Rate of
$B \ra D^* \ell \nu $ }

The concept of the heavy quark symmetry was very important for the
evolution of studies of heavy flavor hadrons. The fact of the fixed
normalization of the $B\ra D$ and $B\ra D^*$ formfactors at zero recoil
in the limit $m_{b,c}\ra \infty$ was of special significance in
applications since suggested the method
for determinations of $|V_{cb}|$ from the exclusive
$B\ra D^*$ semileptonic transition near zero recoil.
To this end
one measures the  differential
rate, extrapolates to the  point of zero recoil and
gets the quantity $|V_{cb}F_{D^*}(0)|$, where $F_{D^*}$ is the axial
$B \ra l \nu D^*$ formfactor.
Since the charm quark is only marginally heavy, it is very important to
estimate the corrections, in particlar, nonperturbative. The exclusive
transition amplitudes are not genuinely short-distance, such transitions
proceed in time intervals $\sim 1/\Lam$. Nevertheless, it turns out
that the large-distance effects appear in these kinematics only
suppressed by $1/m_{c,b}^2 $:
\beq
F_{D^*}(0) = 1 + {\cal O}\left(\frac{\as}{\pi}\right) +
\delta^A_{1/m^2} + \delta^A_{1/m^3}\,+\; ...
\label{8.10}
\eeq

The absence of $1/m_Q$ corrections was first noted
in Ref.~\cite{vshqs}.\footnote{The first preprint version of the
paper \cite{shifheid}
discussing the heavy quark symmetry in QCD was given to me by
M.~Shifman in July 1986.} As far as I understand, the reasoning used by
Shifman and Voloshin  was very simple: let us consider, for example,
the vector $B\ra D$ transition at zero recoil:
\beq
\matel{D}{\bar c
\gamma_0 b}{B}\;=\; \left(M_B+M_D\right) \left\{1+\frac{a}{m_c}-
\frac{a}{m_b}+\,...\right\}
\label{190}
\eeq (short-distance effects
are neglected). The relative magnitude of $1/m_b$ and $1/m_c$ terms is
fixed since at $m_b=m_c$ all corrections must vanish identically.
$T$-invariance, however, says that the coefficients for $1/m_c$ and
$1/m_b$ terms must be equal since $B$ differs from $D$ by only the
value of the heavy quark mass. Thus, both terms must vanish.  This
observation was later studied in more detail in \cite{luke} and is
often called Luke's theorem. It essentially improves the credibility of
this method in spite of certain experimental difficulties.

The task of precision determination of $|V_{cb}|$ from the
exclusive transition requires a
detailed dynamic analysis of various
preasymptotic corrections in Eq.~(\ref{8.10}).
The perturbative part, albeit  technically complicated, is  at least
conceptually transparent.  The theory of power
corrections is more challenging.

The need in evaluation of the $1/m_Q^2$ corrections in Eq.
(\ref{8.10}) for practical purposes was realized quite early
\cite{fn}. In these days the theory of the power corrections
in heavy quarks was immature, our knowledge was scarce, so that it
was hard to decide even the sign of $\delta^A_{1/m^2}$.
The existed opinion stated that the deviations from the symmetry
limit must be very small, and suggestions that they could be as large
as $10\%$ were categorically refuted \cite{fn}.
The situation as it existed by 1994, before the OPE-based heavy
quark expansion was applied to the problem in \cite{vcb,optical} was
summarized in the review lectures \cite{neubtasi}:
\beq
\eta_A=0.986\pm 0.006 \qquad \qquad
\delta^A_{1/m^2} =(-2\pm 1)\% \qquad
\label{c19}
\eeq
($\eta_A$ is a purely perturbative factor introduced for this
transition in HQET),
yielding $F_{D^*}\simeq 0.97$ with a very small error.  This
estimate was promoted to ``one of the most
important and, certainly, most precise predictions of HQET".  Nowadays
we believe that the actual corrections to the symmetry limit are
larger, and the central theoretical value lies rather closer to $0.9$
\cite{vcb,optical}. After heated debates for a few years the estimates
we suggested in April 1994 seem to be accepted in the literature.

Regarding the perturbative calculations {\em per se}, it was later
pointed out
\cite{comment} that the claimed improvement \cite{neubimp} of
the original one-loop estimate was incorrect, and the proper central
value is rather $\eta_A\approx 0.965 $; such a value was confirmed by
complete two-loop ${\cal O}(\as^2)$ computation \cite{czar}.

The existing estimates of the power nonperturbative corrections in
$F_{D^*}$ are based on the sum rules for heavy flavor transitions
\cite{vcb,optical}.  The validity of the sum rules had been questioned
from various perspectives, but now is commonly
accepted.\footnote{Neubert, for example, claimed that the sum rules
were wrong \cite{update,ns} while simultaneously using them. It does
not take long to see why those arguments are invalid.
Nevertheless, so far he has not retracted these claims.} The idea of the
application of the sum rules to $B\ra D^*$ was to consider the
``inclusive'' transition probability into all possible final states and
not limiting them to only $D^*$ we are really interested in, Fig.~1a.
Studying the zero-recoil transition we only fix the spacelike momentum
$\vec{q}$ carried away by the lepton pair, $\vec q=0$. Since such a sum
is an ``inclusive'' probability, it is a short-distance quantity and is
directly expandable in the OPE in powers of $1/m_{c,b}$; the
relation is established applying the usual tools -- analyticity and
unitarity. The idea and technology of the heavy quark sum rules were
recently reviewed in \cite{rev} and discussed in detail in  the
original paper \cite{optical}. Here I mainly dwell on the results.

Schematically, the zero-recoil sum rule for the axial $\bar c \vec
\gamma b$ current responsible for the $B\ra D^*$ transition has the
form \cite{vcb,optical}
\beq
|F_{D^*}|^2 +
\sum_{0<\epsilon _i < \mu}|F_i|^2
\; = \;
\xi_A(\mu) \;-\; \Delta^A_{1/m^2} \;-\;\Delta^A_{1/m^3}\;+
{\cal O}\left(\frac{1}{m_Q^4}\right) \,,
\label{8.16}
\eeq
where
\beq
\Delta^A_{1/m^2} = \frac{\mu_G^2(\mu)}{3m_c^2} +
\frac{\mu_\pi^2(\mu)-\mu_G^2(\mu)}{4}
\left(\frac{1}{m_c^2}+\frac{1}{m_b^2}+\frac{2}{3m_cm_b}
\right)\;.
\label{8.17}
\eeq
Here $F_i$ are the-axial current transition formfactors to excited charm
states $i$ with the mass $M_i=M_{D^*}+\epsilon_i$, and $\xi_A$ is a
short-distance renormalization factor, $\xi_A(\mu)\equiv
\eta_A^2(\mu)$.
Contributions from excitations with $\epsilon$ higher than $\mu$
are dual to perturbative contributions and get lumped into the
coefficient
$\xi _A(\mu )$ of the unit operator, the first term in the right-hand
side of Eq. (\ref{8.16}).

The role of $\mu$ is thus
two-fold:
in the left-hand side it acts as an ultraviolet cutoff in the effective
low-energy theory, and by the same token determines the
normalization
point for the local operators; simultaneously, it defines the infrared
cutoff in the Wilson coefficients.

The $1/m_Q^3$ corrections to the sum rule $\Delta^A_{1/m^3}$ are also
known \cite{rev}. The short-distance renormalization factor
$\xi _A(\mu )$ is calculated perturbatively.  To the first order it was
computed in \cite{optical}, all-order BLM resummation performed in
\cite{blmope}. The most technically involved part of genuine ${\cal
O}(\as^2)$ corrections to $\xi _A(\mu )$  was computed in
\cite{czar}, and the complete result given in \cite{xi}. The
overall short-distance normalization appears to be very small (this is
the result of certain numerical cancellations), $\eta_A(\mu)=\sqrt{\xi
_A(\mu)}\approx 0.99$ for $\mu\simeq 0.6\GeV$ with the uncertainty at a
percent level. The Wilson coefficient for the kinetic operator was also
computed to the next-to-leading order in \cite{xi}; the correction to
it is small.

The sum rule Eq.~(\ref{8.16}) leads to the upper bound:
$$
|F_{D^*}|^2 \simeq \xi_A(\mu)- \Delta^A_{1/m^2} -
\, \sum _{0<\epsilon _i < \mu}|F_i|^2 \;,
$$
\beq
-\delta^A_{1/m^2}\;>\; \frac{1}{2}\Delta^A_{1/m^2} \,\ge\,
\frac{M_{B^*}^2-M_B^2}{8m_c^2} \;\simeq\; 0.035  \;\,.
\label{200}
\eeq
The last relation is a model-independent lower bound for the $1/m^2$
corrections to $F_{D^*}$ at zero recoil \cite{vcb}.

I pause here to make the following remark. The sum rule
Eq.~(\ref{8.16}) is the most obvious way to see the absence of $1/m^2$
IR corrections at zero recoil. The sum of the transition probabilities
has only the calculated $1/m_{c,b}^2$ corrections; no $1/m_Q$
corrections can appear merely because there is no nontrivial local
operator with $D=4$. The excited transition amplitudes are due to the
mass-dependent terms in the Hamiltonian, so they start with $1/m_Q$;
the probabilities are then only $1/m_Q^2$. A slightly different
(fixed $q^2$, that is, a SV rather than zero-recoil) version of this
argument was implied by Voloshin and Shifman in mid-80's when noting
the absence of $1/m_Q$ corrections.  In is curious that the
fact itself of calculability of this inclusive probability was
not discussed in this context, however. A formal QCD derivation not
appealing directly to an effective heavy quark Hamiltonian or heavy
quark symmetry was given in \cite{optical}.

Such a way based on the sum rules is advantageous since clarifies
the subtlety appearing in the presence of QCD-inherent radiative
corrections: the masses used here must be {\em short-distance}, not
pole masses. This differs from the HQET approach. The perturbative
$\mu$-independent factor $\eta_A$ as it is defined in HQET
has an infrared ${\cal O}(\Lam/m_c)$ part {\it via} the dependence on
masses. Likewise the difference $F_D^*-\eta_A$ (or analogous ratio)
does have ${\cal O}(\as/m_c)$ long-distance contribution merely since
$\eta_A$ has it.
The Luke's theorem is not valid as it is literally understood in
HQET.

One can consider the similar sum rules for $\vec{q}\ne 0$, or for other
weak currents. For example, for
the transitions driven by the hypothetical zero-recoil pseudoscalar
current $J_5= \int \,d^3 x\, (\bar c i\gamma_5 b)(x)$
one  obtains the sum rule
\beq
\sum_{\tilde \epsilon_k<\mu} |\tilde F_k|^2
\; = \;
\left( 1/2m_c-1/2m_b\right) ^2
\left( \mu_\pi^2(\mu) -\mu_G^2(\mu)\right) \;>0\;.
\label{8.15}
\eeq
This is the field-theoretic analogue of the difference of the sum rules
(\ref{4.3}) and (\ref{4.4}) and of the inequality discussed earlier.

Returning to $F_{D^*}$, if we want to obtain an actual estimate of the
nonperturbative corrections rather than a bound, we need to know
something about the contribution of the excited states in the sum rule
Eq.~(\ref{8.16}). Unfortunately,
no model-independent answer to this question exists
at present.
The best we can do today is to
assume that the
sum over the excited states
is a fraction $\chi$ of the local term
given by $\mu_\pi^2$ and $\mu_G^2$,
\beq
\sum _{\epsilon _i < \mu}|F_i|^2 \; = \; \chi\:
\Delta^A_{1/m^2}\, ,
\label{z8}
\eeq
where  on general grounds $\chi \sim 1$.
The contribution of the continuum $D\pi$ state can be calculated
\cite{vcb}, however theoretically it is expected to constitute only a
small fraction of the sum over resonant states.
Trying to be optimistic, we rather arbitrarily limit $\chi$ by
unity on the upper side; the larger is $\chi$, the smaller is
$F_{D^*}$.
In this way we arrive at
\beq F_{D^*}\;\simeq \; \eta_A(\mu)\;-\;
(1+\chi)\,\left[\frac{\mu_G^2}{6m_c^2} + \frac{\mu_\pi^2-\mu_G^2}{8}
\left(\frac{1}{m_c^2}+\frac{1}{m_b^2}+\frac{2}{3m_cm_b}\right)
\right] -
\delta^A_{1/m^3}
\, .
\label{z9}
\eeq

Assembling all pieces together we get for $\chi=0.5\pm 0.5$
\beq
F_{D^*}\;\simeq \; 0.91 \;-\;0.013\,
\frac{\mu_\pi^2-0.5\GeV^2}{0.1\GeV^2}\;\pm\;
0.02_{\rm excit}\;\pm\;0.01_{\rm pert}\;\pm\;0.025_{1/m^3}\;\;.
\label{z11}
\eeq
Estimates of the uncertainties in the ${\cal O}(1/m_Q^3)$
corrections and the contributions from the higher excitations
are not very firm and are rather on an optimistic side; they can be
larger. With $1/m_Q^2$ corrections amounting to $\sim 8\%$
noticeably smaller $1/m^3$ effects can be  only a result of accidental
cancellations. Their more elaborated estimates were done in
\cite{blmope}.
Altogether we get
\beq
F_{D^*}\;\simeq \; 0.91 \; \pm 0.06\;,
\label{z31}
\eeq
where the  optimistic uncertainty $\pm 0.1 \GeV^2$ is ascribed to
$\mu_\pi^2$.

Eq.~(\ref{z11}) basically coincides with the estimates given in the
original paper \cite{vcb}.
The QCD-based  analysis definitely favors a significantly larger
deviation of $F_{D^*}(0)$ from unity than those  {\em en vogue} three
or four years ago. Trying to understand the reason behind the
difference, I carefully studied the earlier analyses claiming smaller
effects. They all ascend to the paper \cite{fn};
its subsequent usage was somewhat too
liberal, I would say.
The analysis of
\cite{fn} was not actually a kind of a reasonably accurate evaluation
but rather a general discussion of the scale of possible effects not
pretending on high precision. The effects of the relativistic
spin-orbital interaction in light cloud were practically ignored since
they did not fit the oversimplified quark model used for the estimates.
I think that this was even more essential than the fact itself that the
$1/m_Q$ expansion was performed incorrectly at order $1/m_Q^2$ in
\cite{fn}. It was only in the successive reviews by Neubert where a
tentative evaluation of that paper was gradually promoted to the status
of the accurate prediction, without new relevant input.

\subsubsection{Quantum-mechanical interpretation}

In QM the inequality $\mu_\pi^2>\mu_G^2$ can be illustrated in
different ways. Formally, it expresses the positivity of the Pauli
Hamiltonian \cite{volpp}:
$$
\frac{1}{2m}(\vec\sigma\,i\vec D\,)^2 =
\frac{1}{2m}\left((i\vec D)^{\,2}-\frac{i}{2}\sigma G\right)\;.
$$
More physically, it shows the Landau
precession of a colored, i.e. ``charged" particle in the
(chromo)magnetic field. Let me recall that the particle Hamiltonian in
the magnetic field looks formally like for free particle, $\vec{{\cal
P}^{\,2}}/2m$, with the only difference that the covariant momenta
$\vec{{\cal P}}$ do not commute in the presence of the magnetic field.
This `uncertainty relation' leads to the precession.
Hence,  one has $\aver{p^2}\ge |\vec B|$. Literally
speaking, in the $B^*$ meson the
quantum-mechanical
expectation value of the chromomagnetic  field is suppressed,
$\aver{B_z}=-\mu_G^2/3$. It
completely vanishes in the $B$ meson. However, the essentially
non-classical nature of the `commutator' $\vec{B}$
(e.g.  $\aver{\vec{B}^{\,2}} \ge 3
\aver{\vec{B}}^{2}$), in turn, enhances the bound  which then takes
the same form as in the external classical field.

The basic sum rule Eq.~(\ref{8.16}) also has a transparent
interpretation revealed in \cite{optical}.
From the gluon point of view the  semileptonic
decay of the $b$ quark is an  instantaneous replacement of $b$ by
$c$ quark.
In ordinary QM the overall probability of the produced state to
hadronize to some final state is exactly unity,
which is the first, leading term in the r.h.s. of (\ref{8.16}).
Why then are there any
nonperturbative corrections in the sum rule? The answer is that
the `normalization' of the weak current $\bar c \gamma_\mu \gamma_5 b$
is not exactly unity and depends, in particular, on the external gluon
field. This appears as presence of local higher-dimension
operators in the current.  Indeed, expressing the QCD current in terms
of the nonrelativistic fields used in QM one has
$$
\bar c \gamma_k \gamma_5 b \leftrightarrow \varphi_c^+\left\{
\sigma_k -
\left(\frac{1}{8m_c^2}(\vec\sigma
i\vec D)^2\sigma_k+\frac{1}{8m_b^2}\sigma_k(\vec\sigma
i\vec D)^2\;-
\right.\right.
$$
\beq
\left.\left.
\frac{1}{4m_cm_b}(\vec\sigma i\vec D )\sigma_k(\vec\sigma i\vec D)
\right) +
{\cal O}\left(\frac{1}{m^3}\right)
\right\} \varphi_b\;\;.
\label{8.26}
\eeq
The weak current $\bar c \gamma_5\gamma_k b$, according to
Eq.~(\ref{8.26}) converts the initial wavefunction $\Psi_b$ into
$\tilde\Psi$:
$$
\Psi_B \;\stackrel{\bar c \gamma_5\gamma_k b}{\longrightarrow} \;
\tilde\Psi = \sigma_k \Psi_B -
\left(\frac{1}{8m_c^2}(\vec\sigma
i\vec D)^2\sigma_k+\frac{1}{8m_b^2}\sigma_k(\vec\sigma
i\vec D)^2\;-
\right.
$$
\beq
\left.
\frac{1}{4m_cm_b}(\vec\sigma i\vec D )\sigma_k(\vec\sigma i\vec D)
 + ...\right) \Psi_B\;.
\label{193}
\eeq
Then it is easy to calculate the normalization of $\tilde\Psi\,$:
\beq
\left\|\tilde\Psi\right\|^2 = \left\|\Psi_B\right\|^2 -
\frac{\mu_G^2}{3m_c^2} -
\frac{\mu_\pi^2-\mu_G^2}{4}
\left(\frac{1}{m_c^2}+\frac{1}{m_b^2}+\frac{2}{3m_cm_b}
\right)\:-\:...\;, \qquad \left\|\Psi_B\right\|^2=1\;.
\label{194}
\eeq
The additional terms are just the nonperturbative
correction in the right-hand side of the sum rule.

The similar relation holds for the zero-recoil vector current
$\bar c \gamma_0 b\,$:
\beq
\bar c \gamma_0  b \leftrightarrow \varphi_c^+\left\{
1 -
\frac{1}{8} \left(\frac{1}{m_c^2}+ \frac{1}{m_c^2} -\frac{2}{m_c m_b}
\right) (\vec\sigma i\vec D)^2 +
{\cal O}\left(\frac{1}{m^3}\right)
\right\} \varphi_b
\label{8.26a}
\eeq
and, therefore, the normalization of the resulting wavefunction in
these transitions is
\beq
\left\|\int {\rm d}^3 x \bar c \gamma_0  b(x) \:  \Psi_B \right\|^2 =
\left\|\Psi_B\right\|^2 -
\frac{\mu_\pi^2-\mu_G^2}{4}
\left(\frac{1}{m_c^2}+\frac{1}{m_b^2} -\frac{2}{m_c m_b}
\right)\;-\; {\cal O}\left(\frac{1}{m_Q^3}\right)
\;.
\label{194a}
\eeq

The first two $1/m_Q^2$ terms in the bracket in Eq.~(\ref{8.26}) or
terms $1/m_c^2$, $1/m_b^2$ in Eq.~(\ref{8.26a}) are the result of the
Foldy-Wouthuysen transformation Eq.~(\ref{28}).  Correspondingly, they
were missed in the standard HQET analysis used, for example, in
\cite{fn}. The dominant effect $1/m_c^2$ was thus basically lost. It
would be correctly reproduced in the approach of the Mainz group
\cite{korner}; however, the issue  was not elaborated in enough
detail for practical applications. The above inconsistency was
first noted by A.~Le~Yaouanc in 1994 \cite{leya}. It should be noted
that the Neubert's current numbers for $\delta^A_{1/m^2}$
and $F_{D^*}$ ascending to \cite{update} and
quoted in later reviews, and which are routinely cited by
experimentalists, were derived in \cite{update} using the same
incorrect expansion. The mistake has not been corrected yet. It was the
origin of the claim of Ref.~\cite{update} of insensitivity of the
analysis to the value of $\mu_\pi^2$ which clearly contradicts
Eq.~(\ref{z11}).

The QM interpretation of the sum rules explains why the perturbative
factor $\sqrt{\xi_A(\mu)}$ is the short-distance renormalization
translating the full QCD axial current into the current in the
effective theory.  Often the $\mu$-independent $\eta_A$ is called a
short-distance factor.  It is not quite correct, $\eta_A$ does not
depend on the choice of the separation scale $\mu$ determining which
effects are considered as short-distance and which are still included
into the low-scale physics. It would be possible only if the
perturbative corrections at arbitrary scale vanish, which is not the
case.  In reality $\eta_A$ is a mixture of short-distance and
long-distance effects.  Since, nevertheless, it is defined
perturbatively, it suffers from the IR renormalon uncertainties similar
to those discussed for the pole mass. They are too significant in
$\eta_A$.

\subsection{Summary on $|V_{cb}|$}

I would like to give here a brief summary of the two methods of
extracting $|V_{cb}|$. The most precise at the moment is the value
obtained from $\Gamma_{\rm sl}(B)$: Eqs.~(\ref{w12}) and
Eq.~(\ref{w20}) with the central experimental input values shown there
lead to
\beq
|V_{cb}|\;\simeq\; 0.0419
\cdot
\left(1-0.012\frac{(\mu_\pi^2-0.5\GeV^2)}{0.1\,\rm
GeV^2}\right)\;.
\label{210}
\eeq
The overall relative theoretical uncertainty in this result is
$\delta_{\rm th} \lsim 4\%$. With future refinements implementing
already elaborated strategy, we can expect reducing it down to $2\%$.
It is important that I am speaking here about the {\em defensive}
theoretical accuracy, the error bars that are assumed to cover the
whole interval where the exact number can be. I think that the
theoretical predictions -- to the extent they are not affected by
independent experimental input subject to additional
statistical uncertainties -- should have such a status. The way when the
previous predictions are praised for being ``only $2.5
\sigma$ away'' from {\it a posteriori} updated estimates, seems to be
inappropriate evaluation of the theoretical analysis.\footnote{The idea
that the theoretical predictions can be wrong -- outside the stated
intervals -- in almost a half of cases sounds somewhat humiliating.} To
allow for such a confidence, one should not combine different
theoretical uncertainties in quadrature as it is accustomed with
statistical fluctuations, and I rather added them linearly.

The $B\ra D^*\ell \nu$ zero-recoil rate also provides a good accuracy.
The exact experimental status of the measurements extrapolated to
$\vec{q}=0$ is not completely clear to me at the moment. I heard
different opinions, and leave the final word for the experimental
experts. Using the reported average value and the estimate
$F_{D^*}\simeq 0.9$ a somewhat lower value of $|V_{cb}| \simeq 0.038$
seems to emerge. It is interesting that the central theoretical value,
according to Eq.~(\ref{z11}) exhibits the dependence on $\mu_\pi^2$
similar in magnitude but opposite in sign to Eq.~(\ref{210}):
\beq
|V_{cb}|\;\simeq\; 0.038
\cdot
\left(1+0.014\frac{(\mu_\pi^2-0.5\GeV^2)}{0.1\,\rm GeV^2}\right)\;.
\label{211}
\eeq
The theoretical uncertainty $\delta_{\rm th}$  here constitutes
probably $\delta_{\rm th} \approx 6\%$, however this estimate relies on
additional mild theoretical assumptions. It is not clear how at the
moment it can be decreased. Speaking of feelings, personally I think
that $F_{D^*}$  larger than $0.95$ or below $0.85$ would be
unnatural. Even the end points of this interval look extreme. However,
the existing analyses do not allow to forbid them confidently, and the
feeling itself may be merely a result of a historically formed common
belief rather than based on scientific facts.

It is remarkable that
the values of $|V_{cb}|$ that
emerged from exploiting two theoretically complementary
approaches are
very  close. The progress was not for free: it became possible
only
due to essential refinements of the theoretical tools in the  last
several  years, which prompted us, in  particular, that the
zero-recoil $B\ra D^*$ formfactor $F_{D^*}$ is probably close to $0.9$,
significantly lower than previous expectations. The decrease in
$F_{D^*}$
and more accurate experimental data which became
available shortly after,
reduced the gap between the exclusive and inclusive
determinations of $|V_{cb}|$.

A few years ago such an accuracy and agreement would be considered as
an important success. Nowadays we are more demanding to the
heavy quark theory. The difference in the two values of $|V_{cb}|$ from
a certain perspective may look as a hint for the discrepancy:
the central value of
$|V_{cb}|$ from $B\ra D^*$ decay seems to be somewhat lower than
that from
$\Gamma_{\rm sl}(B)$. Since both theoretical values
depend to a certain extent on the precise magnitude of $\mu_\pi^2$,
it is tempting
to think that the actual
value of $\mu_\pi^2$ is somewhat larger than the ``canonical" 0.5
GeV$^2$. Increasing it
by about $0.2\GeV^2$  makes  the two results much closer. This would
not really contradict known facts. I hasten to add, though, that the
existing experimental error bars are such that speculations about
adjusting $\mu_\pi^2$ are premature. Moreover,   theoretical
uncertainties in the exclusive formfactor also preclude us from  the
above adjustment of $\mu_\pi^2\,$  ($F_{D^*}$ can well be, say, $0.87$
even at the canonical value $\mu_\pi^2=0.5 \GeV^2$).  The ways to
sharpen the theoretical predictions of $\Gamma_{\rm sl}$ are more
or less clear. Future accurate measurements will, hopefully, allow one
to directly measure -- through comparison with $\Gamma_{\rm sl}(B)$ --
the exclusive formfactor with accuracy better than that achieved by
today's theory.  Thus, we  will get  a new source of information on
intricacies of the strong  dynamics in a so far rather poorly known
regime.

Changing the parameters like $\mu_\pi^2$ {\it per se} within reasonable
intervals is not tightly limited at the moment. However, it may
well imply  other related changes -- the size of $1/m_c$ corrections
for formfactors at $\vec{v} \ne 0$, delaying onset of duality with
smaller values of $\mu_G^2(0.8\GeV)$, {\it etc}. Not all these links
are properly taken into account account in the analyses. The
comprehensive study is still to be done. One element is already
clear. In most of the existing theoretical analyses it is essentially
assumed that the approximate duality
between the  actual hadronic amplitudes and
the
quark-gluon ones sets in already at the excitation energies $\sim
0.7$ to $1 \GeV$. While there are no
experimental indications so far that this is not the case (at least, in
the semileptonic physics), the proof is not known either. If that is
not true, and duality starts only above $1\GeV$, most probably one
would have to abandon the idea of accurate determination of $|V_{cb}|$
from the exclusive $B\ra D^{(*)}$ transitions. The only  option still
open will be the inclusive semileptonic decays where the energy
release is large, $\sim 3.5\GeV$. Of course, in such a pessimistic
scenario the theoretical precision in $|V_{cb}|$ will hardly  be
better than $5\%$.

\subsubsection{Comments on the literature}

Since the question of the accurate determination of $V_{cb}$ is very
topical, there is a wide spectrum of opinions here. Speaking to the
experimental colleagues, it would not be fair to avoid mentioning
alternative suggestions. Over the last few years with the developing of
the genuinely QCD-based methods, general convergence of central values
in the estimates is clearly observed. The controversy mainly shifted to
the complicated issue of estimating theoretical uncertainties.

Those who are familiar with ongoing debates can notice that my
assignments for the theoretical uncertainties in $\Gamma_{\rm sl}$ are
on the lower side of the spectrum. For the $B\ra D^* \ell\nu$ zero
recoil rate they are closer to common estimates (with the
exception of Neubert who pretends to a significantly
more accurate prediction). The
smaller uncertainties in $\Gamma_{\rm sl}$ are anticipated {\it a
priori}. First, $\Gamma_{\rm sl}$ is a genuinely short-distance quantity
while $F_{D^*}$ is not, starting $1/m_c^2$ effects. Second, the
normalization of $F_{D^*}$ rests on the heavy quark symmetry and thus
suffers from downgrading the expansion parameter: it is $1/m_c$ rather
than $1/m_b$. The task of the QCD-based HQE was to make full use
of these advantages.

Let me first dwell on $B\ra D^* \ell\nu$. In general, the statements of
a good accuracy here have been merely transferred from the old reviews
\cite{neubtasi,neubpr} claiming three times smaller corrections than
today and, correspondingly, small error bars. While the central
value quoted by the author went, in a few iterations, down to the
current prediction for $F_{D^*}$, a similar due revision did not affect
the suggested error bars.  A closer look shows that they are not better
motivated than before.

The central value quoted by Neubert now does not differ at a
discussible level from the original estimates of the heavy flavor
sum rules \cite{vcb}. The analysis is based anyway on the very same
relations and assumptions. It is stated, in addition, that the paper
\cite{update} where the sum rule analysis was reproduced, has been
complimented by some symmetry relations; this was called later the
``hybrid'' approach by the author. In fact, the sum rules manifestly
respect the heavy quark symmetries, and the latter cannot provide any
further constraint. Careful studying of the analysis of
\cite{update} may seem complicated due to presence of numerous
parameters and notations
as well as contours on plots called upon to
make numerical conclusions looking convincingly. It can be done
nevertheless, and shows that the analysis of the $1/m^2$ terms merely
was not correct.  The flaws were briefly discussed in \cite{rev} and
were conceptually related to missing the contributions from the
Foldy-Wouthuysen transformation.

The basic points which led Neubert to the small uncertainties were the
following:\\
~$\bullet$ discarding $1/m_c^3$ and higher-order corrections
altogether\\
~$\bullet$ {\it ad hoc} fixing the value $\mu_\pi^2=0.4\GeV^2$
in the analysis\\
~$\bullet$ the claim of
insensitivity of the result \footnote{The insensitivity was illustrated
by varying $\mu_\pi^2$ in a very narrow interval from $0.36\GeV^2$ to
$0.5\GeV^2$. This allowed to bury the impact of changing $\mu_\pi^2$
under uncertainties in other model parameters.} to $\mu_\pi^2$ related
to missing the term $(\mu_\pi^2-\mu_G^2)/m_c^2$.\\ Curing these flaws
returns one to the estimates given in Sect.~3.4.

Another flaw of the estimates by Neubert is improper account for the
radiative corrections. The formfactor $F_{D^*}$ was judiciously split
into two parts. The first was ``purely perturbative'' factor
$\eta_A$, and all the rest was just called the nonperturbative
corrections:
\beq
\delta^A_{1/m^2} \,+\, \delta^A_{1/m^3} \,+\,...\;\equiv
\; F_{D^*}\,-\,\eta_A \;;
\label{210a}
\eeq
$\eta_A$ was defined
as a sum of the formal perturbative series calculated without any
infrared cutoff. Such a quantity cannot, however, be consistently
defined even if our knowledge of the perturbative coefficients were
practically unlimited. The associated numerical uncertainties are quite
significant. Even adopting an optimistic convention for the uncertainty
with the {\it ad hoc} factor $1/\pi$ in front of the imaginary part of
the Borel-resummed series (like in Eq.~(\ref{103a})), one gets for the
$1/m_Q^2$ and $1/m_Q^3$ irreducible uncertainties
\beq
\delta_{\rm IR}^{1/m^2}\;\simeq \;
\frac{2}{27} \, {\rm e}\, ^{5/3}\,
\left(\Lam^{\overline{\rm MS}}\right)^2 \left(\frac{1}{m_c}+
\frac{1}{m_b}\right)^2\; \simeq \; 0.022
\left(\frac{\Lam^{\overline{\rm MS}}}{250\MeV}\right)^2
\label{212}
\eeq
\beq
\delta_{\rm IR}^{1/m^3}\;\simeq \; \frac{{\rm e}\, ^{5/2}}{72}\,
\left(\Lam^{\overline{\rm MS}}\right)^3 \left(\frac{11}{m_c^3}
+\frac{5}{m_c^2 m_b} +\frac{5}{m_c m_b^2}
+ \frac{11}{m_b^3} \right)\;
\simeq \;
0.014 \left(\frac{\Lam^{\overline{\rm MS}}}{250\MeV}\right)^3
\;,
\label{213}
\eeq
respectively (higher-order IR renormalons are smaller). Thus, the
uncertainties in both ``perturbative'' and ``nonperturbative''
corrections defined in such a way somehow appeared smaller than their
irreducible intrinsic uncertainty.

To avoid confusion I must note that Eq.~(\ref{212}) was obtained in
\cite{ns}.
This, however, contradicted the claimed high accuracy of calculating
$\eta_A$ \cite{neubimp}. Hence, already in the next
papers the numerical estimate for this uncertainty was skillfully
reduced by a significant factor. It was achieved by simple two-step
manipulations: the value of $\delta_{\rm IR}^{1/m^2}$ was first
expressed via the square of the relative IR uncertainty in the mass,
Eq.~(\ref{103a}).  Then a rather low numerical uncertainty in the mass
was used.
In this way $\delta_{\rm IR}^{1/m^2}$ seemed to emerge only $0.006$, to
the author's satisfaction \cite{update,neubblm2}. Using the direct
formulae, however, one gets about 4 times larger uncertainty.  For this
reason the two expressions for $\delta_{\rm IR}^{1/m^2}$ in $\eta_A$
and $\delta_{\rm IR}$ in $m_Q^{\rm pole}$ were never quoted
together, and the uncertainty $\delta_{\rm IR}^{1/m^2}$ was always
numerically evaluated by Neubert in this ``indirect way'' {\it via}
more or less arbitrarily defined $\delta m_Q^{\rm pole}$ rather than
using the known explicit expression.

The practical inconsistency of the approach adopted in
\cite{update} is well known. Physically it leads to double counting of
the large-distance domain -- once as a part of the perturbative
effects, then in the nonperturbative corrections. Numerically it is
quite significant in the case of $F_{D^*}$. For example,
Ref.~\cite{lig} pointed out that the bounds used by Neubert for
$F_{D^*}-\eta_A$ are too strongly offset by the perturbative
corrections. In fact, the theoretical inconsistency of such an
approach were even noted by the author himself in \cite{update}, where
the results were stated conditionally valid provided the corrections
would not alter them significantly (according to \cite{lig}, they just
do). This qualification has never appeared later in the numerous
author's papers quoting \cite{update}.

In any case, it is easy to see that the bounds of the type used in
\cite{update} are
justified only if the perturbative factor $\xi_A^{1/2}(\mu)$ is
used instead of
$\eta_A$, see Eqs.~(\ref{8.16}, \ref{200}). This
short-distance factor is well calculated now \cite{xi}, and is
significantly larger than the value used by Neubert. It is not clear
whether the author plans to revise those inconsistent estimates, and
what will be the central value of the updated numbers. Taking
everything at face value, the central value for $F_{D^*}$, according to
Neubert, must have been close to $0.94$.\vspace*{.2cm}

Now, let us turn to $\Gamma_{\rm sl}$. In spite of the accurate
experimental data and a well-developed theoretical description,
determination of $V_{cb}$ in this way often unjustifiably
discarded being ascribed inflated
uncertainties. As I mentioned before, the origin was the misconcept of
the pole mass flourishing in HQET a few years ago. In particular, it
was tacitly assumed that the quark pole mass has an unambiguous value.
Then it was observed that:\\
\, (a) It is difficult to accurately extract
$m_b^{\rm pole}$ from experiment. In any particular calculation one can
identify effects left out, which can change its value by $\sim
200\MeV$. This uncertainty leads to a theoretical error $\delta_{\rm
I}\simeq 10\%$ in $\Gamma_{\rm sl}(B)$.\\
\, (b) When routinely
calculating $\Gamma_{\rm sl}(B)$ in terms of the pole masses, there are
significant higher order corrections $\delta_{\rm II} \simeq 10\%$.

Thus the conclusion was made:  $\Gamma_{\rm sl}(B)$ cannot be
calculated with accuracy better than
$\delta_{\rm I} + \delta_{\rm II} \sim 20\%$, and, correspondingly, at
best $\delta |V_{cb}|/|V_{cb}| \sim 10\%$ \cite{update}.

Both observations (a) and (b) above are correct, beyond any doubt.
If one more step is taken, however, the conclusion is invalidated --
one should take into account the fact
that the origin of these two uncertainties is actually the same, and
therefore they practically cancel each other.

Since the nature of the heavy quark in the QCD-based HQE was clarified
in \cite{pole}, the numerical relevance of these theoretical facts was
illustrated in a number of publications
\cite{upset,bbbsl,beauty,crad,five}. Nowadays only the above two
problems allegedly impair the precise calculation of $\Gamma_{\rm
sl}$: the danger of large perturbative
corrections and the uncertainty in $m_b$.
The anatomy of large corrections and their cancellation was described
recently, for example, in the review \cite{rev}. Here I would like
to give a complimentary illustration.

The HQE in QCD expresses the width in terms of the short-distance mass
$m_b(\mu)$ with $\mu\simeq 1\GeV$. Such a mass is well defined and has
a certain value. We do not know it precisely, of course. The HQE
assumes, nevertheless, that it can be extracted from
experiment with theoretically unlimited accuracy. Let us adopt here the
determination from the moments of ${\rm e^+ e^-}\ra b\bar{b}$ cross
section discussed in Sect.~3.2.3. Thus, we calculate the semileptonic
width directly via $V_{cb}$ and $\sigma_{{\rm e^+ e^-}}^{b\bar b}(s)$.
Reference
to the quark mass is made only for computational purposes. Let us also
neglect the small nonperturbative corrections at this point and fix the value of
$m_b-m_c$. Both $\Gamma_{\rm sl}$ as a function of $m_b(\mu)$ and the
mass $m_b(\mu)$ itself as expressed via
$\sigma_{{\rm e^+ e^-}}^{b\bar b}(s)$
undergo perturbative corrections. How important is their overall impact?

To see this, we can first neglect the perturbative corrections
altogether. It is important that it must be done at both stages of
extracting $V_{cb}$ simultaneously: it is not legitimate to keep ${\cal
O}(\as)$ correction in one part and discard it in another part of a
single relation. If we neglect them consistently,
we would get a certain value which we can denote
$|V_{cb}^{(0)}|$. It is somewhat smaller than quoted in
Eq.~(\ref{w12}). Already here the two competing effects are apparent:
we would use a smaller $m_b\simeq 4.60\GeV$ but, at the same time,
discard negative perturbative corrections to the width.

Now we include ${\cal O}(\as)$ corrections. Again, we must use the same
(not running in the first order) value of $\as$ at both stages. We get
a new value, $|V_{cb}^{(1)}|$. It appears that
\beq
   |V_{cb}^{(1)}| / |V_{cb}^{(1)}| \;\simeq\; 1.06\;.
\label{220}
\eeq
The effect of the leading perturbative corrections changes
$|V_{cb}|$ at a $5\%$ level only! This numerical result is easy to
interprete also in terms of the usual calculations expressed {\it via}
the pole masses: the first-order radiative effects shift
$m_b^{\rm pole}$ from $4.60\GeV$ up to $4.82\GeV$, but bring in the
perturbative suppression factor about $0.8$.

One can incorporate the BLM improvement, that is, account for running
of $\as$, again in strictly the same way in $\Gamma_{\rm sl}$ and
$\sigma_{{\rm e^+ e^-}}^{b\bar b}(s)$. The impact of, {\it e.g.,} the
second-order corrections changes $|V_{cb}|$ by less than a percent
\cite{upset}.  And this magnitude of the ${\cal
O}(\as^2)$ effects is quite expected
if the ${\cal O}(\as)$ correction was only a few
percent. The all-order BLM resummation does not change this conclusion.

The complete ${\cal O}(\as^2)$ analysis of extracting
$m_b(\mu)$ from $\sigma_{{\rm e^+ e^-}}^{b\bar b}(s)$ has not been done
yet, and we cannot literally extend this comparison to the non-BLM
effects at order $\as^2$. Nevertheless, it is difficult to imagine how
a significant effect exceeding a percent level can emerge from it.

The usually cited `naive' evaluation of the magnitude of the perturbative
corrections in the width differs from the above in one but quite
important aspect. It forgets that the pole mass $m_b$ routinely used
there has to be perturbatively calculated anew in every order, with the
numerical effect similar to Eq.~(\ref{numbers}). Instead it is treated
as a fixed value.  It is reasonable to ask -- which exactly? The answer
is not clear, since only the ``correction factor'' is
quoted.  And the corrections look significant indeed.
It only remains unclear to a
reader how with these huge higher-order correction the final number
quoted for $|V_{cb}|$ differs from what had been obtained without any
refinements \cite{vcb} by less than $1\%$, and even neglecting the QCD
corrections altogether yields a $5\%$ accuracy.  The reason is
explained above.

\subsubsection{Analyticity and unitarity constraints on the
formfactor}

Extracting $|V_{cb}|$ from the exclusive decays implies extrapolating
data to zero recoil. Near zero
recoil
statistics in the decays $B\ra D^* \ell\nu$ is
very limited, and the result for the
differential decay rate at this
point is  sensitive  to the
way one extrapolates the experimental data to
$\vec
q=0$. Most simply  this is
done through linear extrapolation.
Noticeable curvature of the
formfactor
would change  the experimental value for $|V_{cb}F_{D^*}(0)|$.
Under the circumstances it is natural to try to get independent
theoretical information on
the $q^2$-behavior of the amplitude.

Some time ago it was emphasized \cite{rafael} that
additional constraints on the $q^2$-behavior follow from
 analytic properties of the  $B\ra D^*$
formfactor considered as
a function of the momentum transfer $q^2$, combined with certain
unitarity
bounds. In particular, in the dispersion integral
\beq
F(q^2)\;=\; \frac{1}{\pi}\, \int \; ds \,\frac{\Im F(s)}{s-q^2}
\label{n2}
\eeq
the contribution of the physical $s$-channel domain
$q^2>(M_B+M_{D^{(*)}})^2$  is bound since $|F|^2$
describes the exclusive production of $B \bar D$ and cannot exceed
the
total $b \bar c$ cross section. The integral (\ref{n2}) receives
important contribution from the domain below the open $b\bar c$
threshold, where $F(q^2)$ has narrow pole-like
singularities corresponding to a few lowest $b\bar c$ bound states.
Introducing a set of unknown residues and making plausible
estimates of
the positions of these bound states, on the one hand, and adding a
small non-resonant subthreshold
contribution, on the other hand,
\res{lebed} suggested an {\em  ansatz }
for the
formfactor in the whole decay domain. Since the residues are
unknown
and  the momentum transfer in the actual decays
$B\ra
D^{(*)}\ell\nu $ varies in a rather narrow range,
the advantages of this parameterization  over the standard
polynomial fit are not clear at the moment.

A much more stringent
relation between
the slope and the curvature of the formfactor (i.e., $F'(q^2)$ and
$F^{\prime\prime}(q^2)$ at  zero recoil) was claimed in \cite{neubcap}.
Unfortunately, this relation was hastily  incorporated in some
experimental analyses.  It was shown in \cite{crad,rev} that the
analysis of Ref.~\cite{neubcap} was erroneous, and correcting the flaws
kills the predictivity of the relations discussed there. This
conclusion was later confirmed in \cite{glebslope}.

A different few-parameter ansatz for the shape of the effective $B\ra
D^*$ formfactor was also suggested in \cite{glebslope}. Although it
seems to be one of reasonable possible ansatse, its model-independence
was exaggerated. The authors relied too heavily on the approximate
heavy quark symmetry in the threshold region. The indirect form in
which these constraints were imposed hid this fact. Additionally, the
shape is sensitive to the exact structure of all the subthreshold
$b\bar{c}$ states, which were taken from potential models whose
reliability for the higher excited states is suspicious.  A
detailed consideration reveals that one should allow for larger
corrections to the parameterization suggested in \cite{glebslope}.

\subsubsection{On the $1/m_Q$ corrections to the widths and duality
violation}

The question of the asymptotic expansion of the
inclusive widths at the nonperturbative level was
not too obvious, as illustrated in Sect.~3.1.
Even though the HQE answered it, this question
is still occasionally raised in the literature. The suggestions of
possible $1/m_Q$ nonperturbative effects in different form still
flash every now and
then \cite{lms,gleb,jin,alt,flash2}. The experimental evidences
that ${\rm BR_{sl}}(B)$ and the lifetime $\tau_{\Lambda_b}$ seem to be
lower than suggested by theory additionally provoke such speculations.

There are a few reasons behind such a doubt. One of
them lies in purely technical aspects. The traditional approach of
HQET to the semileptonic widths \cite{cgg} is to calculate only the
differential semileptonic width $\Gamma_{\rm sl}(q^2)$ where the lepton
invariant mass $q^2$ is fixed. The total width is finally obtained
integrating $\Gamma_{\rm sl}(q^2)$:
\beq
\Gamma_{\rm sl}\;=\;
\int_0^{(M_B-M_D)^2} \; {\rm d}q^2\:  \Gamma_{\rm sl}(q^2) \label{222}
\eeq
(the case of $b\ra u$ is simply obtained by setting $M_D=m_c=0$). At
not too large $q^2$ this differential
width indeed can be easily calculated including the
leading nonperturbative corrections.
For example, up to $1/m^2$ effects it is
given by
$$
\Gamma_{\rm sl}(q^2) = \frac{G_F^2|V_{cb}|^2}{96\pi^3}
\left\{
\frac{(m_b^2-m_c^2)^2 + q^2 (m_b^2+m_c^2)- 2q^4}{m_b^3}
\left[(m_b^2-m_c^2-q^2)^2 - 4m_c^2 q^2
\right]^{\frac{1}{2}}
\right.
$$
\beq
\left.
+\; m_b^3\, {\cal O}\left(
\frac{\Lam}{\left(m_b-m_c-\sqrt{q^2}\right)}
\right)
\right\}
\;.
\label{223}
\eeq

However, at $q^2$ approaching $(m_b-m_c)^2$ the $1/m_Q$ expansion
breaks down and the differential width cannot be evaluated. In this
soft domain no literal $1/m_Q$ expansion is available.
For example, with both $b$ and $c$ quarks heavy enough, at maximal
$q^2$ the actual hadronic width differs from the quark width by the
universal factor
\beq
\frac{\Gamma_{\rm had}(q^2)}{\Gamma_{\rm quark}(q^2)}\;=\;
1+\frac{3}{2} \frac{\La (m_b-m_c)}{m_b m_c}
\label{224}
\eeq
which explicitly includes $1/m_c$ correction. Although decreasing $q^2$
somewhat diminishes the ratio, even after the $D^{**}$ threshold the
deviation is still of a similar order, {\it i.e.} scales like $1/m_Q$.

At the same time, the overall contribution to the semileptonic width
from this domain of large $q^2$ is {\it a priori} at the relative level
of $\Lam/(m_b-m_c)$, so it has to be calculated to address even $1/m_Q$
effects. Moreover, the upper limit of integration over $q^2$ is clearly
dictated by the hadron rather than quark masses; this is particularly
manifest for $b\ra u$ where $M_D \ra m_\pi\simeq 0$.  Naively one could
have expected the width to depend, at least partially, on the hadron
rather than quark masses.

This problem of the HQET approach underlies the reason why, for
example, the paper \cite{cgg} allowed for the possibility to have
$1/m_Q$ nonperturbative effects even in the semileptonic widths. (For
the nonleptonic widths such an approach is not applicable at all.) In
this language the QCD statement of vanishing such nonperturbative
effects is quite nontrivial: while the width $\Gamma_{\rm sl}(q^2)$
cannot be literally calculated in the above domain, the integral over
$q^2$
\beq
\int_{q^2_{\rm min}} ^{(M_B-M_D)^2} \; {\rm d}q^2\:
\Gamma_{\rm sl}(q^2)
\label{225}
\eeq
can be, provided $m_b-m_c-\sqrt{q^2_{\rm min}} \gg \Lam$. This fact
automatically emerges in the approach suggested in \cite{buv,bs,prl}.
To get it in the HQET-type approach required usage of the sum rules
for the heavy quark transitions \cite{optical}, and was
demonstrated recently in \cite{five}.

It is worth emphasizing that the QCD-specific sum rules guarantee that
the leading nonperturbative
effect affecting the phase space in the semileptonic decay
cancels against the corrections to the hadronic formfactors, the fact
which is difficult to ensure in {\it ad hoc} quark models. This
cancellation was apparently missed in \cite{jin}. As a matter of fact,
the cancellation persists even in the perturbative effects up to
momenta $\sim m_b/5$.

The paper \cite{lms} claimed that there is an independent $d=4$
operator in HQET capable to affect the semileptonic widths producing
$1/m_b$ nonperturbative
corrections. Such a statement is unacceptable. I think that the
confusion was merely a matter of ill-formulated expansion procedure.

A similar deficiency affects in my opinion the analysis
\cite{flssl}. The calculation of the width included a set of the
nonperturbative parameters among which there was $\La$ governing $1/m_b$
corrections explicitly introduced in the analyses. The inconsistency was
again rooted to relying on the hadron masses. The QCD states, on the
contrary, that for the $B$ meson width $\Gamma_{\rm sl}(B)$ the $B$
meson mass $M_B$ is no more relevant than the mass of $\Lambda_b$, or
$M_{B_s}$, or any their combination!

I cannot avoid mentioning that lobbying a possibility of $1/m_Q$
corrections got recently new impetus. Resurrecting the previous
misconceptions Grinstein {\it et al}. \cite{gleb} claimed establishing
the OPE-violating $1/m_Q$ terms in the inclusive widths on the example
of the `t~Hooft  model, the large-$N_c$ two-dimensional QCD. I cannot
dwell on details here, but can assure the reader that these results are
wrong.

To summarize, there must be no reason so far for admitting a
possibility of $1/m_Q$ corrections in the inclusive widths, even though
it may not be obvious in certain approaches. \hspace*{.1cm}

A more subtle problem is violations of local duality which can affect
the widths, {\it e.g.} semileptonic, at a certain level.
Reliable information on
this aspect of QCD is scarce. The only model specifically designed to
address the issue is presented in Ref.~\cite{inst}.
It predicts negligible deviations from duality in the semileptonic
decays of $B$ mesons.
Irrespectively of the model, they must be small. In this
respect I would disagree with Bj who tried to argue
in favor of naturalness of significant effects \cite{bjhaw}. The
arguments examined in more detail rather suggest
a small scale for the effects. Of course, it cannot be interpreted as a
rigorous proof.

The violation of duality must fast decrease with energy. It is worth to
recall that in a similar four-fermion hadronic decay width of $\tau$
lepton one only has $E=m_\tau\simeq 1.78\GeV$. Nevertheless, it is
often believed that the duality holds there at the level of $1$ or
$2\%$. (More realistically, the scale of the violation is around $5$ to
$7\%$.\footnote{Since the effects of duality violation must oscillate
as a function of $m_\tau$ \cite{inst}, at a certain mass they
nevertheless may vanish.}) In $b\ra c\,\ell\nu$ the energy release is
$3.5\GeV$. Moreover, there must be an additional suppression which can
be revealed by varying $m_c$ as a free parameter.

In $b\ra c\, \ell\nu$ the energy release is the smallest when $m_c$
increases and approaches $m_b$. In this limit (it is just the SV limit)
there is an exact duality at least up to $1/m_b^2$ terms, the fact
emphasized first by Shifman  and Voloshin in the mid 80's in one of the
first papers \cite{vshqs} established the base of the HQE. The effect
can be evaluated similar to $F_{D^{(*)}}$ and is below a percent level.

In another extreme limit $m_c\ra 0$ the heavy quark symmetry does not
apply, however the energy release approaches $5\GeV$. At such a scale
it also must be negligible. It is hardly possible that the duality
violation can have a strong peak at an intermediate $m_c$.
Comparing to $\tau$ leptons, we thus have an essentially larger energy
parameter and an important suppression of the duality-violating effects
by the heavy quark symmetry.
It is hardly reasonable in this situation to allow
a much stronger effect compared to `unprotected' hadronic $\tau$
decays.

It is interesting to discuss another argument by Bj who notes that in
$\rm e^+e^-$ annihilation one still has a significant resonance
structure up to $E=\sqrt{s}\approx 2.5\GeV$. In B decays the energy is
safely above this domain, but in general one has to be careful relying
on this experience. For the kinematics of hadronic transitions are
quite different. In $\rm e^+e^-$ we have a point-like production of
the initial quarks while in $B$ decay the final state quarks start
evolving at large separation $\sim \Lam$ after the decay. This
`multiperipheral' nature in the heavy quark decays determines the
difference in the nonperturbative
effects which start with $1/m_Q^2$ {\it vs.}
$1/E^4$ in the annihilation.

This peculiarity is reflected in the duality-violating effects as
well, in particular in $\tau$ decays. Considering only the vector
channel decay rate one experimentally observes larger, sizeable
violations of duality. They are strongly washed out as soon as the
axial channel is added when more low-energy resonances contribute. In
this respect $\Gamma_{\rm sl}(B)$ has a clear advantage: a variety of
final states with different angular momenta is much more rich here even
without producing extra quark pairs.

The scale of onset of duality is typically governed by the masses of the
first prominent resonances. In $\tau$ decays they are $M_\rho$ and
$m_\pi$, $M_{a_1}$ for $V$ and $A$, respectively.  Such scale is lower
in the semileptonic decays. For example, the $P$ wave excitations in $b\ra c$
have the mass gap $\lsim 500\MeV$. It is directly related to the richer
quantum number structure of the decay. Even radiatively excited states
are expected below $M_D+1\GeV$.

As a result, a closer look at the potential problems with duality
rather suggests that it should work better in $\Gamma_{\rm sl}(B)$ than
in $\Gamma_{\rm had}(\tau)$. I think that a more dedicated analysis is
to be done to reveal reasons -- if any -- why the former may be more
vulnerable. So far all arguments suggest that $\Gamma_{\rm sl}(B)$ is
safer.

\subsection{Lepton Spectra and Decay Distributions.\\
Model-insensitive Determination of $|V_{ub}|$ }

The inclusive differential distributions in semileptonic (or
similar radiatively-induced) decays are treated in the HQE similarly
to the inclusive widths \cite{dpf,prl,koyrakh}. An important fact
pointed out in \cite{prl} is that the QCD-based OPE automatically leads
to an analogue of nonrelativistic motion of the heavy quark inside the
hadron. This ``Fermi motion'' was phenomenologically introduced long
ago, first in \cite{ap} and then elaborated further to the status of a
well-formulated model \cite{acm}. A similar phenomenological approach
was later used in a number of papers \cite{paschos}. A detailed
description of the Fermi motion in the framework of the $1/m_Q$
expansion was later given in \cite{motion,bsg} and in
\cite{randall}.

The ``Fermi motion'' in QCD has certain peculiarities which are absent
in the phenomenological models \cite{motion,roman}. The
distribution over the `primordial' Fermi momentum $F(\vec{p}\,)$ is
replaced by a certain distribution function $F(x)$, where $x\le 1$
measures the momentum of the $b$ quark in the units of $M_B-m_b$.
First distinction is that $F(x)$ is one-dimensional;
one can define only the distribution over a certain projection of the
momentum.

Second, $F(x)$ {\em depends} essentially on the final state quark mass
(more exactly, on its velocity). While it is the same for $b\ra u\,\ell
\nu$ and $b\ra s+\gamma$ where the final quark is ultrarelativistic, it
is rather different for $c$ quark in $b\ra c\,\ell \nu$ where it is
closer to a nonrelativistic particle.

Finally, $F(x)$ is actually normalization-point dependent.

Although $F(x)$ depends on the final state quark mass, there are still
certain relations between the moments of $F(x)$ at arbitrary mass. The
moments are expressed via the expectation values of the local heavy
quark operators; these relations are examples of the generic sum rules
for the decays of heavy flavors \cite{optical}. In particular, the
`center of gravity' of the distribution over the primordial momentum
points at the heavy quark mass $m_b$. The second moment of $F(x)$, its
dispersion, is proportional to the kinetic expectation value
$\mu_\pi^2$. The third moment is proportional to the Darwin term,
{\it etc}. Let me mention one fact \cite{roman,bsg} which is often
missed: the AC$^2$M$^2$ model \cite{acm} does correspond to a certain
QCD distribution function $F(x)$, the so-called `Roman' function which
however is not Gaussian. Moreover, the value of $\mu_\pi^2$ in this
model is {\em not} given by $p_F$ itself
but is a more complicated function of $p_F$ and $m_{\rm sp}$, the two
parameters of the AC$^2$M$^2$ model.

It appears that the effects of Fermi motion are very important for a
possible determination of $V_{ub}$. We know
certainly that $|V_{ub}| \neq 0$ holds since (i) the decays
$B \ra \pi \ell \nu$, $\rho \ell \nu$ have been identified and
(ii) $B \ra X\,\ell \nu $ has been observed with lepton energies $E_\ell$
that
are accessible only if $X$ does {\em not} contain a charm hadron:
$E_\ell \geq (M_B^2 - M_D^2)/2M_B = 2.31 \GeV$.
Translating these findings into reliable numbers
for
$|V_{ub}|$
is a much more difficult task theoretically: On the one hand
the exclusive decays $B \ra \pi \ell \nu$, $\rho \ell \nu$ depend on
bound-state effects in an essential way.
On the other hand in analyzing the endpoint spectrum for
charged leptons in the inclusive decays $B \ra X\, \ell \nu$ one
encounters
different sorts of systematic problems. Only a fairly small
fraction
of the charmless semileptonic decays $B \ra X_u \ell \nu$, namely
around $10\%$
or so, produce a charged lepton with an energy {\em beyond} that
possible for $B \ra X_{c} \ell \nu$.
The data suggest that $|V_{ub}|/|V_{cb}| \sim 0.1$, however the actual
estimates vary within a factor of two.
In addition, the $b\ra c$ rate is so
much bigger than that for $b\ra u$ that leakage from it due to
measurement errors becomes a serious background problem;
furthermore the endpoint region is particularly sensitive to
the nonperturbative dynamics.

Since the inclusive semileptonic width $\Gamma_{\rm sl}(b\ra u\ell\nu)$
is reliably calculated theoretically, Eq.~(\ref{22r}), the best way to
determine $|V_{ub}|$ would be to accurately determine
${\rm BR_{sl}}(b\ra X_u\ell\nu)$ where $X_u$ denotes the charmless
hadronic states. It is not yet completely clear how reliably this
fraction can be measured in experiment. Of course, the problem is not
statistics {\it per se}.

The most direct way to disentangle $b\ra u\,\ell\nu$  from $b\ra
c\,\ell\nu$ without tagging the secondary charm decay would be to study
the invariant mass of hadrons in the final state $M_X$:
\beq
\frac{{\rm d}}{{\rm d}M_{X}} \Gamma (B \ra X\, \ell \nu )\;, \qquad
M_X^2 = \left(\sum_i P_{\rm hadr}^{(i)}\right)^2\;.
\label{4mx}
\eeq
For the free quark decay one has $M_X^2\simeq 0$ in $b\ra u$ and
$M_X^2= m_c^2$ for the $b\ra c$ transitions. In actual decays the mass
can take values exceeding $m_\pi$ and $M_D$, respectively. The increase
in mass can originate both in the perturbative processes when a hard
gluon is emitted in the decay, or through soft bound-state or
hadronization processes. The leading soft effects emerge
due to same physics which gives rise to
the Fermi motion. They are quite significant. For example, the
average invariant mass square of hadrons in $b\ra u$ gets the
nonperturbative correction $\sim \Lam \cdot m_b$ \cite{WA}:
\beq
\aver{M_X^2}_{\rm nonpert} \:=\: \frac{7}{10} m_b\left(M_B-m_b\right)
\,+\,{\cal O}\left(\Lam^2\right)\;\simeq\;
1.5\GeV^2
\label{290}
\eeq
(the suggestion of \cite{cgg} about the absence of these relative
$1/m_Q$ corrections to $M_X^2$ was erroneous). The perturbative
corrections lead to $\aver{M_X^2}_{\rm pert} \sim \frac{\as}{\pi}
m_b^2$, however in the $B$ decays this increase is still smaller than
the nonperturbative corrections.

The above estimates already show that the broadening of the
$M_X$-distribution is quite essential; whether its impact is
insignificant or dramatic for discriminating the two
semileptonic decay channels, was not clear {\it a priori} and
required numerical analysis. It was done in \cite{dist}.
The charm quark mass happens to be just
near the borderline separating the domains of `heavy' quark
$m_c \gg \sqrt{\Lam m_b}$
where the effect of broadening is small, and of `light' quark $m_c \ll
\sqrt{\Lam m_b}$ when the kinematic difference between the inclusive
$b\ra u$ and $b\ra c$ decays is buried under the hadronization effects.

To quantify the effect of the strong interactions let us introduce,
following Refs.~\cite{dist,mx} the fraction of the $b\ra u$ events
with $M_X$ below a certain cutoff mass $M_{\rm max}$:
\beq
\Phi(M_{\rm max}) = \frac{1}{\Gamma (b\ra u)}
\int _0^{M_{\rm max}} {\rm d}M_X \:\frac{{\rm d}\Gamma}{{\rm d}M_X}\;.
\label{18mx}
\eeq
Clearly, $\Phi(0)=0$ and $\Phi(M_B)=1$ independently of strong
interactions. The main question to theory is whether it can calculate
accurately enough $\Phi(M_{\rm max})$  with $M_{\rm max}\lsim 1.5\GeV$.

The dedicated analysis was carried out in \cite{dist,mx}, and the
conclusion appeared to be quite optimistic -- the strong interaction
effects in $b\ra u \,\ell\nu$
are not expected to populate the domain above $M_X=1.5\GeV$ too
significantly.  The theoretical predictions for the spectrum can be
found in the above papers. Later a similar analysis was undertaken in
\cite{wisemx}.

As anticipated, the mass spectrum is broad and extends even
beyond $M_D$ -- yet only a small fraction does so, namely
$\sim 10\% $. Due to measurement errors there will be a tail from
$b\ra
c$ transitions {\em below} $M_D$. To avoid this leakage one  can
concentrate on recoil masses below a certain value  $M_{\rm max}$.
The actual choice of $M_{\rm max}$ is driven by  competing
considerations: the lower $M_{\rm max}$, the less leakage from
$b\ra c$ will occur -- yet the smaller the relevant statistics. Even
more important, the accuracy of the theoretical
predictions deteriorates at lower cutoff mass. The predictions and their
major uncertainties are illustrated in Figs.~5 taken from \cite{mx}.

\begin{figure}
\vspace{5.3cm}
\includegraphics{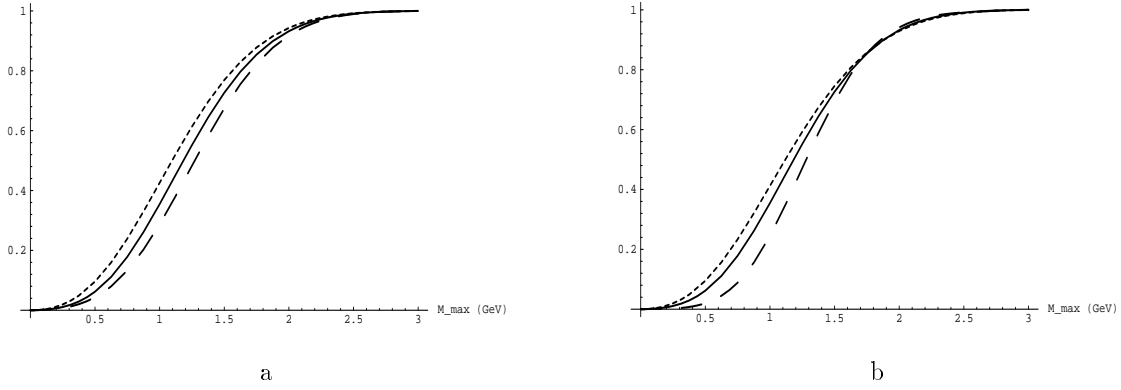}
\caption{\footnotesize
The integrated fraction of the $b\ra u \,\ell \nu$ events $\Phi(M_x)$.
({\bf a}): Dependence on $m_b$. Long-dashed, solid,
and short-dashed lines are $m_b(1\GeV)=4.58\GeV$, $4.63\GeV$
and $4.68\GeV$.
({\bf b}): Dependence on $\mu_\pi^2$. Long-dashed, solid,
and short-dashed lines correspond to $\mu_\pi^2 =0.3\GeV^2$,
$0.5\GeV^2$ and $0.7\GeV^2$.}
\end{figure}

This analysis showed that it seems important to reach a value of $M_x
\simeq 1.5\GeV$ in the reliable measurement of the fraction
$\Phi(M_{\rm max})$. If the necessary discrimination of $b\ra c$ decays
can be achieved at such masses, a rather precise model-insensitive
determination of $|V_{ub}|$ and/or $|V_{ub}|/|V_{cb}|$ will be
possible. Those who are interested in further details can find them in
Ref.~\cite{mx}.

I have to note that the conclusions of paper \cite{wisemx} were less
definite, and its predictions suffered from larger uncertainties. The
treatment of both the perturbative
and nonperturbative corrections was somewhat different;
I consider some elements of the analysis not quite consistent.

Sharpening the theoretical predictions for $\Phi(M_{\rm max})$ would
require, in particular, a more refined treatment of the
perturbative corrections.
In the $b\ra u \,\ell\nu$ or $b\ra s+\gamma$ distributions they are not
very simple since include strong bremsstrahlung corrections that must
be dealt with carefully. This will not be a stumbling block, however. A
certain approximation, the ``APS'' (Advanced Perturbative Spectrum) was
elaborated in \cite{bsg} which is feasible and, at the same time, must
yield more than enough precision in accounting for the perturbative corrections
to the distributions even in these transitions. If future studies
demonstrate the possibility of accurate measurements of the hadronic
invariant mass in the semileptonic decays and the anticipated accuracy becomes
clear, such refined calculations will be done.

Concluding this discussion, I want to remind that it is very desirable
to study such distributions {\em separately} for $B^{\pm}$ and $B^0$
mesons. The same refers also to studying the end-point region of the
lepton spectra. In this way it will be possible to directly study and
detect the so-called weak annihilation,
the nonperturbative effect
sensitive to the flavor of the spectator quark \cite{WA}. Weak
annihilation is a relatively small $1/m_b^3$ effect in the total
widths.  However, in the semileptonic transitions it mainly originates
from the end-point domain, $E_\ell\approx M_B/2$ and small $M_X$. Thus,
in these studies its relative effect is enhanced. It would have an
independent theoretical interest as well providing valuable information
for the HQE about the expectation values of the four-fermion operators,
in particular, their nonfactorizable pieces \cite{WA}.

\section{Challenges in the HQE}

Concluding my lectures, I cannot avoid mentioning places where the HQE
seems to have problems when compared to experiment. It is probably
premature to speak of a direct contradiction; nevertheless,
today's question marks carry the seeds of tomorrow's advances.
Basically there are two problems
where our theoretical understanding is lagging behind.
Both are related to nonleptonic decays.

\subsection{Semileptonic Branching Ratio of the $B$ Mesons and $n_c$}

The theoretical attitude to this problem changes with time.
Twenty years ago the parton model gave a prediction ${\rm
BR_{sl}}(B)\simeq 13\;$--$15\%$, which was accurate enough according to
the existed standards. The variation reflected mostly the choice of
quark masses. While the rates of $c\ra c\ell\nu$ and $b\ra c \bar{u}d$
were affected by the choice of $m_c$ in a more or less same way, the
rate of the $b\ra c \bar{c}s$ channel decreased much faster when
larger masses (closer to the `constituent' rather than short-distance
`current' ones) were used. Therefore, ${\rm BR_{sl}}(B)$ increases for
larger masses. It should be remembered that the difference of $m_b$ and
$m_c$ is fixed, so the choices of $m_c$ and $m_b$ are always
correlated. Even though the heavy quark symmetry had not been
formulated, the latter fact was clearly realized at least in the early
80's \cite{rueckl}.

The experimental situation was not very definite and indicated a rather
large ${\rm BR_{sl}}(B)$ which reasonably fitted the `larger' mass
option \cite{altp}. Since then it became more or less standard to use
the larger quark masses, and the parton model prediction for ${\rm
BR_{sl}}(B)$ was accepted to be $13\,$--$\,14\%$, even though a smaller
value could be obtained as well. Since the impact of the
nonperturbative
corrections was completely unknown and presumably larger, it was not
seriously discussed even when the better data became available.

The situation changed when in 1992 Bigi {\it et al.} showed that there
are no $1/m_Q$ nonperturbative corrections to the inclusive widths, both
semileptonic and nonleptonic. Moreover, even
the smaller $1/m_b^2$ nonperturbative effects
were readily calculated \cite{buv,bs,koytau}. They appeared to be
suppressed, in particular, as a result of certain cancellations. The
overall effect $\Delta_{\rm nonpert}{\rm BR_{sl}}(B)$ was found to be
about $-0.5\%$; the nonperturbative effects could not be blamed for a
discrepancy any more \cite{baffling}.

It prompted a more careful analysis of the perturbative corrections to the
widths. In particular,
the ${\cal O}(\as)$ corrections to the nonleptonic $b\ra
c\bar{u}d$ width were calculated accounting for the non-zero $m_c$
\cite{bagan}; this additionally enhanced the nonleptonic width.
Later the account for the charm mass was also completed for
$b\ra c\bar{c}s$ \cite{godz,volcc,lss}. Altogether, these ${\cal
O}(\as)$ corrections further decreased ${\rm BR_{sl}}(B)$ down to
$11\,$--$\,12\%$. Although this shift naively seems very significant
and may raise concerns about the convergence of the perturbative
corrections, it actually is not dramatic if one starts with more
appropriate short-distance masses, the choice forgotten for historical
rather than rational reasons. It is important to keep in mind that the
relative increase in the nonleptonic width was particularly
significant in the $b\ra c\bar{c}s$ channel.

The issue of the semileptonic branching ratio
must be considered in conjunction with the charm yield $n_c$,
the number of charm states emerging from $B$ decays.
To measure $n_c$ one assigns charm multiplicity {\em one}
to $D$, $D_s$, $\Lambda _c$ and $\Xi _c$ and {\em two} to
charmonia. Zero is assigned to the charmless hadronic final state.
It is obvious that
\beq
n_c \simeq 1 +\mbox{BR}(B \ra c \bar c s \,\bar q)
-\mbox{BR}(B \ra \mbox{no charm})\, .
\label{8.32}
\eeq
Such a joint analysis was motivated already in \cite{buv}: the energy
release in $b\ra c\bar{c}s$  is rather small, and this can lead to
significant duality-violating and higher-order effects. The stability
of the perturbative expansion also downgrades. Measuring $n_c$ allows
one to effectively exclude this channel from the theoretical
calculations.

The data on $n_c$ were not quite consistent for some time, but finally
seem to converge to -- unfortunately -- a smaller, more difficult
for theory value. If we take experiment and theory at face value, the
allowed intervals do not overlap.
The semileptonic fraction seems
too low, about $10.6\%$, which is possible to accommodate only at $n_c
\gsim 1.25$. Such a large number is not compatible with the CLEO
measurements.
It is not clear how it must be interpreted. At this School I heard
suggestions from the experimental colleagues whose opinion I highly
respect, that one may not take too closely the discrepancy within two,
or even within three quoted error bars. Taking such an optimistic (for
a theorist) attitude, one could have been relaxed. I cannot, however
make such a decision myself and have to discuss numbers as they are
officially quoted.

In my opinion, it is difficult to accommodate theoretically the value
of ${\rm BR_{sl}}(B)$ below $11.2\%$ if $n_c$ does not exceed $1.20$.
Those who say that it is easy to obtain it in the SM, probably stretch
uncertainties too liberally. Of course, even if eventually experiment
arrives at the current central values, we will not have to revise the
SM. One possibility would be that the second-order perturbative
corrections in
the nonleptonic modes are grossly underestimated and too significant.
Another -- probably, more realistic -- is the violation of
local duality in the nonleptonic widths; I'll return to it a bit later.
{\it A priori}, one could expect significant, $\sim 15\%$
nonperturbative effects in the widths. The actual problem lies in the
fact that the leading corrections are calculated and happen to be
small.\footnote{The coefficient for the leading nonperturbative
correction is calculated only in the so-called leading logarithmic
approximation. There are reasons to think that the additional
terms in the complete ${\cal O}(\as)$ coefficient can be as large. }

\subsection{Lifetimes of Heavy Flavor Hadrons}

As stated before, differences between meson and baryon decay widths
arise already in order $1/m_Q^2$. The perturbative corrections to the
lifetime ratios are completely absent. The lifetimes of the various
mesons get differentiated effectively first in order $1/m_Q^3$. A
detailed review can be found in \cite{stone2,BELLINI}.

Because the charm quark mass is not much larger than typical
hadronic scales one can expect to make only semi-quantitative
predictions on the {\em charm} lifetimes, in particular for the
charm baryons. The agreement of the predictions with the
data is good. I would even say it is too good keeping in mind that the
$1/m_c$ expansion can hardly be justified.

As far as the {\em beauty} lifetimes are concerned the $1/m_b$
expansion is to be applicable.
Table \ref{TABLE20} contains the world averages
of published data
together with the
predictions. The latter were actually made before data (or data of
comparable sensitivity) became available.

Data and predictions on the meson lifetimes are completely and
non-trivially  consistent. Yet even so, a  comment is in order for
proper  orientation. The numerical prediction is based on the
assumption of factorization at a typical hadronic scale which is
commonly taken as the one where $\as(\mu_{\rm hadr})\simeq 1$.
While there is no justification
for factorization at $\mu\sim m_b$, there exists ample
circumstantial evidence in favor of approximate factorization  at a
typical hadronic scale -- from the QCD sum rule calculations, to
lattice evaluations, to $1/N_c$ arguments. More to  the point,  the
validity of factorization can be probed in semileptonic  decays of $B$
mesons in an independent way, as  was pointed out in \cite{WA}.

The possible effect of the nonfactorizable contribution has been
discussed in detail in \cite{WA,Ds}. They include also the nonvalence
gluon mechanism discussed long ago in \cite{fritmink} in the simplified
language of the quark model.
Significant nonfactorizable contributions
would in general lead to large effects of weak annihilation in $D_s$
mesons where experimentally such effects are quite small. Of course,
we cannot reliably treat $\Gamma_D$ in the $1/m_Q$ expansion, and
$\Gamma_{D_s}-\Gamma_{D^+}$ is sensitive to a particular combination of
the expectation values of a few four-fermion operators. Nevertheless,
it is a serous indication that the effects due to nonfactorizable
expectation values must be suppressed. Thus, the estimates of the
preasymptotic corrections to the widths based on factorization, most
probably, are valid within a $50\%$ accuracy. The actual uncertainty in
$f_B^2$ is not much smaller; it includes the subleading $1/m_b$ effects
which formally belong to neglected $1/m_b^4$ corrections to
$\Gamma_B$.

The recent paper \cite{baek} attempted to estimate the nonfactorizable
expectation values in $B$ mesons in QCD sum rules. Although the
calculations were carried out in a too simplified manner and cannot be
really considered as predictions, they showed that there hardly exists
room for large nonfactorizable contributions, which could yield drastic
effects conjectures in \cite{ns2}. In particular, I must note that the
possible interval $0.8$ to $1.4$ for $\tau_{B^-}/\tau_{B^0}$ claimed
by Neubert \cite{neubhaw} is unacceptable. While one cannot guarantee
{\it a priori} that the onset of the $1/m_Q$ expansion is safely below
$m_b$ and the latter can be applied to actual $b$ hadrons, it
is certain that such large deviations {\em cannot} be obtained in the
framework of the self-consistent $1/m_Q$ expansion of the width itself.
The quoted interval emerged merely as a result of reckless
manipulation with {\it ad hoc} introduced
hadronic parameters. Their values saturating such corrections would
violate certain bounds \cite{boost} which are analogous to the
unitarity constraints. In any case, these flavor-dependent $1/m_Q^3$
effects under discussion originate only in particular quark decay
channels. It is clear that the preasymptotic effects must constitute
only a small fraction of the widths in order to be amenable to the
$1/m_Q$ expansion.

\begin{table}[t]
\caption{QCD Predictions for Beauty Lifetimes
\label{TABLE20}}
\begin{center}
\begin{tabular} {|l|l|l|l|}
\hline
Observable &QCD Expectations ($1/m_b$ expansion)& Ref. &
Data from \cite{BELLINI}\\
\hline
\hline
$\tau (B^-)/\tau (B_d)$ & $1+
0.05(f_B/200\, {\rm MeV} )^2 $ & \cite{mirage} & $1.04 \pm 0.04$ \\
\hline
$\bar \tau (B_s)/\tau (B_d)$ &$1\pm {\cal O}(0.01)$ &
\cite{stone2}
&  $ 0.97\pm 0.05$ \\
\hline
$\tau (\Lambda _b)/\tau (B_d)$&$\gsim 0.9 $ & \cite{stone2} &
$0.77\pm 0.05$ \\
\hline
\end{tabular}
\end{center}
\end{table}

The agreement of the data on $B$ meson lifetimes with experiment is
obscured by the apparent conflict for $\tau_{\Lambda_b}/\tau_{B_d}$.
To predict the $1/m_Q^3$ corrections to $\tau_{\Lambda_b}$ in the
$1/m_b$ expansion one needs to evaluate the baryonic expectation values
of two operators,
\beq
\matel{\Lambda_b}{\bar b b\,\bar{u}\gamma_0 u }
{\Lambda_b}\;,\qquad \matel{\Lambda_b}{\bar b \frac{\lambda^a}{2}b\,
\bar{u}\gamma_0  \frac{\lambda^a}{2}u}{\Lambda_b}\;.
\label{305}
\eeq
They do not have the usual factorizable contribution, and their values
are rather uncertain. Nevertheless, it was shown that their
contributions cannot be too large \cite{boost} and the maximal
effect in $\Gamma_{\Lambda_b}$ does not exceed $10\,$--$\,12\%$.
To achieve the larger corrections one would have to go beyond a usual
description of baryons when light quarks are ``soft''.  This agrees
with the fact that the constituent quark model estimates typically
yield about $3$ to $5\%$ enhancement \cite{rosner}.  A similar
conclusion has been reached by the authors of Ref. \cite{BARI} who
analyzed the relevant baryonic matrix elements through QCD sum rules.

We see that there are indications of disagreement between theory and
experiment, at the level of at least two quoted experimental error
bars. The opinion is sometimes expressed that before a disagreement
reaches three standard deviations, it is premature to start worrying. I
am not sure that it fully applies. On the one hand, the literal
theoretical expectations for $\tau_{\Lambda_b}/\tau_{B_d}$ concentrate
closer to $0.95$, and the quoted limit $0.9$ is not a central
theoretical value.  On the other hand, these theoretical predictions
were known already for a number of years \cite{stone2}, so I hope that
the experiments have analyzed
obvious possible biases in the data and corrected those which could
generate the enhanced gap.
Neither argument is convincing by itself, but together they give me a
feeling that the disagreement does exist.

If it will be firmly established in the future experiments, I think
that in the framework of the SM the most probable explanation is the
violation of duality. As a matter of fact, its significance in
nonleptonic widths is theoretically expected {\it a priori}. Indeed, the
expansion parameter for the widths is not $m_b$ directly but rather the
energy release which is noticeably smaller in $b\ra c$. Moreover, the
preasymptotic corrections depend on the concrete form of the weak
interaction involved. For the four-fermion interaction they are
enhanced:  the large-$5$ arguments \cite{five} mentioned earlier
suggest that the actual scale parameter is smaller, $\sim E_{\rm
rel}/5$. In the semileptonic decays this does not deteriorate the
expansion since it is automatically protected by the heavy quark
symmetry when $m_c$ increases.
Even if we keep the invariant mass of leptons $\sqrt{q^2}$ fixed, the
corresponding quark width remains correct at least up to $1/m_Q$ terms
for arbitrary, for example, maximal $q^2$ when the energy release is
tiny, see Eq.~(\ref{224}).

For the nonleptonic decays the heavy quark symmetry does not generally
apply, and at insufficient energy release one expects significant
violations of duality.
A few years ago, therefore, we would not be too much
concerned with present discrepancies. Now, on the other hand, we
have evaluated the leading terms in the $1/m$ expansion of the width
and did not find indications for significant effects. Since
violation of duality is conceptually related to the asymptotic nature
of the OPE \cite{shiftasi,inst}, we could have expected that duality
works at a few percent level as well.

\subsection{$1/m_Q$ Expansion and Duality Violation}

The problem of duality violation attracts more and more attention of
those who study the heavy quark theory; a recent extensive discussion
was given in \cite{inst}. The expansion in $1/m_Q$ is asymptotic. There
are basically two questions one can ask here: what is the onset of
duality, i.e. {\em when} does the
expansion start to work? The most straightforward approach was undertaken in
\re{bm}, and no apparent indication toward an increased energy scale was
found. Another question, {\em how} is the equality of the QCD
parton-based predictions with the actual decay rates achieved, was barely
addressed. Though a relevant example of such a problem is easy to
give.

The OPE ensures that no terms $\sim 1/m_Q$ can be in the widths in QCD
and the
corrections start with $1/m_Q^2$. However, the OPE {\em per se} cannot
forbid a
scenario where, for instance,
\beq
\frac{\delta \Gamma_{H_Q}}{\Gamma_{H_Q}} \;\sim\ \;C \;
\frac{\sin{(m_Q \rho)}}{m_Q\rho}
\;,\;\;\; \rho \sim \Lam^{-1}
\;.
\label{44crad}
\eeq
In the actual strong
interaction, $m_b$ and $m_c$ are fixed and not free parameters, so,
from the practical viewpoint these types of corrections are
not too different -- but the difference is profound in the theoretical
description! It reflects specifics of the OPE in Minkowski space, and such
effects can hardly be addressed, for example, in the lattice
simulations. Their control requires a deeper understanding of the
underlying QCD dynamics beyond the knowledge of first few
nonperturbative condensates.

In fact, the literal corrections of the type of \eq{44crad} are hardly
possible; the power of $1/m_Q$ in realistic scenarios is larger, and
these duality corrections must be eventually exponentially suppressed
though, probably, starting at a higher scale \cite{inst}. But a theory
of such effects is still in its embryonic stage and needs an additional
experimental input as well.

Nevertheless, we know that the duality-violating terms in the HQE are
not completely arbitrary, and they must obey certain constraints. For
example, they cannot be monotonous and positive or negative but
oscillate, {\it etc}. All these exact properties are often missed in
the theoretical discussions.

It is worth adding that the folklore about violations of duality is
often rather superficial. Even the used terminology is not always
consistent. It was written recently, for example, in \cite{bjhaw} that
the assumptions behind the $1/m_Q$ expansion for the widths, in
particular, for the nonleptonic widths, include a certain factorization
of the decay amplitudes. In fact, such a factorization is not required
at all, like it is not required in the asymptotic expansion of $R_{\rm
e^+e^-}(s)$. The only assumption is the usual
factorization of contributions of large-distance and short-distance
physics, the OPE. Moreover, in principle it is enough to assume it only
in Euclidean domain. A stronger assumption of such a factorization
for actual decay processes in Minkowski kinematics at fixed energy
would lead to additional constraints on violation of local duality.

It has been quite popular to state that the quark-hadron duality sets
in only when the number of hadrons in the final state becomes large
\cite{dualnum}. This is a wrong criterium. The
semileptonic decays in the SV kinematics are a nice example. In the
limit
\beq
m_b, \,m_c\;\gg \Lam\;, \qquad m_b-m_c \gg \Lam\,, \qquad
\frac{m_b-m_c}{m_b+m_c} \sim v \ll 1
\label{280}
\eeq
the exact quark-hadron duality perfectly holds with the theorem-like
rigorousness while the total decay rate is saturated by only two final
states, $D^*$ and $D$ (even only one state can be retained if special
weak currents are considered).
Simultaneously, the yield of the excited states  can be arbitrary
small: it is driven by two parameters $\left(\Lam/m_c\right)^2 \ll 1$
and $\left(\frac{m_b-m_c}{m_b+m_c}\right)^2 \ll 1$. This fact was
actually noted in \cite{vshqs} in the mid 80's and was among the first
steps in the development of the heavy quark symmetry and dynamic
$1/m_Q$ expansion. It is really surprising why this classic result used
to be ignored even in the relatively recent papers.

On the other hand, the duality violation has peculiarity in decays of
heavy quarks. Apart from the usual assumption of local duality, the
practical applications of the OPE expansion relies here on the property
of {\it global} duality, which can also be violated to a certain
extent. An explanation of what is global duality would lead me too far
astray; it can be found in Ref.~\cite{optical}, Sect.~IV.A where this
notion was introduced. Let me only note that since then this term has
often been abused in the literature. For example, it is tavtological to
discuss the violations of global duality in $R_{\rm e^+e^-}$ or in the
hadronic $\tau$ decay width (more accurately, it would be merely
verifying the dispersion relation). Everything there is local duality
and the global duality is a very simple statement, a direct consequence
of analyticity and unitarity.  \vspace*{-.07cm}

I come to the end of my lectures without a formal concluding section.
Theory of heavy quarks has become a well-developed field of QCD. Not
only it benefits from the power of theoretical methods elaborated in
QCD over 25 years of evolution; the application of the heavy quark
expansion provides feedback for the theory of QCD itself. A few slices
of the heavy quark theory I discussed here demonstrate that it is an
actively evolving field where many interesting questions still are to
be understood. Without doubts, the data from the new generation of
experiments will provide the new impetus; we all are looking
forward to new fascinating surprises.
\vspace*{-.07cm}\\

{\bf Acknowledgments:} \hspace{.1em} It is a pleasure to thank the
organizers of the School in Varenna for the nice and creative
atmosphere and interesting program.  My special thanks to I.~Bigi and
L.~Moroni, and to M.~Artuso and S.~Stone for informal discussions.
Working on the heavy quark theory I always benefitted from close
collaboration with colleagues from Minnesota M.~Shifman, A.~Vainshtein
and M.~Voloshin. The presentation of Sect.~3.2.3 was to a large extent
shaped by discussions with K.~Melnikov and O.~Yakovlev; I also thank
A.~Czarnecki for useful comments.  The discussions with S.~Brodsky,
Yu.~Dokshitzer and A.~Mueller on certain aspects of QCD are gratefully
acknowledged.

This work was supported in part by NSF under the grant number
PHY~96-05080 and by RFFI grant 96-15-96764.

\end{document}